\documentclass[%
    prx,
    twocolumn,
    superscriptaddress,
    longbibliography,
    nofootinbib,
]{revtex4-2}

\usepackage{graphicx}
\usepackage{booktabs}
\usepackage{amsmath}
\usepackage{amssymb}
\usepackage{algorithmic}
\usepackage{hyperref}

\usepackage{dcolumn}
\usepackage{bm}

\hypersetup{
  colorlinks = true,
  urlcolor = blue,
  linkcolor = blue,
  citecolor = blue
}

\usepackage[ruled,vlined,linesnumbered]{algorithm2e}
\DontPrintSemicolon

\SetKwComment{Comment}{$\triangleright$ }{}
\SetKwInput{KwInput}{Input}
\SetKwInput{KwOutput}{Output}

\SetKw{Continue}{continue}
\SetKw{Break}{break}
\SetKw{Return}{return}

\newcommand*{\ftname}{FT-Weave}

\begin{document}


\title{\textbf{FT-Weave: Real-Time Compilation Framework for Reconfigurable Fault-Tolerant Quantum Architectures} 
}%

\author{Wan-Hsuan Lin}
 \email{Contact author: wlin@quera.com}
 \affiliation{University of California, Los Angeles, CA 90095, USA}
 \affiliation{%
 QuEra Computing Inc., Boston, MA, USA
}
\author{Milan Kornja\v{c}a}%
\affiliation{%
 QuEra Computing Inc., Boston, MA, USA
}%
\author{Chen Zhao}%
\affiliation{%
 QuEra Computing Inc., Boston, MA, USA
}%
\author{Sheng-Tao Wang}%
\affiliation{%
 QuEra Computing Inc., Boston, MA, USA
}%
\author{Jason Cong}%
\affiliation{University of California, Los Angeles, CA 90095, USA}%

\date{\today}

\begin{abstract}
Fault-tolerant quantum computing (FTQC) is essential for large-scale quantum computation, 
but realizing useful application throughput requires coordinating resource preparation, assignment, routing, and logical execution under strict hardware and timing constraints.
Many FTQC compilation approaches construct offline schedules using nominal or fixed magic-state factory throughput. 
Such schedules cannot respond to stochastic resource-preparation and teleportation outcomes, leading to execution stalls and hardware underutilization.
In this work, we introduce \ftname, a stage-aware real-time FTQC compilation framework that jointly coordinates resource preparation, resource assignment, teleportation routing, and correction handling.
By adapting to runtime resource availability and hardware constraints, \ftname\ allows preparation, communication, and logical execution to overlap.
We instantiate \ftname\ on two representative neutral-atom, early FTQC architectures: transversal STAR and a $T$-state cultivation architecture.
Under the evaluated hardware and latency model, 
\ftname\ achieves a speedup of up to $3\times$ over a baseline compilation flow for simulations of the two-dimensional transverse-field Ising model.
In the case study, 
we further find that maximizing exposed concurrency does not necessarily minimize execution time. 
Although fine-grained asynchronous execution can reduce local idle time, its smaller optimization windows and increased routing contention can outweigh these gains. 
Together, these results show that effective runtime coordination, rather then exposed parallelism alone, determines how efficiently FTQC resources translate into application throughput: \ftname{} provides a blueprint for solving this real-time orchestration problem across resource protocols and architectures.
\end{abstract}

\maketitle


\section{Introduction}
\label{sec:intro}

\begin{figure*}[htbp]
    \centering
    \includegraphics[width=\linewidth]{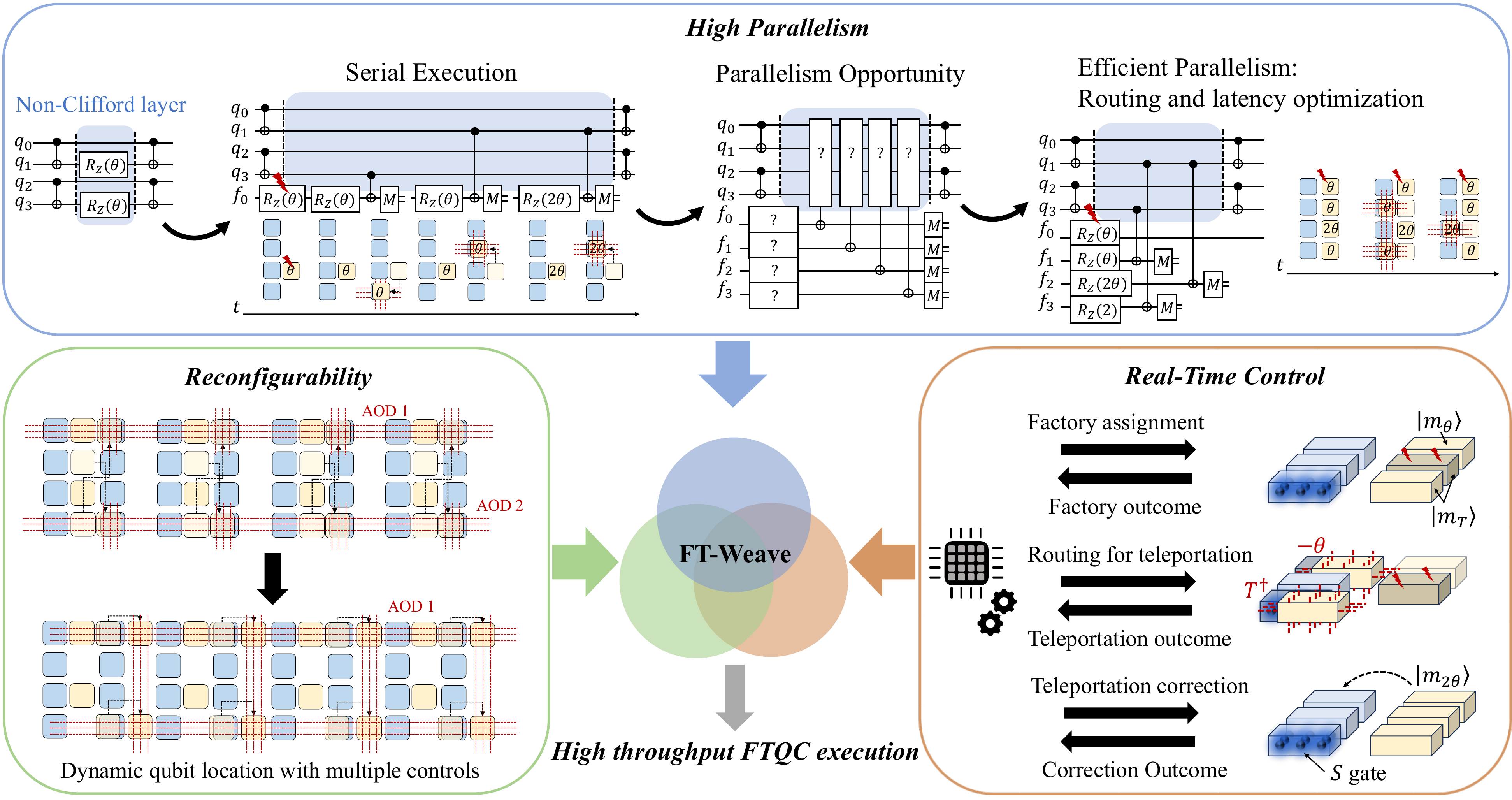}
    \caption{\textbf{Three interdependent capabilities for efficient fault-tolerant quantum computing (FTQC) execution.}
    (Top) High parallelism overlaps resource preparation, routing, and logical execution, whereas stage serialization leaves hardware underutilized.
(Bottom left) Reconfigurability adapts qubit layouts and control regions to changing computation and communication demands.
(Bottom right) Real-time control responds to stochastic preparation and teleportation outcomes by updating resource assignment, routing, and correction handling.
Because a bottleneck in any one capability can limit the others, reconfigurability, real-time control, and high parallelism must be optimized jointly, motivating the unified \ftname\ compilation framework.
    }
    \label{fig:realtime_parallelism}
\end{figure*}

Fault-tolerant quantum computing (FTQC) is critical for executing large-scale quantum computation reliably.
Recent advances in quantum hardware~\cite{google2024,Manetsch_2025,ransford2025helios98qubittrappedionquantum,evered2026highfidelityentanglinggatesnonlocal}, quantum error correction~\cite{Williamson_2026,gidney2024magicstatecultivationgrowing,claes2025cultivatingtstatessurface,chen2025efficientmagicstatecultivation,sahay2026foldtransversalsurfacecodecultivation,PhysRevLett.134.090603,eickbusch2025demonstratingdynamicsurfacecodes,zhao2026ultrahighratequantumerrorcorrection,correlated_decoding}, 
and logical-qubit demonstrations~\cite{Bluvstein_2023,google2024,paetznick2024demonstrationlogicalqubitsrepeated,Bluvstein_2025,zhang2025demonstrating,dasu2026computingencodedlogicalqubits} are bringing early fault-tolerant quantum computing (EFTQC) closer to reality.
Scaling these systems, however, requires more than increasing physical-qubit counts: architectures, compilers, and runtime systems must coordinate resource preparation, communication, logical operations, and error correction throughout execution.

As illustrated in \autoref{fig:realtime_parallelism}, efficient FTQC execution depends on three tightly coupled capabilities: \emph{reconfigurability}, \emph{real-time control}, and \emph{high parallelism}.
Reconfigurability adapts hardware resources and communication patterns to changing demands; real-time control coordinates these activities in response to stochastic outcomes; and high parallelism enables concurrent resource preparation, routing, and logical execution. 
Bottlenecks in any one can limit end-to-end performance, so the three capabilities must be optimized jointly.

First, reconfigurability enables parallel execution by adapting qubit layouts and communication resources to changing execution demands.
Neutral-atom platforms are particularly well suited to this model because they support flexible qubit arrangements and dynamic control~\cite{Bluvstein_2022,Bluvstein_2023,Evered_2023}.
Movable qubits, reconfigurable interaction regions, and dynamically steerable laser resources can redistribute computation and communication resources during execution~\cite{Bluvstein_2025,cain2026shorsalgorithmpossible10000,ismail2025transversal,Sales_Rodriguez_2025}.
Without this flexibility, static layouts and fixed communication paths can constrain resource movement and limit exploitable concurrency.


Second, real-time control is necessary because resource-preparation and teleportation outcomes are not known at compile time.
Beyond the real-time feedback already required for quantum error correction~\cite{RevModPhys.87.307,PhysRevX.11.041058,caune2024demonstratingrealtimelowlatencyquantum}, FTQC runtimes must track resource availability, assign prepared states to logical operations, route them to their destinations, and handle teleportation corrections.
Because these stochastic events determine which operations can proceed, execution plans must adapt continuously at runtime.

Third, high parallelism is necessary to convert available hardware resources into application throughput~\cite{Litinski2019gameofsurfacecodes}. 
Overlapping logical operations, resource preparation, routing, and error correction can hide preparation latency, but only when scheduling, resource allocation, and communication are coordinated. Otherwise, contention and serialization can leave available hardware underutilized. 

As summarized in \autoref{fig:realtime_parallelism}, these three capabilities are interdependent. 
Greater parallelism increases communication pressure; 
reconfigurability can alleviate this pressure but introduces additional control choices; 
and real-time control determines when and where resources are prepared, moved, and consumed. 
Consequently, optimizing any one capability in isolation may merely shift the bottleneck elsewhere. 
Efficient FTQC execution therefore requires an integrated orchestration layer that jointly coordinates preparation, assignment, communication, and logical execution under stochastic outcomes and reconfigurable connectivity.

Prior system-level work addresses individual parts of this orchestration problem, but does not provide such an integrated solution. 
Resource-estimation studies increasingly model realistic resource-generation constraints, including stochastic magic-state production, but do not translate runtime outcomes into hardware-level placement, routing, and control decisions~\cite{awasthi2026pricepayoffnondeterminismfault,leblond_lattice_surgery,flasq}.
Architecture-specific compilers generate execution schedules and hardware instructions , but typically optimize a fixed subset of decisions, such as scheduling, placement, routing, and resource management, within a particular execution model~\cite{dascot,lsqca,qbrige,beverland2022assessingrequirementsscalepractical,zhu2026o3lsoptimizinglatticesurgery}.
Other runtime systems explicitly support stochastic resource production, but focus on particular protocols or application settings, such as distillation or continuous-angle  rotation~\cite{magicpool,rescq,puremagic,Harvest,vqe_star,kurita2026generalcircuitcompilationprotocol,93zr-1ykb}.
These solutions remain closely tied to particular combinations of resource-generation protocol, communication mechanism, and architecture, and typically address only a subset of the runtime decisions required for end-to-end execution.

Despite their architectural differences, these schemes repeatedly encounter the same decisions: 
what resources to prepare, which factories should serve each logical operation, how prepared resources should be routed, and how hardware resources should be allocated as execution unfolds. 
What is missing is a common compilation abstraction that coordinates these decisions across the complete execution flow while allowing architecture-specific protocols and constraints to be incorporated through well-defined interfaces.

In this paper, we present \ftname, a stage-aware real-time compilation framework that provides a blueprint for solving the FTQC orchestration problem under high parallelism, stochastic execution, and reconfigurable connectivity.
\ftname\ decomposes execution into four recurring stages: resource preparation, resource assignment, routing, and correction handling.
It further partitions compilation into static and dynamic phases: the static compiler determines long-term plans and stage dependencies, while the runtime resolves decisions driven by resource availability and stochastic outcomes.
This separation defines common interfaces between stages while allowing each architecture to supply its own protocols, constraints, and optimization policies.

We instantiate \ftname\ on two representative neutral-atom EFTQC architectures~\cite{beyond_NISQ,Preskill2018quantumcomputingin}: transversal STAR~\cite{PRXQuantum.5.010337,PhysRevX.15.021057,starv3,ismail2025transversal} and a $T$-state cultivation architecture~\cite{gidney2024magicstatecultivationgrowing,chen2025efficientmagicstatecultivation,claes2025cultivatingtstatessurface,sahay2026foldtransversalsurfacecodecultivation}.
Although STAR prepares angle-dependent rotation states and cultivation prepares $T$ states through a staged protocol, both require online coordination of stochastic preparation outcomes, resource assignment, routing, teleportation, and correction handling.
Expressing both within the same stage interfaces shows that \ftname\ can accommodate distinct resource protocols without redesigning the overall compilation flow.

Besides providing a blueprint to execution orchestration,
\ftname\ provides a common testbed for evaluating execution policies and architectural trade-offs.
By varying runtime policies and hardware configurations, we quantify how resource preparation, assignment, routing, and hardware constraints interact to determine end-to-end performance.
Under the evaluated hardware and latency model, \ftname\ achieves a speedup of up to $3\times$ over the baseline compilation flow for simulations of the two-dimensional transverse-field Ising model (TFIM).
In the transversal STAR study, coordinated synchronous execution can outperform fine-grained asynchronous execution: although asynchronous control exposes additional concurrency, its smaller optimization windows can increase routing contention and coordination overhead.
Together, these results show that runtime coordination, rather than parallelism alone, can determine how effectively nominal hardware resources translate into application throughput.
They also demonstrate the utility of \ftname\ for hardware--software co-design.

The remainder of this paper is organized as follows.
\autoref{sec:flow} introduces the compilation framework and its static--dynamic partition; \autoref{sec:eftqc} reviews the target EFTQC architectures; \autoref{sec:compilation_eftqc} describes the architecture-specific compilation strategies; \autoref{sec:ft_weave_output} defines the compiled execution; \autoref{sec:execution_policy} compares synchronous and asynchronous policies; and \autoref{sec:result} presents the evaluation.
\section{\ftname \ Compilation Framework}
\label{sec:flow}

\begin{figure}
    \centering
    \includegraphics[width=\linewidth]{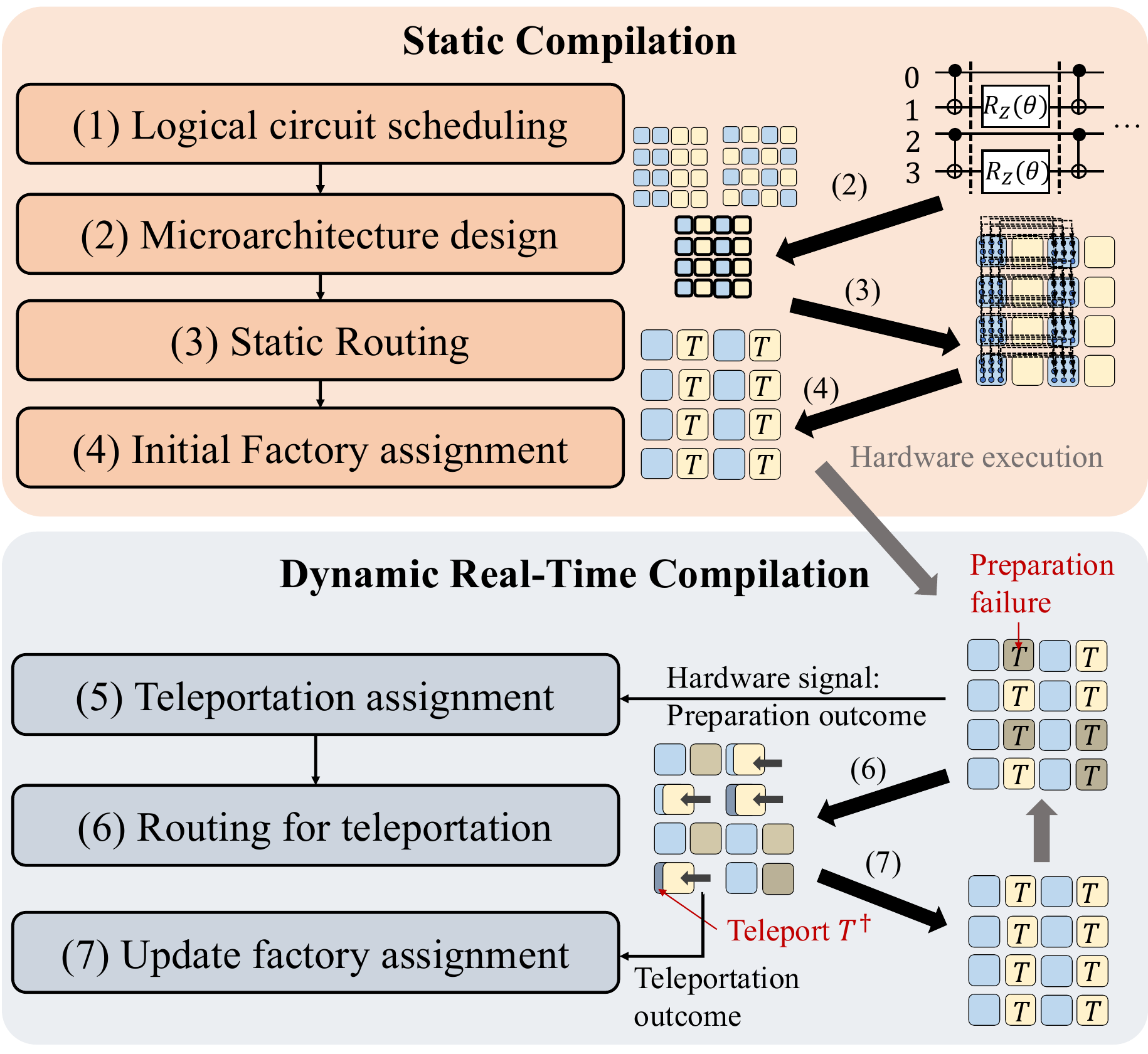}
    \caption{\textbf{Overview of the \ftname\ compilation framework.}
\ftname\ partitions compilation into four static and three dynamic steps.
The static phase performs
(1) \emph{logical circuit scheduling}, which groups logical operations into Clifford and non-Clifford layers;
(2) \emph{microarchitecture design}, which determines the organization of data blocks, routing space, and factory resources;
(3) \emph{static routing}, which determines communication paths known before execution; and
(4) \emph{initial factory assignment}, which associates upcoming non-Clifford operations with candidate factory outputs.
In the running example, the scheduled circuit contains both Clifford and non-Clifford layers, and the high resource demand leads the compiler to select a layout that interleaves data and factory regions.
During execution, preparation outcomes initiate the runtime feedback loop.
The dynamic phase then
(5) assigns an available prepared resource state to a requesting logical operation;
(6) computes a feasible route for teleportation; and
(7) incorporates the teleportation outcome and any resulting correction demand into the runtime state, revising future factory assignments.
}
    \label{fig:compilation}
\end{figure}

Rather than targeting a specific FTQC protocol, 
\ftname\ provides a stage-aware real-time compilation framework for architectures that rely on probabilistic resource preparation and runtime coordination. 
As shown in \autoref{fig:compilation}, 
\ftname\ organizes FTQC execution around four recurring stages: 
resource preparation, resource assignment, routing, and correction handling. 
These stages define interfaces between compilation and execution rather than prescribing their implementation, 
allowing architecture-specific protocols and optimization policies to be incorporated within a common compilation flow.

\ftname\ further partitions the decisions within each stage between static and dynamic compilation phases. 
Decisions determined by the circuit and architecture are optimized offline, 
whereas decisions that depend on probabilistic execution outcomes are deferred to runtime. 
This partition preserves long-range optimization opportunities while allowing the execution plan to adapt to hardware feedback.
Throughout this section, we use \emph{execution stages} to refer to the four conceptual components introduced above—resource preparation, resource assignment, routing, and correction handling—\emph{compilation phases} to distinguish static and dynamic decision-making, and \emph{compilation steps} to denote the seven numbered operations in \autoref{fig:compilation}.

\paragraph{Static compilation.}
The static compilation phase constructs the base execution plan before program execution begins. 
Because circuit operations and their dependencies are known in advance, 
their scheduling and predictable communication structure can be determined offline.
The four static compilation steps in~\autoref{fig:compilation} make this process concrete. 
In (1) \emph{logical circuit scheduling}, 
the compiler topologically orders logical operations and groups mutually independent operations into executable Clifford and non-Clifford layers.
In (2) \emph{microarchitecture design}, the compiler maps these operations onto a physical organization that specifies the locations of data blocks, routing space, and factory resources. 
In (3) \emph{static routing}, the compiler determines communication paths that are known before execution, such as those used for deterministic Clifford operations.
Finally, in (4) \emph{initial factory assignment}, upcoming non-Clifford operations are associated with candidate factory outputs, such as the resource states used to implement the two $R_Z(\theta)$ operations in the example.
Because resource preparation is stochastic, these assignments are provisional: the static compiler establishes an optimized initial plan without assuming that every assigned resource will be available when needed or that each teleportation will succeed.

The example in \autoref{fig:compilation} illustrates these steps.
The CNOT operations are organized into Clifford layers, 
while the two $R_Z(\theta)$ operations form a non-Clifford layer whose execution requires runtime coordination.
Given the high demand for non-Clifford resource states, 
the compiler selects a microarchitecture that interleaves data and factory regions to reduce routing overhead and initially associates the non-Clifford operations with nearby candidate resource states.
Together, these offline decisions establish the execution structure from which runtime adaptation begins.

\paragraph{Dynamic compilation.}
During hardware execution, \ftname\ transitions to the dynamic compilation phase. 
Runtime decisions depend on stochastic events, including resource-preparation outcomes and repeat-until-success teleportation outcomes.
Because these events determine which resources are available, where they must be routed, and which operations can execute next, the associated assignment, routing, and scheduling decisions cannot be fully resolved before execution.

Hardware outcomes close the feedback loop between the quantum processor and the \ftname\ runtime.
The runtime maintains the current execution state, 
including prepared-resource locations, preparation outcomes, routing availability, and pending logical operations, and updates the execution plan as new outcomes arrive.
The three dynamic compilation steps in \autoref{fig:compilation} illustrate this process.
When the hardware reports the preparation outcome, 
(5) \emph{teleportation assignment} selects an available prepared resource for the logical operation.
Given this assignment, (6) \emph{routing for teleportation} finds a feasible path from the selected resource to the data block while respecting the routing resources currently occupied by other operations. 
After teleportation, the measurement outcome may complete the logical operation or generate a corrective operation with a new resource demand.
Then, (7) \emph{update factory assignment} incorporates the outcome and any resulting resource demand into the controller state, revising subsequent factory-to-operation assignments.
The resulting execution forms a closed loop: preparation outcomes trigger resource assignment, assignments determine routes, and teleportation outcomes update future resource demand.

\paragraph{Real-time latency.}
Because dynamic decisions lie on the execution critical path, the latency of the complete hardware--software--hardware feedback loop can directly affect end-to-end circuit runtime.
We therefore define real-time latency to include: 
(i) acquisition of measurement results from the quantum hardware; 
(ii) signal propagation and control-system transport between the hardware and runtime controller; 
(iii) decoding and other classical processing required to interpret the measurement record and determine the relevant logical outcome; 
(iv) real-time compilation, including state updates, resource reassignment, routing, and generation of the next executable instructions; 
and (v) delivery of those instructions through the control stack back to the hardware.
Although decoding may dominate this latency budget in some systems, all five components contribute to the feedback delay.
Reducing this latency is important because delayed feedback can increase qubit idle time and reduce execution throughput.

\paragraph{Execution policies and architectural specialization.}
A key feature of \ftname\ is the separation between the execution framework and the policies used to implement it.
The stage abstraction supports both synchronous layer-oriented execution and event-driven asynchronous execution.
Because resource preparation, resource assignment, routing, and correction handling are expressed through the same stage interfaces, different execution policies can be implemented and evaluated under common hardware assumptions and a common runtime model.
This allows \ftname\ to isolate the trade-off between execution parallelism and coordination overhead.
The same modularity supports different FTQC architectures by specializing the implementation of individual execution stages rather than redesigning the overall compilation flow.

\section{EFTQC Execution Models and Architecture}
\label{sec:eftqc}

\begin{figure}[t]
\centering
\includegraphics[width=\linewidth]{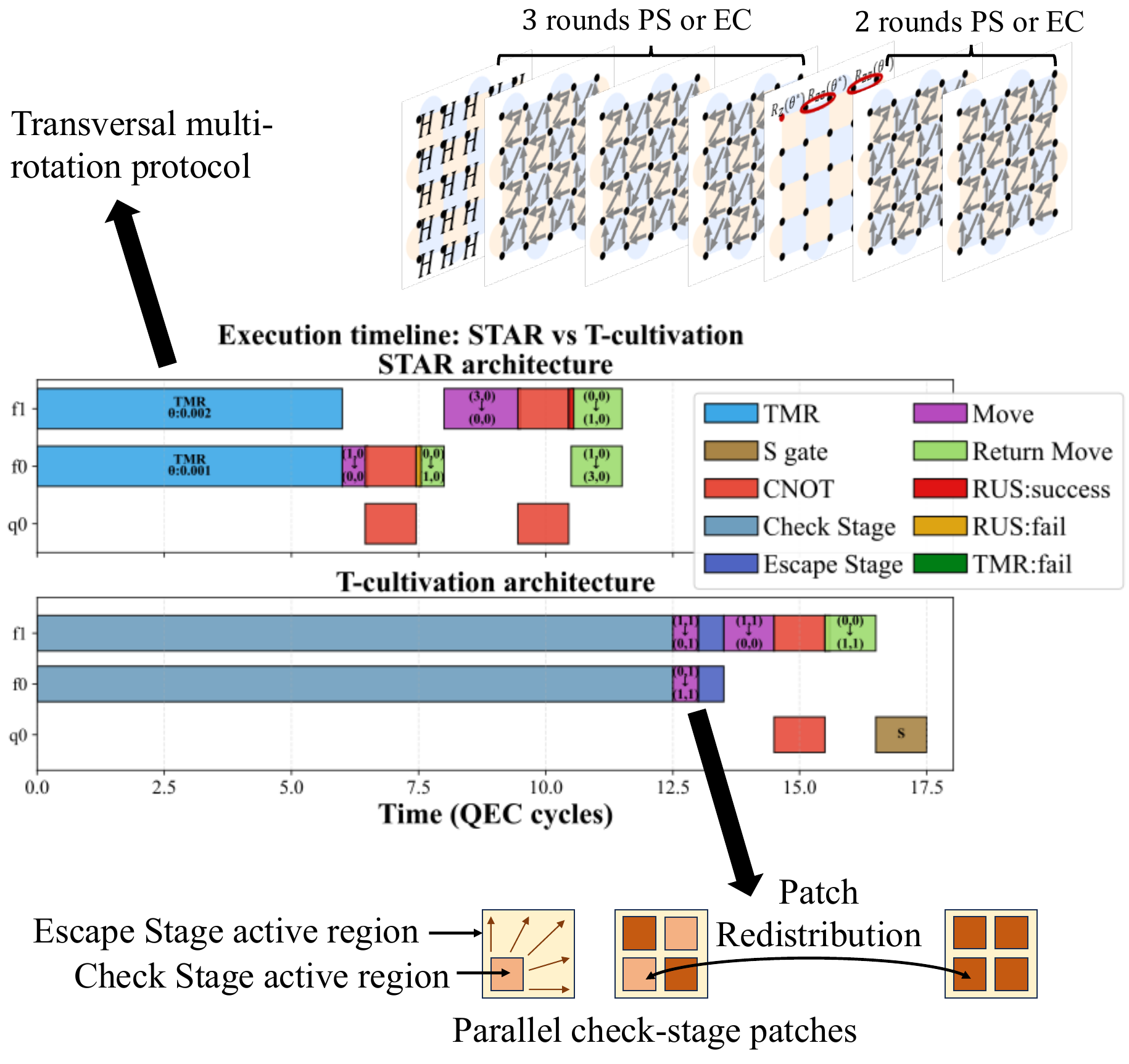}
\caption{
\textbf{Execution abstraction for EFTQC architectures.}
Both the STAR and $T$-state cultivation architectures use probabilistic resource preparation followed by teleportation-based consumption. 
In STAR (top), the transversal multi-rotation (TMR) protocol prepares an angle-dependent resource state on a logical patch.
In the execution model used here, each TMR attempt takes six QEC cycles. 
After preparation, RUS teleportation consumes the resource state, 
and an unsuccessful teleportation requires the preparation of a corrective state with a different angle, creating runtime-dependent resource demand. 
In $T$-state cultivation (bottom), the check stage uses a small footprint and takes 12.5 QEC cycles, whereas successful states proceed to the escape stage, 
which expands each surviving state to the target code distance and takes 0.5 QEC cycles in the configuration shown. 
Because the check stage has a lower success probability and a smaller spatial footprint than the escape stage, multiple check-stage patches can operate in parallel, and successful patches can be redistributed among available escape regions.
These differing space--time characteristics create dynamically changing resource availability and require runtime coordination of resource preparation, assignment, routing, and correction handling.}
\label{fig:eftqc_arch}
\end{figure}

EFTQC aims to execute practically useful quantum algorithms on the first generation of fault-tolerant quantum processors~\cite{beyond_NISQ,Preskill2018quantumcomputingin}.
In this regime, quantum simulation is widely regarded as one of the most promising application domains because it can deliver scientific value with relatively modest logical resources~\cite{Campbell_2022,daley2022practical,lin2022heisenbergeftqc}.
Quantum simulation also provides a particularly relevant benchmark for studying EFTQC execution: 
its structured circuits expose substantial algorithm-level parallelism 
while retaining the real-time challenges of probabilistic resource preparation, routing, and teleportation-based resource consumption.
Accordingly, we use quantum simulation as the primary application benchmark throughout this work.

Although surface-code-based fault tolerance provides a path toward universal quantum computation~\cite{PhysRevA.71.022316,PhysRevA.86.032324,harry2025resource_estimation,Zhou_2025}, 
preparing high-fidelity non-Clifford resource states remains a dominant performance bottleneck, 
often requiring thousands of physical qubits and many QEC cycles per resource state~\cite{Litinski2019magicstate,PhysRevA.86.052329,Gidney2019efficientmagicstate,PhysRevA.95.032338}.
Recent EFTQC architectures therefore focus on reducing the overhead of resource generation.
In this work, we consider two representative neutral-atom EFTQC architectures: 
the transversal STAR architecture~\cite{PRXQuantum.5.010337,PhysRevX.15.021057,starv3,ismail2025transversal} and a $T$-state cultivation architecture~\cite{gidney2024magicstatecultivationgrowing,chen2025efficientmagicstatecultivation,claes2025cultivatingtstatessurface,sahay2026foldtransversalsurfacecodecultivation}. 

\subsection{STAR Architecture}
The STAR architecture implements logical $R_Z(\theta)$ rotations by preparing angle-dependent logical resource states through the transversal multi-rotation (TMR) protocol~\cite{PRXQuantum.5.010337,PhysRevX.15.021057}. 
Unlike conventional approaches that synthesize arbitrary rotations into sequences of $T$ gates~\cite{rzsynthesis}, 
STAR directly prepares the target rotation state, substantially reducing the non-Clifford overhead.
As illustrated in \autoref{fig:eftqc_arch}, 
a logical rotation state is prepared by applying transversal physical rotations across a logical patch together with multiple rounds of syndrome extraction (SE) before and after the rotation. 
These SE rounds both suppress logical errors and probabilistically project
the transversally rotated state onto the desired logical subspace.
Because the overlap with this accepted subspace depends on the applied
physical rotation angle, 
which is chosen according to the target logical
rotation $\theta$, 
the preparation success probability is angle-dependent, as presented in \autoref{sec:appendix:exp_setting}.
The detailed relation between the physical rotation, logical rotation angle,
and preparation success probability is derived in
Refs.~\cite{starv3,ismail2025transversal}.
Consequently, TMR preparation constitutes a major component of the execution latency.

Prepared rotation states are consumed through repeat-until-success (RUS) teleportation. 
Depending on the teleportation outcome, 
the protocol applies either the desired $R_Z(\theta)$ operation or an undesired $R_Z(-\theta)$ operation;
the latter requires a corrective $R_Z(2\theta)$ rotation. 
As a result, the demand for logical resource states evolves dynamically during execution and cannot be determined completely before runtime.
In addition to resource preparation, 
logical patch movement also contributes significantly to execution latency. 
Neutral-atom platforms are therefore particularly attractive for STAR because movable logical patches and dynamically reconfigurable interaction regions naturally support adaptive factory assignment, long-range communication, and movement-aware scheduling~\cite{ismail2025transversal}. 
These capabilities enable efficient redistribution of logical patches and runtime reassignment of factories, 
motivating the compiler optimizations presented in the following sections.

Prior compilation work has explored online scheduling and lookahead preparation for STAR workloads~\cite{rescq,vqe_star}. 
However, these approaches primarily target STAR-specific execution models and assume lattice-surgery-based communication.
In contrast, \ftname{} treats STAR as one instantiation of a stage-aware real-time compilation framework, separating architecture-specific policies for resource preparation, resource assignment, routing, and correction handling from the overall compilation flow. 

\subsection{Magic-State Cultivation}

Magic-state cultivation (MSC)~\cite{gidney2024magicstatecultivationgrowing,chen2025efficientmagicstatecultivation,claes2025cultivatingtstatessurface,sahay2026foldtransversalsurfacecodecultivation} provides a lower-overhead alternative to conventional magic-state distillation by injecting noisy $|T\rangle$ states into small code patches and progressively increasing their fidelity through multiple rounds of cultivation and post-selection. 
Instead of immediately preparing large logical patches, the protocol gradually expands successful states to larger code distances, 
reducing the space--time overhead of producing high-fidelity magic states.

The preparation workflow consists of injection, cultivation, and escape stages. 
The injection and cultivation stages operate on relatively small logical patches and have comparatively low success probabilities, 
whereas the final escape stage expands surviving patches to the target code distance using substantially larger logical patches. 
This difference in spatial footprint motivates provisioning many compact early-stage factories in parallel while reserving fewer large escape factories. 
As illustrated in \autoref{fig:eftqc_arch}, successful patches are dynamically redistributed from cultivation factories to available escape factories, 
where they complete the preparation process before being consumed through standard gate teleportation.

Consequently, efficient execution requires not only coordinating probabilistic resource preparation but also dynamically assigning successful patches to escape factories and moving them between stages.
Reconfigurable neutral-atom architectures are well suited to these runtime operations because logical patches can be relocated efficiently during execution.
Although MSC prepares universal $|T\rangle$ states rather than angle-specific rotation states, it shares the same high-level execution pattern as STAR: probabilistic resource preparation followed by teleportation-based resource consumption.

\subsection{Neutral-Atom Execution Architecture}
\label{sec:neutral_atom_arch}

We instantiate both execution models on a reconfigurable neutral-atom architecture in which logical patches can be transported between resource-preparation and execution regions.
This reconfigurability supports runtime factory reassignment, redistribution of successfully prepared resources, and long-range interactions.
In the transversal STAR architecture, atom transport additionally enables transversal Clifford operations by bringing corresponding physical qubits into interaction proximity~\cite{ismail2025transversal}.

Neutral-atom arrays employ two complementary trapping mechanisms:
static spatial light modulator (SLM) traps and movable acousto-optic deflector (AOD) traps.
Atoms stored in SLM traps remain fixed, whereas AOD traps transport selected atoms across the array.
To perform a transport-mediated operation, selected atoms are transferred to AOD traps, moved to their target locations, and subsequently returned or relocated as required.

AOD movement is constrained by the coupling among its horizontal and vertical control tones.
During one AOD activation, atoms located at the Cartesian product of the activated tones are picked up simultaneously, and atoms sharing a tone move together along the corresponding axis.
Consequently, movements executed in parallel must satisfy three constraints:
(1) their trajectories must not intersect;
(2) atoms picked up from the same source row (column) must be deposited into the same destination row (column); and
(3) the activated tones must not unintentionally pick up atoms that are not scheduled to move.
These constraints determine which transports can be performed simultaneously and therefore directly affect routing and transversal-operation latency~\cite{tan2024enola,zac}.
They motivate the movement and routing optimizations introduced in \autoref{sec:compilation_eftqc}.
\section{Compilation Techniques for EFTQC Architectures}
\label{sec:compilation_eftqc}

\begin{figure*}[t]
  \centering
  \includegraphics[width=\linewidth]{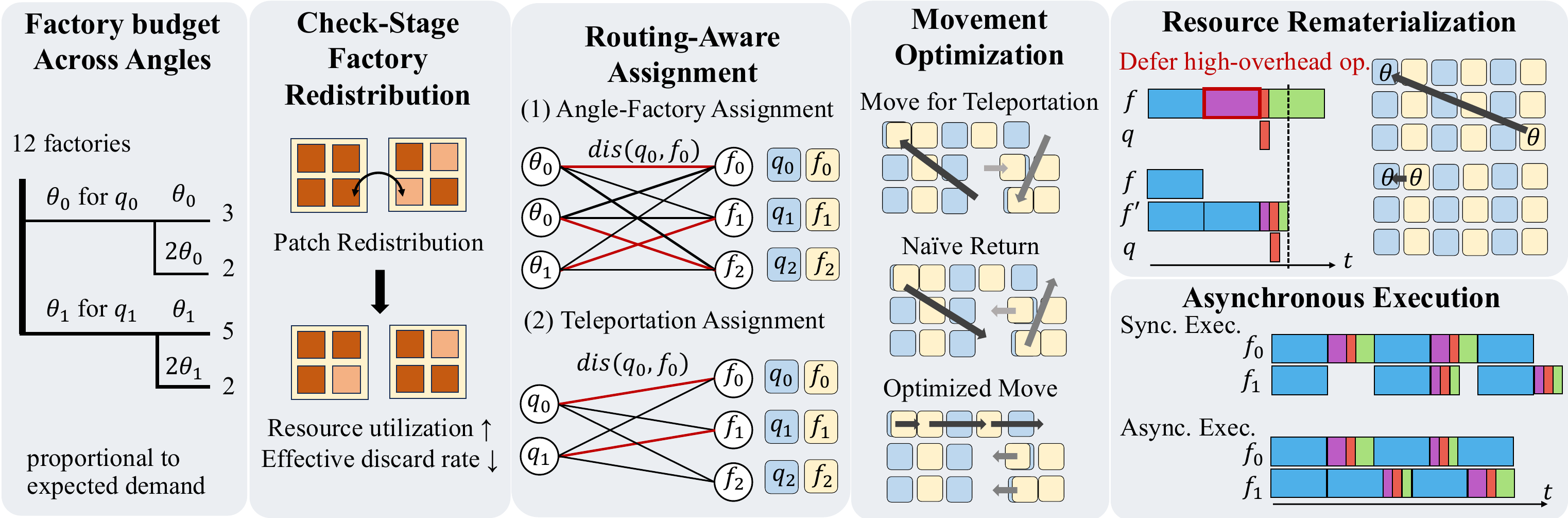}
  \caption{\textbf{Overview of the compilation techniques.}
From left to right, the figure summarizes techniques for resource preparation and correction handling, resource assignment, routing, and execution policy.
Dynamic angle collection distributes the available factory budget across current rotation requests and speculative RUS correction angles.
Check-stage factory redistribution reassigns successful intermediate patches to available escape factories, improving resource utilization and reducing the effective discard rate.
Routing-aware assignment matches angle-demand slots to factories before preparation and prepared resource states to compatible logical-qubit requests after preparation while accounting for movement cost.
Movement optimization reduces both forward teleportation-routing and factory-return-routing overhead; different arrow colors denote distinct movement batches.
Resource rematerialization defers selected high-overhead teleportations when preparing a replacement in a later round is expected to be faster than routing the current resource.
Finally, asynchronous execution overlaps preparation and teleportation to expose additional execution parallelism.}
  \label{fig:eftqc_compile_technique}
\end{figure*}

\autoref{fig:eftqc_compile_technique} summarizes the architecture-specific compilation techniques used to instantiate \ftname\ for the two neutral-atom EFTQC architectures considered in this work. 
Highly parallel EFTQC execution tightly couples resource preparation, assignment, routing and correction handling.
\ftname\ addresses these interactions through a collection of stage-specific optimizations while preserving the common compilation flow introduced in \autoref{sec:flow}.

The techniques span the four execution stages defined by \ftname. 
Dynamic angle collection and check-stage factory redistribution optimize resource preparation, 
routing-aware factory assignment coordinates resource assignment, 
and movement and routing optimization reduces the cost of routing prepared resources. 
The correction-handling stage feeds execution outcomes back into preparation and assignment.
For example, in STAR, teleportation outcomes generate corrective rotation demands that are passed to dynamic angle collection and subsequent resource assignment. 
Resource rematerialization further couples preparation and routing by allowing the runtime to regenerate a resource when doing so is more efficient than routing an existing one. 
Together, these techniques illustrate how architecture-specific optimizations can be implemented within the common stage interfaces of \ftname.
Beyond these compilation techniques, the final panel of \autoref{fig:eftqc_compile_technique} compares synchronous and asynchronous execution. 
We discuss these execution policies separately in \autoref{sec:execution_policy}.

\subsection{Dynamic Angle Collection}
\label{sec:compilation_eftqc:angle_collection}

For STAR, resource preparation and correction handling are coupled because teleportation outcomes may generate additional corrective rotations.
Unlike conventional $T$-state factories, which repeatedly prepare a fixed resource state, STAR factories must prepare angle-dependent resource states whose demand evolves dynamically throughout execution.
Accordingly, the runtime allocates a shared pool of factories across both the currently required rotation angles and the possible correction angles that may arise along subsequent RUS branches.
Because the requested current and speculative preparations can exceed the available factory capacity, the runtime must decide how much preparation capacity to allocate to each active request and speculative correction branch.

\ftname\ employs a hierarchical angle-collection strategy that allocates the global factory budget across active rotation requests and their RUS correction sequences.
The runtime first partitions factories among active rotation requests according to their estimated demand.
Within each request, the allocated budget is further divided between the immediate angle and speculative correction angles based on estimated demand.
This hierarchical allocation enables parallel preparation across multiple correction levels while maintaining capacity for the active workload.
It corresponds to the allocation illustrated in \autoref{fig:eftqc_compile_technique}, where the factory budget is divided first among active requests and then across their associated correction levels.
The detailed allocation procedure is presented in \autoref{appendix:angle_collection}.

\subsection{Check-Stage Factory Redistribution for Cultivation-Based Architecture}
\label{sec:compilation_eftqc:check_stage_redistribution}
For cultivation-based FTQC, \ftname\ specializes the resource-preparation stage to exploit transferable intermediate cultivation states.
The second panel of \autoref{fig:eftqc_compile_technique} shows how the runtime reassigns successful check-stage patches to available escape factories.
Because these intermediate states are interchangeable, a successful patch need not remain within its original factory pipeline.
This enables the runtime to rebalance resources dynamically, 
improving utilization and reducing discard overhead.
This optimization illustrates how an architecture-specific resource-preparation policy can be incorporated within \ftname's common stage abstraction.

We formulate this redistribution as a minimum-weight bipartite matching problem. 
One partition consists of successful intermediate states, 
while the other contains available escape factories. 
The edge weight represents the physical movement cost between a state and a candidate factory. 
Solving the resulting assignment minimizes the total redistribution overhead. 

We solve the matching using the Jonker--Volgenant algorithm~\cite{jonker1988shortest,2dassignment}, 
which has worst-case cubic complexity in the number of nodes.
In practice, redistribution is performed independently within local factory regions, 
allowing multiple small matching instances to be solved in parallel and satisfying the low-latency requirements of online execution.
Because the nodes represent logical patches and factories rather than physical qubits, each instance is limited to a few hundred nodes.
Assignment problems of this scale can be solved efficiently on a field-programmable gate array (FPGA) in hundreds of microseconds~\cite{10.1145/3546072}. 

\subsection{Routing-Aware Factory Assignment}
\label{sec:compilation_eftqc:factory_assignment}

Within the resource-assignment stage, 
\ftname\ performs routing-aware assignment both before and after probabilistic resource preparation.
The first assignment shapes where resources are likely to become available, 
while the second adapts that plan to where resources actually become available after stochastic preparation.
As illustrated in \autoref{fig:eftqc_compile_technique}, 
both decisions use the same minimum-weight maximum matching, 
with stage-specific graph construction and a common Jonker--Volgenant solver. 
The detailed cost functions are described in \autoref{appendix:factory_assignment_details}.

\paragraph{Angle Assignment.}
Angle assignment determines which rotation angle each factory prepares next.
\ftname\ matches idle factories to the angle-demand slots produced by dynamic angle collection, 
including both current requests and speculative lookahead demands.
Current requests are prioritized to avoid delaying execution.
The factory-to-demand mapping is solved as a minimum-cost perfect matching matching problem, 
with edge weights combining request priority with the estimated movement cost from each factory to the target logical patch.
This is the upper matching graph in the third column of \autoref{fig:eftqc_compile_technique}, 
where distance-aware edge costs guide the assignment of angle demands to factories.

\paragraph{Teleportation Assignment.}
Once preparation outcomes become available, teleportation assignment matches successfully prepared resource states to requesting logical qubits. 
Successfully prepared resources implementing the same rotation are interchangeable, 
allowing the runtime to choose the assignment that minimizes movement after preparation outcomes are known.
Thus, \ftname\ groups requests by rotation angle,
and within each angle class, 
prepared resources are matched to requesting qubits,
which corresponds to the lower matching graph in the third column of \autoref{fig:eftqc_compile_technique}.
This flexibility is particularly useful for Hamiltonian-simulation workloads, 
where many qubits request identical rotations and therefore provide a larger assignment space.

\subsection{Movement and Routing Optimization}
\label{sec:compilation_eftqc:move_optimization}

For the neutral-atom EFTQC architectures considered in this work, \ftname\ specializes the routing stage to account for the physical constraints of atom transport.
\ftname\ therefore optimizes both forward routing for resource delivery and return routing for factory relocation. 
Although both phases move the same class of resources, 
they operate under fundamentally different occupancy conditions, as illustrated in \autoref{fig:eftqc_compile_technique}. 
Forward routing occurs while the computational array is densely occupied, 
whereas after teleportation the return path contains substantially more free space. 
\ftname\ therefore optimizes the two phases differently.

\paragraph{Forward routing.}
During forward teleportation routing, \ftname\ organizes the required atom movements into parallel AOD activations while avoiding collisions. 
Because the array is densely occupied during this phase, 
movements are first partitioned by their source--destination transfer signatures, defined by shared source and destination rows or columns.
These partitions are then batched using a minimum chain decomposition of the resulting movement partial order, 
minimizing the number of sequential movement batches within each partition. 
Detailed algorithms are provided in \autoref{appendix:movement_optimization}.

\paragraph{Return routing.}
After teleportation, factories are relocated to empty sites for subsequent preparation. 
Because the array is much sparser,
\ftname\ can exploit additional scheduling flexibility through three optimizations. 
First, \emph{destination reassignment} allows factories to move to any suitable empty site rather than their original locations.
Second, \emph{movement rebatching} combines return movements that become mutually compatible.
Finally, \emph{relay decomposition} breaks long relocations into shorter transfers that overlap more effectively with other movements.
Together, these optimizations reduce return-routing critical-path latency.
In \autoref{fig:eftqc_compile_technique}, the ``Optimized Move'' example illustrates the resulting schedule, whereas the ``Na\"ive Return'' example directly reverses the teleportation movements.
Detailed routing and batching algorithms are provided in \autoref{appendix:movement_optimization}.

\subsection{Resource Rematerialization}
\label{sec:compilation_eftqc:operation_dropout}
An initially prepared resource is not always worth consuming immediately. 
If delivering it incurs high movement cost or provides little useful parallelism, discarding the resource and preparing a replacement at a factory closer to the target can reduce overall execution latency.
As illustrated in \autoref{fig:eftqc_compile_technique}, \ftname{} addresses such cases through resource rematerialization: rather than immediately consuming a resource whose delivery incurs high routing overhead, the runtime may defer teleportation and prepare a replacement in a later preparation round.
The decision compares the latency of executing the current teleportation, including movement and gate latency, with the latency of another preparation round.
The released factories can then participate in subsequent preparation and assignment decisions.
By trading additional resource preparation for reduced movement, rematerialization couples the preparation and routing stages, analogous to rematerialization in classical compilers.
Detailed decision rules are presented in~\autoref{appendix:operation_dropout}.

\section{\ftname{} Compiled Execution}
\label{sec:ft_weave_output}

The output of \ftname\ consists of two complementary representations that specify the compiled execution: 
an \emph{execution timeline}, which determines when operations occur, 
and a \emph{movement schedule}, which determines how qubits are transported to realize them. 
Together, these representations capture the temporal and spatial decisions made by the compiler and can be lowered to architecture-specific hardware instructions without further logical-level scheduling or routing decisions.

We introduce these two outputs using a representative STAR execution. 
The same example also illustrates how the compilation  techniques described in \autoref{sec:compilation_eftqc}  jointly improve the resulting schedules.

\paragraph{Execution timeline.}

\begin{figure*}[htbp]
  \centering
  \includegraphics[width=\linewidth]{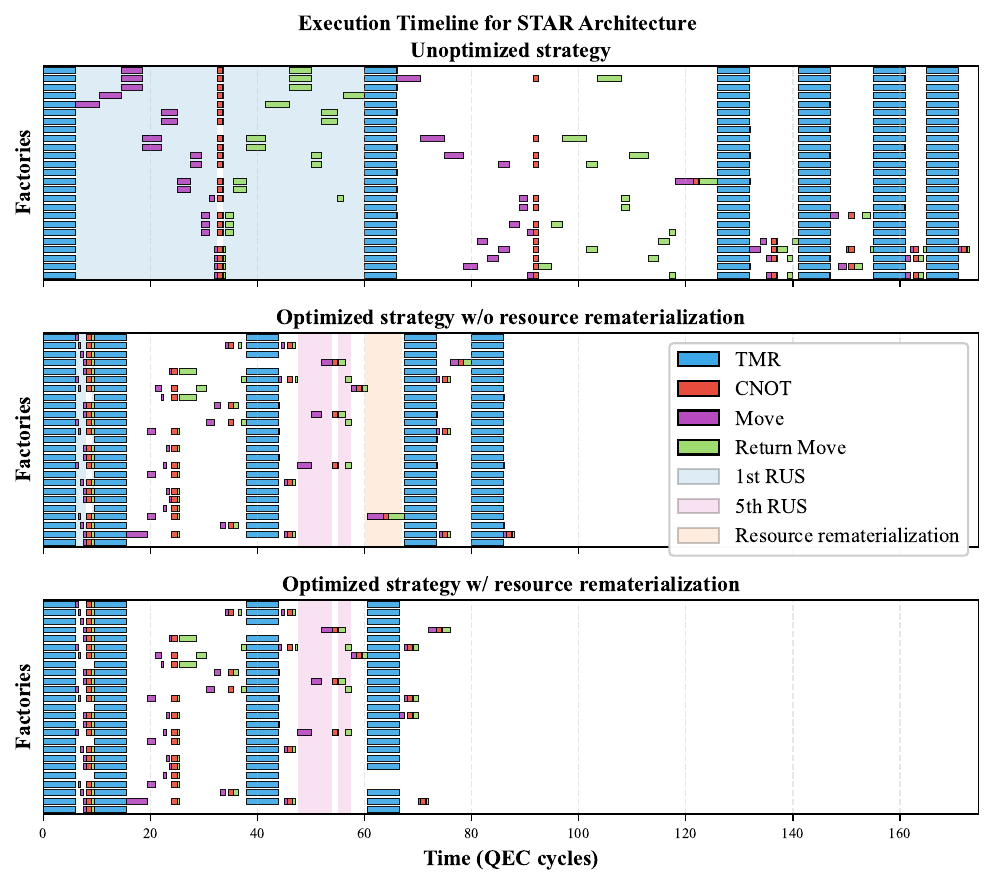}
  \caption{\textbf{Example of FT-Weave timeline output for three compilation strategies applied on STAR benchmark }
The same workload—a layer of identical-angle $R_Z$ rotations on 25 logical qubits executed using 25 factories and one AOD—is compiled using three strategies: an unoptimized baseline (top), routing-aware assignment and movement optimization (middle), and the same optimizations with resource rematerialization (bottom).
Each row tracks the activity of one factory over QEC cycles; colors denote TMR preparation, CNOT gates, forward movement, and return movement, as indicated in the legend.
The shaded regions identify the first and fifth RUS rounds and a resource-rematerialization decision examined in \autoref{sec:compilation_eftqc}.
} 
  \label{fig:strategy_comp_timeline}
\end{figure*}

The execution timeline represents the temporal schedule produced by \ftname\ at QEC-cycle granularity. 
In the STAR example in \autoref{fig:strategy_comp_timeline}, 
each row tracks one factory's activity, including resource preparation, routing, teleportation, and idle periods.
The figure compares three timelines for the same logical workload, 
enabling the effects of the compilation techniques to be observed directly.

In the baseline timeline, 
long transport paths and limited movement concurrency introduce extended idle intervals. 
Applying routing-aware assignment and movement optimization changes both the spatial routes and their temporal overlap, producing a more compact schedule. 
The final timeline illustrates how resource rematerialization further changes the execution plan by selectively deferring high-overhead teleportations.

\paragraph{Movement schedule.}

\begin{figure*}[htbp]
  \centering
  \includegraphics[width=\linewidth]{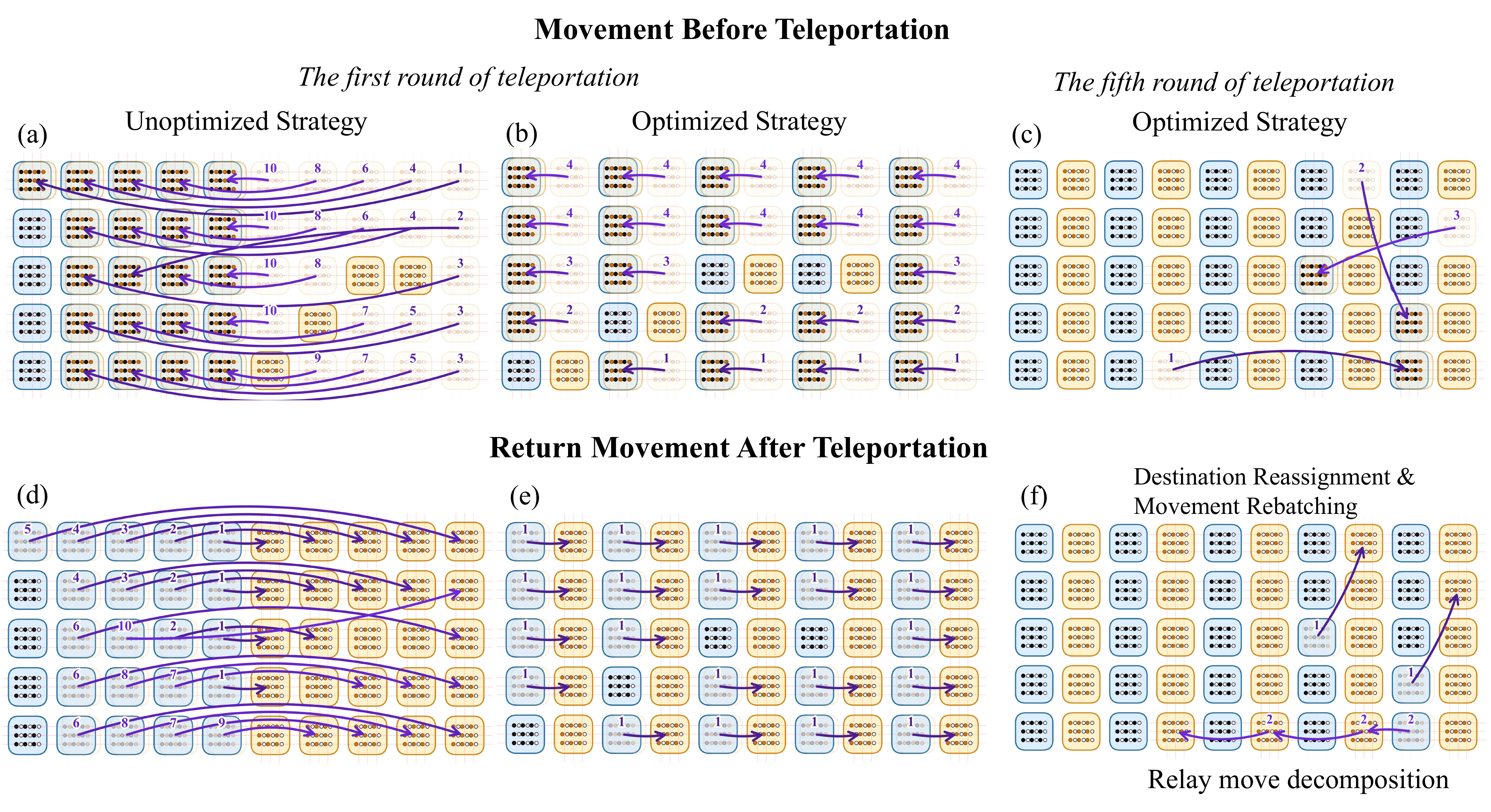}
  \caption{\textbf{Example movement schedules produced by \ftname.}
Blue and yellow boxes denote logical-qubit and factory patches, respectively; each patch is depicted as a $3\times3$ array of nine data qubits in this distance-3 surface-code schematic.
Purple arrows indicate atom movements, annotated numbers specify their execution order, and dashed red lines indicate the AOD row and column selections.
The upper row shows forward movement before teleportation, and the lower row shows the corresponding return movement.
The first two columns compare the (a,d) unoptimized and (b,e) optimized schedules for the first teleportation round, highlighted in blue in \autoref{fig:strategy_comp_timeline}.
The third column shows (c) the optimized forward and (f) return schedules for the fifth teleportation round, highlighted in pink in \autoref{fig:strategy_comp_timeline}.
}
  \label{fig:strategy_opt_vs_unopt}
\end{figure*}

Each movement interval in the execution timeline is associated with a spatial movement schedule that specifies the source and destination locations, compatible movement groups, and the sequence of AOD activations required to realize that interval.
\autoref{fig:strategy_opt_vs_unopt} visualizes representative schedules from \autoref{fig:strategy_comp_timeline}.

The first teleportation round illustrates how routing-aware assignment and movement batching change the spatial schedule, 
producing shorter routes and fewer sequential movement steps in this example.
The corresponding return schedule further benefits from destination reassignment and rebatching, 
avoiding the repeated back-and-forth transfers of the baseline. 
The later teleportation round illustrates the additional effect of relay decomposition, 
which allows long-distance relocations to be divided and overlapped with other movements.

Together, the execution timeline and movement schedule form the compiled execution produced by \ftname. 
The timeline exposes temporal behavior such as latency, utilization, and concurrency, 
while the movement schedule captures spatial behavior such as routing distance, movement depth, and AOD usage. 
These outputs therefore provide both the executable schedule and the basis for evaluating how compilation techniques, execution policies, and architectural choices affect system-level performance.

\section{Execution Policy}
\label{sec:execution_policy}

\ftname\ supports both synchronous and asynchronous execution policies through the same real-time control interface. 
Both policies maintain the same runtime state for logical operations, factories, and movement resources and invoke the same compilation techniques from \autoref{sec:compilation_eftqc}.
They differ in when hardware outcomes are observed and new scheduling decisions are made. 
Synchronous execution coordinates decisions at global preparation and teleportation boundaries, 
whereas asynchronous execution reacts to individual hardware completion events. 
This distinction exposes a fundamental trade-off: 
synchronous execution provides larger optimization windows for global coordination and batching, 
while asynchronous execution can expose finer-grained parallelism and reduce waiting between execution stages.

\begin{algorithm}[htpb]
\caption{Synchronous Execution}
\label{alg:sync}
\KwInput{Scheduled logical operations and hardware configuration}
\KwOutput{Execution trace $\mathcal{T}$}

Initialize runtime state from the scheduled operations and hardware configuration\;

$\mathcal{Q}\leftarrow$ currently enabled rotation requests\;
$\mathcal{F}_{\mathrm{succ}}\leftarrow\emptyset$\;
$\mathcal{T}\leftarrow[\ ]$\;

\While{$\mathcal{Q}\neq\emptyset$}{
    $\mathcal{F}_{\mathrm{idle}}\leftarrow$ currently available factories\;

    $\mathcal{A}\leftarrow
    \textsc{CollectAngles}(\mathcal{Q},|\mathcal{F}_{\mathrm{idle}}|)$\;
    \Comment{\autoref{sec:compilation_eftqc:angle_collection},\autoref{alg:angle_collection}}
    \label{line:sync_collect}

    $\mathcal{M}_{\mathrm{prep}}\leftarrow
    \textsc{AssignFactory}
    (\textsc{Prepare},\mathcal{F}_{\mathrm{idle}},\mathcal{A})$\;
    \Comment{\autoref{sec:compilation_eftqc:factory_assignment},\autoref{alg:factory_assignment}}
    \label{line:sync_prepare_assign}
        \tcp{Controller $\rightarrow$ quantum hardware}
        \textsc{DispatchPreparation}($\mathcal{M}_{\mathrm{prep}}$)\;
        \label{line:sync_dispatch_prep}

        \tcp{Quantum hardware $\rightarrow$ controller}
        $(\mathcal{F}_{\mathrm{new}},\mathcal{F}_{\mathrm{fail}})
        \leftarrow\textsc{ReceiveOutcome}
        ()$\;
        \label{line:sync_receive_prep}

        $\mathcal{F}_{\mathrm{succ}}\leftarrow
        \mathcal{F}_{\mathrm{succ}}\cup\mathcal{F}_{\mathrm{new}}$\;

        $\textsc{RecordTrace}(\mathcal{T})$\;\label{line:record_trace}

    \While{$\mathcal{F}_{\mathrm{succ}}\neq\emptyset$}{
        $\mathcal{M}\leftarrow
        \textsc{AssignFactory}
        (\textsc{Teleport},\mathcal{F}_{\mathrm{succ}},\mathcal{Q})$\;
        \Comment{\autoref{sec:compilation_eftqc:factory_assignment},\autoref{alg:factory_assignment}}
        \label{line:sync_tele_assign}


        $\mathcal{B}_{\mathrm{fwd}}
        \leftarrow
        \textsc{OptimizeForwardMovement}(\mathcal{M})$\;
        \Comment{\autoref{sec:compilation_eftqc:move_optimization},\autoref{alg:forward_movement}}
        \label{line:sync_forward_move}

        $\mathcal{B}'_{\mathrm{fwd}}\leftarrow
        \textsc{Rematerialize}(\mathcal{B}_{\mathrm{fwd}})$\;
        \Comment{\autoref{sec:compilation_eftqc:operation_dropout},\autoref{alg:rematerialization}}
        \label{line:sync_remat}


        \If{$\mathcal{B}'_{\mathrm{fwd}}=\emptyset$}{
            \Break\;
        }

        $\mathcal{F}_{\mathrm{ret}}\leftarrow$
        factories associated with $\mathcal{B}'_{\mathrm{fwd}}$\;

        $\mathcal{B}_{\mathrm{ret}}
        \leftarrow
        \textsc{OptimizeReturnMovement}
        (\mathcal{F}_{\mathrm{ret}})$\;
        \Comment{\autoref{sec:compilation_eftqc:move_optimization},\autoref{alg:return_movement}}
        \label{line:sync_return_move}

        \tcp{Controller $\rightarrow$ quantum hardware}
        \textsc{DispatchTeleportation}
        ($\mathcal{B}'_{\mathrm{fwd}},\mathcal{B}_{\mathrm{ret}}$)\;
        \label{line:sync_dispatch_tele}

        \tcp{Quantum hardware $\rightarrow$ controller}
        $(\mathcal{O}_{\mathrm{succ}},\mathcal{O}_{\mathrm{fail}})
        \leftarrow\textsc{ReceiveOutcomes}()$\;
        \label{line:sync_receive_tele}

        $\textsc{Update}
        (\mathcal{Q},
         \mathcal{F}_{\mathrm{succ}},
         \mathcal{O}_{\mathrm{succ}},
         \mathcal{O}_{\mathrm{fail}})$\;
        \label{line:sync_update}

        $\textsc{RecordTrace}(\mathcal{T})$\;
    }
    $\textsc{ReleaseFactory}()$\; \label{line:release_factory}
}

\Return{$\mathcal{T}$}\;
\end{algorithm}

\subsection{Synchronous Execution}
\autoref{alg:sync} summarizes the synchronous real-time controller.
In synchronous execution, 
the controller operates at global preparation and teleportation boundaries. 
At the beginning of each round, it collects the currently enabled rotation requests, determines the available factory budget, and jointly assigns preparation demands to idle factories (\autoref{line:sync_collect}--\autoref{line:sync_prepare_assign}). 
The controller then dispatches all preparation operations and waits for every attempt in the round to complete before processing the outcomes or making teleportation assignments. 
Because the complete set of preparation outcomes is available before teleportation assignment, 
the runtime can jointly match successful resources to pending rotations and batch their movements using the optimization techniques in \autoref{sec:compilation_eftqc} (\autoref{line:sync_receive_prep}--\autoref{line:sync_return_move}).
The \textsc{RecordTrace} statement in \autoref{line:record_trace} appends the preparation instructions and the execution outcome to the execution trace.

After routing optimization and rematerialization, 
the selected teleportations are dispatched as a group. 
Because the synchronous controller plans the complete teleportation round before dispatch, forward and return movements are both scheduled before execution.
The \textsc{Update} operation in \autoref{line:sync_update} then updates the
logical-instruction state.
For a completed rotation, the executed request is removed from $\mathcal{Q}$ if teleportation succeeds, 
and any newly enabled dependent logical operations are added.
When the teleportation outcome requires a subsequent RUS correction, 
the completed request is replaced by the corresponding correction-angle request.
Consumed factories are removed from $\mathcal{F}_{\mathrm{succ}}$. 
Consequently, $\mathcal{Q}$ always represents the currently enabled rotation requests rather than a static copy of the input circuit.
After all successful resources have been processed, 
The \textsc{ReleaseFactory} statement in \autoref{line:release_factory} resets factories that do not hold a useful resource state and marks them as available for the next global preparation round.

\begin{figure*}[htbp]
    \centering
    \includegraphics[width=\linewidth]{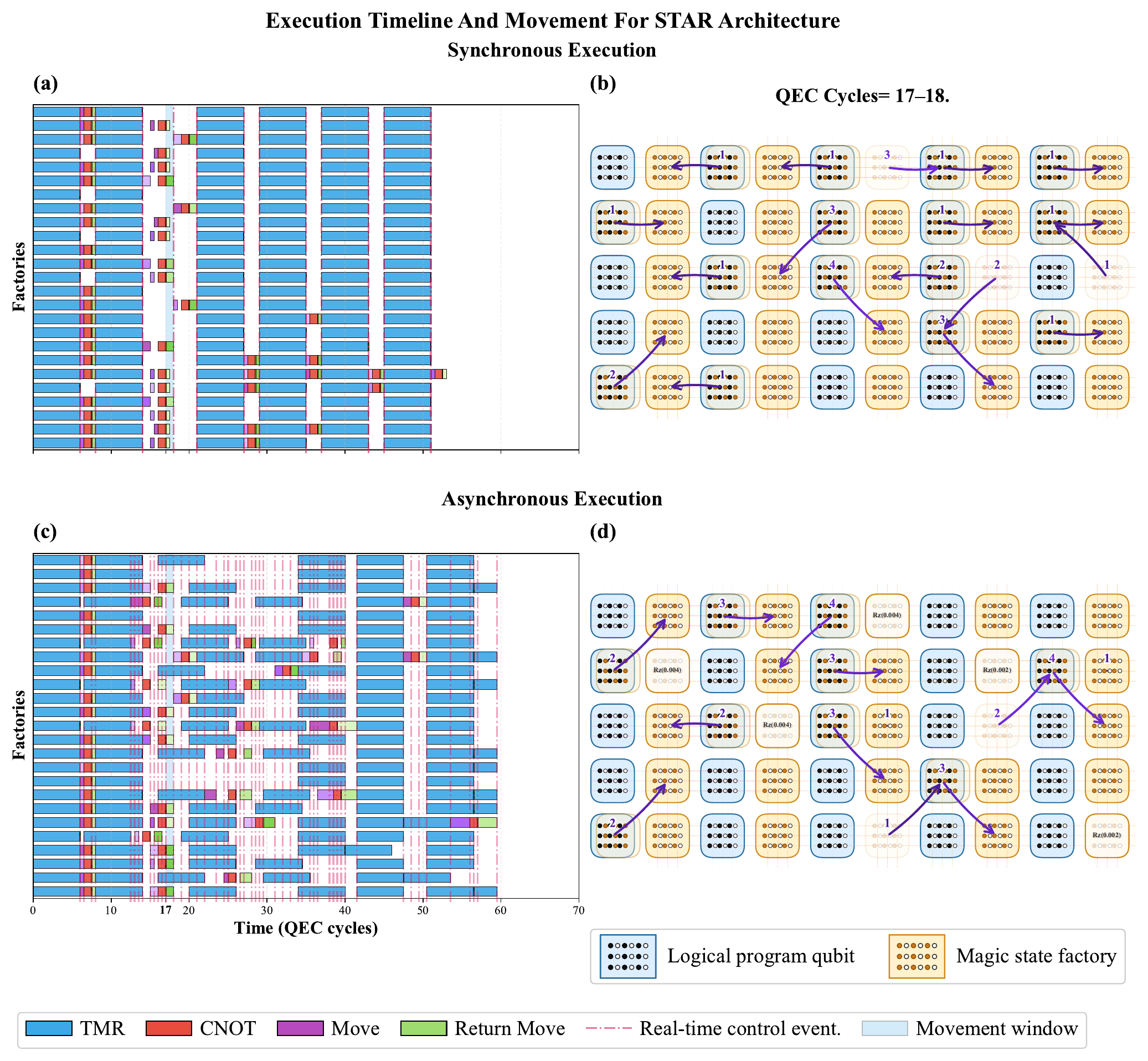}
    \caption{\textbf{Synchronous and asynchronous STAR execution.}
A layer of identical-angle $R_Z$ rotations on 25 logical qubits is executed using 25 factories and four AODs.
The left column compares the (a) synchronous and (c) asynchronous execution timelines, with each row tracking the activity of one factory over QEC cycles.
Colors denote TMR preparation, CNOT gates, forward movement, and return movement; magenta dash-dotted lines mark real-time control events, and the blue shading identifies the movement window visualized in the right column.
The right column shows the corresponding movement schedules during QEC cycles 17--18 for (b) synchronous and (d) asynchronous execution.
Under synchronous execution, global preparation and teleportation boundaries leave some factories idle while prepared resources are routed.
Under asynchronous execution, preparation in some factories overlaps with teleportation movement in others whenever the required AOD resources are available.}
    \label{fig:star_exec_sync}
\end{figure*}

\autoref{fig:star_exec_sync} illustrates this synchronized behavior.
Synchronization creates idle time because factories that complete or fail preparation early cannot immediately begin new work. 
This produces the clear preparation--teleportation boundaries visible in the left panel of the figure.
In exchange, the controller obtains a global view of preparation outcomes, enabling coordinated teleportation assignment and larger movement batches.

\subsection{Asynchronous Execution}

\begin{algorithm}[t]
\caption{Asynchronous Execution}
\label{alg:async}
\KwInput{Scheduled logical operations and hardware configuration}
\KwOutput{Execution trace $\mathcal{T}$}

Initialize runtime state from the scheduled operations and hardware configuration\;
$\mathcal{Q}\leftarrow$ currently enabled rotation requests\;
$\mathcal{T}\leftarrow[\ ]$\;
$\mathcal{F}_\mathrm{idle}\rightarrow$ currently available factories\;
$\mathcal{F}_\mathrm{succ}\rightarrow$ factories with resource states\;
$\mathcal{F}_\mathrm{ret}\rightarrow$ factories requiring return movement\;
$\mathcal{\zeta}_\mathrm{idle}\rightarrow$ currently available AODs\;

\While{$Q\neq\emptyset$ or in-flight operations remain}{
    \tcp{Quantum hardware $\rightarrow$ controller}
    $e\leftarrow\textsc{ReceiveSignal}()$\;
    \label{line:async_receive}
    Update $\mathcal{F}_\mathrm{idle}$, $\mathcal{F}_\mathrm{succ}$, $\mathcal{F}_\mathrm{ret}$, $\mathcal{Q}$, and $\mathcal{\zeta}_\mathrm{idle}$ based on $e$\;
    \label{line:async_update}
    $\textsc{RecordTrace}(\mathcal{T})$\;
    \If{$\mathcal{F}_\mathrm{idle}\neq\emptyset$ \textbf{and} $\mathcal{\zeta}_\mathrm{idle}\neq\emptyset$}{
        $\mathcal{A}\leftarrow
        \textsc{CollectAngles}(\mathcal{Q},|\mathcal{F}_\mathrm{idle}|)$\;
        \Comment{
        \autoref{sec:compilation_eftqc:angle_collection},
        \autoref{alg:angle_collection}}

        $\mathcal{M}_{\mathrm{prep}}\leftarrow
        \textsc{AssignFactory}
        (\textsc{Prepare},\mathcal{F}_{\mathrm{idle}},\mathcal{A})$\;
        \Comment{
        \autoref{sec:compilation_eftqc:factory_assignment},
        \autoref{alg:factory_assignment}}
        \If{$\mathcal{M}_{\mathrm{prep}}\neq\emptyset$}{
            \tcp{Controller $\rightarrow$ quantum hardware}
            \textsc{DispatchPreparation}($\mathcal{M}_{\mathrm{prep}}$)\;
        \label{line:async_prepare}
            Update $\mathcal{F}_\mathrm{idle}$ and $\mathcal{\zeta}_\mathrm{idle}$\;
            $\textsc{RecordTrace}(\mathcal{T})$\;
        }
        
    }

    \If{$\mathcal{F}_\mathrm{succ}\neq\emptyset$ \textbf{and} $\mathcal{\zeta}_\mathrm{idle}\neq\emptyset$}{
        $\mathcal{M}\leftarrow
        \textsc{AssignFactory}
        (\textsc{Teleport},\mathcal{F}_{\mathrm{succ}},\mathcal{Q})$\;
        \Comment{
        \autoref{sec:compilation_eftqc:factory_assignment},
        \autoref{alg:factory_assignment}}

        $\mathcal{B}_{\mathrm{fwd}}\leftarrow
        \textsc{OptimizeForwardMovement}(\mathcal{M})$\;
        \Comment{
        \autoref{sec:compilation_eftqc:move_optimization},
        \autoref{alg:forward_movement}}

        $\mathcal{B}'_{\mathrm{fwd}}\leftarrow
        \textsc{Rematerialize}(\mathcal{B}_{\mathrm{fwd}})$\;
        \Comment{
        \autoref{sec:compilation_eftqc:operation_dropout},
        \autoref{alg:rematerialization}}
        \If{$\mathcal{B}'_{\mathrm{fwd}}\neq\emptyset$}{
            \tcp{Controller $\rightarrow$ quantum hardware}
            \textsc{DispatchTeleportation}($\mathcal{B}'_{\mathrm{fwd}}$)\;
            Update $\mathcal{F}_\mathrm{succ}$ and $\mathcal{\zeta}_\mathrm{idle}$\;
            $\textsc{RecordTrace}(\mathcal{T})$\;
            \label{line:async_teleport}
        }
    }

    \If{$\mathcal{F}_\mathrm{ret}\neq\emptyset$ \textbf{and} $\mathcal{\zeta}_\mathrm{idle}\neq\emptyset$}{
        $\mathcal{B}_{\mathrm{ret}}\leftarrow
        \textsc{OptimizeReturnMovement}
        (\mathcal{F}_{\mathrm{ret}})$\;
        \Comment{
        \autoref{sec:compilation_eftqc:move_optimization},
        \autoref{alg:return_movement}}

        \tcp{Controller $\rightarrow$ quantum hardware}
        \textsc{DispatchReturn}($\mathcal{B}_{\mathrm{ret}}$)\;
        \label{line:async_return}
        Update $\mathcal{F}_\mathrm{ret}$ and $\mathcal{\zeta}_\mathrm{idle}$\;
        $\textsc{RecordTrace}(\mathcal{T})$\;
        \textsc{ReleaseFactory}()
    }
}
\Return{$\mathcal{T}$}\;
\end{algorithm}

Asynchronous execution replaces global stage boundaries with the event-driven controller summarized in  \autoref{alg:async}.
Whenever the hardware reports a relevant event (\autoref{line:async_receive}),
such as the completion of preparation or teleportation or the release of an AOD resource,
the runtime updates its execution state and immediately determines which new operations have become executable (\autoref{line:async_update}).
For example, a successful preparation makes the prepared resource available for teleportation (\autoref{line:async_teleport}), 
a failed preparation returns the factory to the pool eligible for another preparation attempt (\autoref{line:async_prepare}), 
and a teleportation outcome either completes the current rotation or enables the corresponding RUS correction operation.

Dispatch is nonblocking: preparation, teleportation, and return movement remain in flight while the controller continues processing other events.
Consequently, a factory whose preparation fails can immediately begin another attempt, a successfully prepared resource can begin teleportation while other factories remain in preparation, and return movement can proceed independently when movement resources become available.
This event-driven overlap is visible in the asynchronous timeline of \autoref{fig:star_exec_sync}, where preparation, teleportation, and return movement can proceed concurrently across different factories whenever the required movement resources are available.

This additional concurrency, however, comes at the cost of scheduling inefficiency.
Synchronous execution observes the complete set of preparation outcomes before assigning teleportations, 
providing a larger optimization window for jointly matching prepared resources to pending requests and batching compatible movements.
In asynchronous execution, preparation outcomes arrive independently, 
causing prepared resources to become available at different times and often producing smaller assignment and movement batches.
Concurrent forward and return movements may also compete for the same AOD resources, 
increasing routing contention, 
while decisions based only on the currently available factories and demands can lead to less favorable assignments for subsequent operations.
Moreover, finer-grained hardware interaction requires the controller to process more frequent completion and measurement signals, update runtime state, and dispatch newly enabled operations, 
increasing classical scheduling and communication overhead.

Synchronous and asynchronous execution therefore expose a fundamental trade-off between \emph{coordination} and \emph{concurrency}: 
synchronization provides larger optimization windows for resource assignment and movement batching, 
whereas asynchronous control exposes finer-grained opportunities to overlap preparation, teleportation, and movement.
Which policy minimizes end-to-end latency depends on the workload, factory organization, available movement resources, and real-time control overhead.
Because both policies are implemented through the same \ftname\ compilation techniques and hardware model, 
\ftname\ provides a common framework for isolating and systematically evaluating these execution-policy trade-offs.
We quantify their impact on end-to-end execution performance in \autoref{sec:result}.

\section{Evaluation}
\label{sec:result}

We evaluate \ftname{} on the STAR and $T$-state cultivation architectures using a distance-9 surface code. 
As a representative EFTQC workload, we simulate the two-dimensional transverse-field Ising model with periodic boundary conditions using second-order Trotterization. 
We set all Hamiltonian coefficients to unity and simulate a single Trotter step.
This setup captures a common structure of lattice Hamiltonian simulation, 
consisting of repeated layers of local interactions and single-qubit rotations that expose substantial algorithm-level parallelism while requiring continuous non-Clifford resource generation and consumption.
We consider problem sizes ranging from $4\times4$ to $10\times10$ logical qubits and, 
unless otherwise specified, provision one resource-state factory per logical qubit.
This proportional provisioning rule scales the factory count with the workload size while allowing us to study how effectively the runtime translates the available parallelism into execution throughput. 
The one-to-one ratio is an evaluation choice rather than a constraint of \ftname.

Execution time is measured in QEC cycles, where one QEC cycle corresponds to a syndrome-extraction round.
Following the execution model described in \autoref{sec:appendix:exp}, 
logical Clifford operations require one QEC cycle, 
while each STAR TMR preparation requires six QEC cycles~\cite{ismail2025transversal}. 
For the $T$-state cultivation architecture based on the MSC-5 protocol, 
the check and escape stages require 12.5 and 0.5 QEC cycles, respectively~\cite{sahay2026foldtransversalsurfacecodecultivation}. 
Logical-patch movement incurs additional latency proportional to the Manhattan distance between the source and destination patches. 
These parameters instantiate one representative neutral-atom execution model rather than a universal hardware configuration; consequently, 
the quantitative contribution of individual optimizations depends on the assumed operation and movement latencies.
The timing and movement models are configurable within \ftname{} and can be adapted to different hardware parameters and transport mechanisms.
We vary the number of AODs from one to five, 
beyond which we observe little additional improvement under this model, 
to capture regimes ranging from movement-constrained to near-saturated execution.
Additional implementation details, including preparation-success models and timing parameters, are provided in \autoref{sec:appendix:exp_setting}.

Using this setup, we evaluate how compiler and runtime optimizations affect end-to-end execution, 
how real-time coordination interacts with available hardware controls, 
and how the role of real-time compilation differs between FTQC architectures. 
The quantitative results are specific to this execution model.
Our broader goal is to expose system-level interactions that are not captured by nominal factory space--time estimates alone.

\begin{figure*}[htbp]
    \centering
    \includegraphics[width=\linewidth]{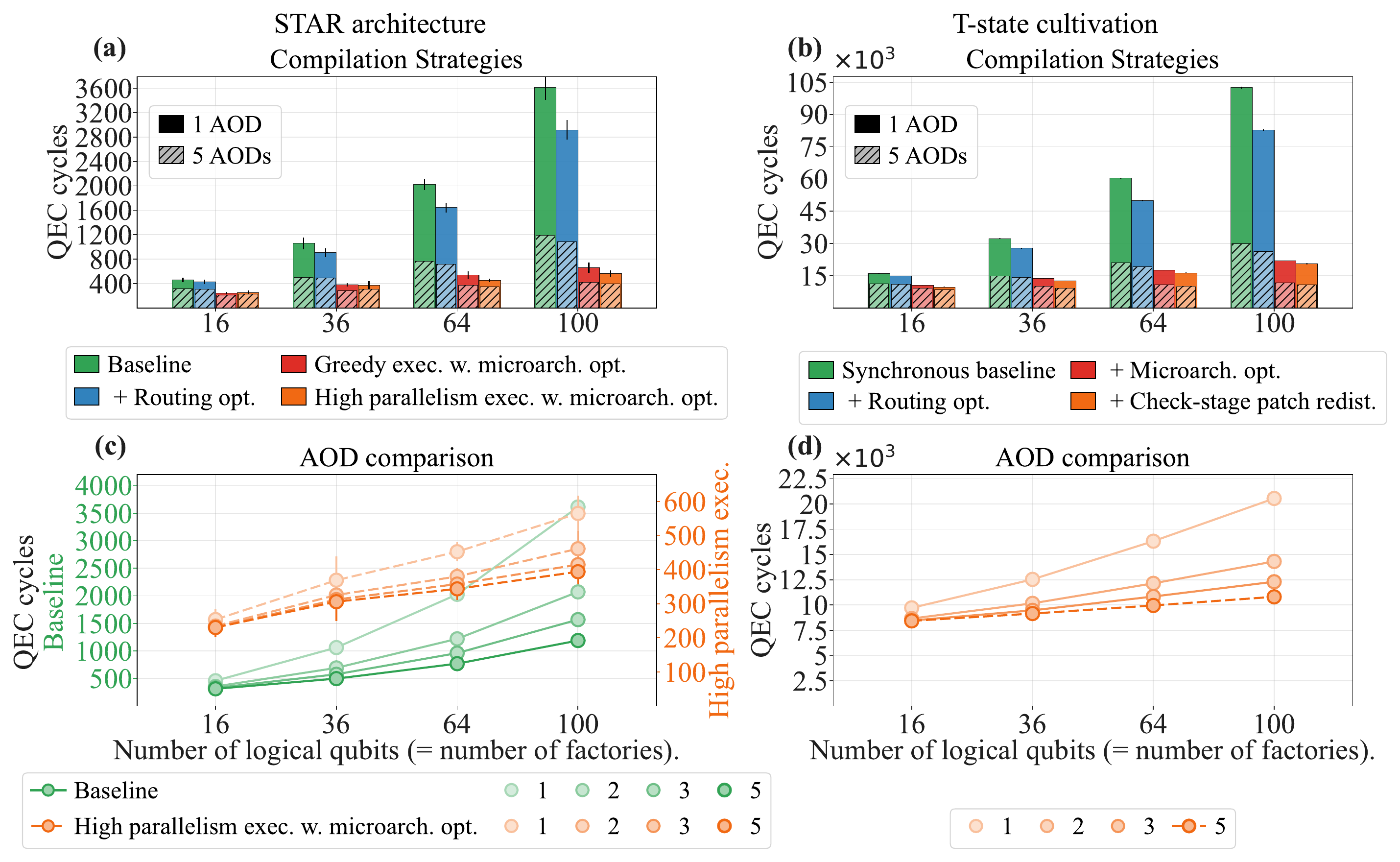}
    \caption{
    \textbf{Impact of \ftname{} compilation strategies and AOD movement parallelism on execution time.}
    Results are shown for STAR (left) and the $T$-state cultivation architecture (right) at surface-code distance $d=9$ for TFIM systems containing 16, 36, 64, and 100 logical qubits, with an equal number of factories.
    The top row shows the cumulative effect of the compilation strategies; dark and light shades denote results with one and five AODs, respectively.
    (a) For STAR, the \emph{baseline} uses synchronous execution with a separated microarchitecture in which factories and logical qubits occupy different regions as shown in \autoref{fig:compilation}.
    \emph{+Routing} adds the forward- and return-movement optimizations in \autoref{sec:compilation_eftqc:move_optimization}.
    \emph{Greedy exec.}~additionally uses the alternating-column microarchitecture, dynamic angle collection, routing-aware factory assignment, resource rematerialization, and asynchronous execution.
    \emph{High parallelism exec.}~replaces asynchronous execution with coordinated synchronous execution while retaining the other optimizations.
    (b) For $T$-state cultivation, the configurations are cumulative: the synchronous baseline uses the separated microarchitecture; \emph{+Routing} adds movement and routing optimization; \emph{+Microarch.~opt.}~interleaves factories and logical qubits; and \emph{+Check-stage patch redist.}~reassigns successful intermediate states to available escape factories.
    The bottom row shows execution time as the number of independently controlled AODs increases from one to five.
    (c) The STAR panel compares the baseline using the left green axis with the high-parallelism configuration using the right orange axis.
    (d) The cultivation panel shows the fully optimized configuration with the axis scaled by $10^3$.
    Bars and markers show the mean, and error bars show the standard deviation over 10 trials.
    }
    \label{fig:star_t_runtime_comp}
\end{figure*}

\begin{figure}[htbp]
    \centering
    \includegraphics[width=\linewidth]{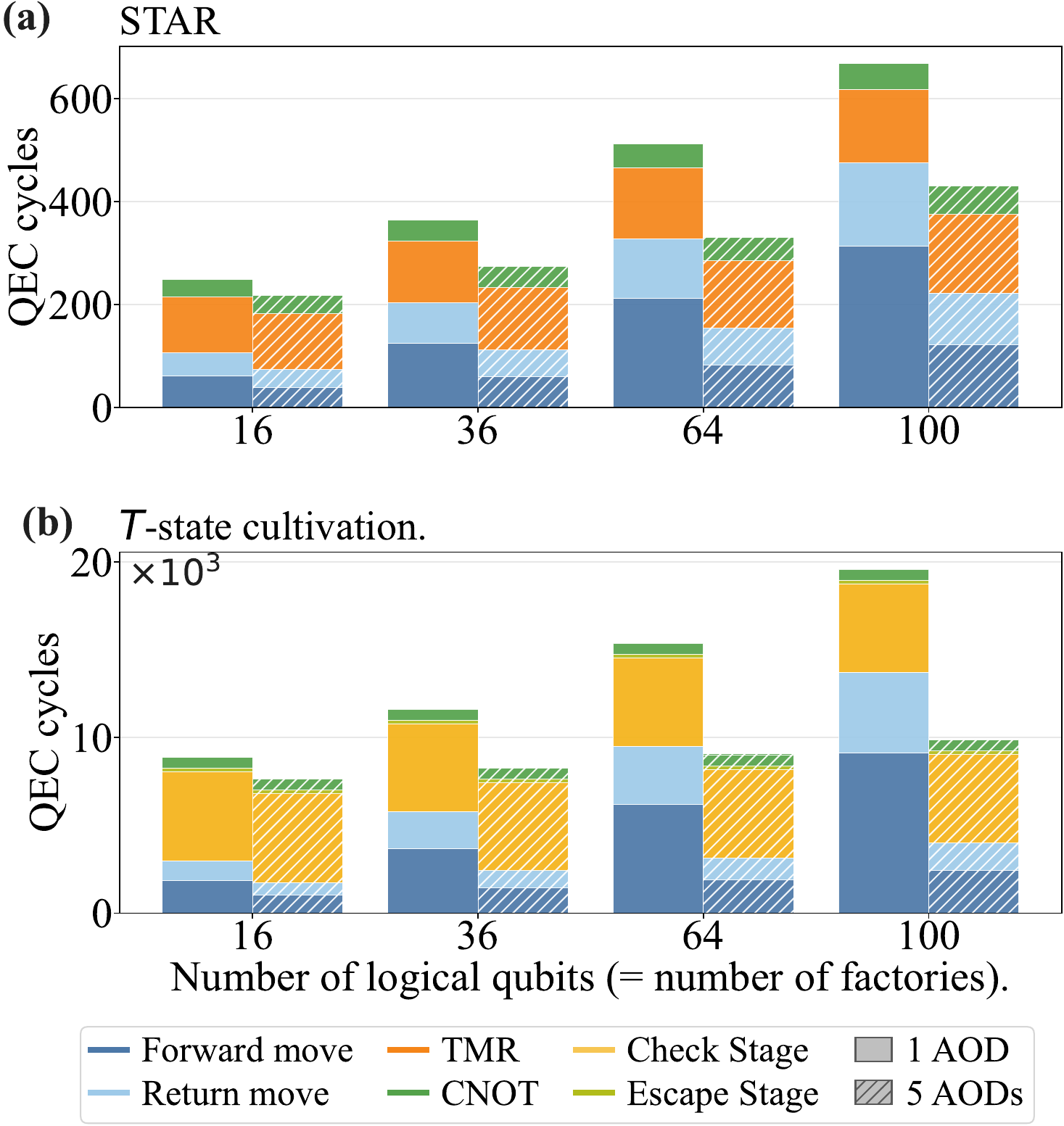}
    \caption{
       \textbf{Runtime decomposition for (a) STAR and (b) the $T$-state cultivation architecture under varying AOD parallelism at surface-code distance $d=9$.}
    Results are shown for TFIM systems containing 16, 36, 64, and 100 logical qubits, with an equal number of factories.
    Within each pair, the solid bar corresponds to one AOD and the hatched bar corresponds to five AODs.
    The stacked components show the time spent on forward movement, return movement, and CNOT execution; the STAR profile additionally includes TMR preparation, whereas the cultivation profile includes the check and escape stages.
    Execution time is reported in QEC cycles average across 10 runs, with the bottom-panel axis scaled by $10^3$.
    For STAR, movement primarily delivers prepared rotation resources to logical qubits and returns factory qubits.
    For cultivation, movement additionally redistributes successful intermediate states between the check and escape stages.
    Additional AODs reduce movement time through greater transport parallelism, illustrating how the two architectures place different demands on \ftname{}'s real-time assignment, routing, and resource-redistribution mechanisms under the evaluated timing model.
    }
    \label{fig:runtime_profile}
\end{figure}

\subsection{Impact of Compiler Optimizations}

\autoref{fig:star_t_runtime_comp} compares the cumulative impact of the proposed compiler optimizations on STAR and $T$-state cultivation-based architectures.
For STAR, the baseline combines synchronous execution with an unoptimized microarchitecture in which logical qubits and factories occupy separate regions on opposite sides of the lattice, 
as illustrated in \autoref{fig:strategy_opt_vs_unopt}. 
Applying the movement and routing optimization from \autoref{sec:compilation_eftqc:move_optimization} reduces execution time by 15\% with one AOD and 4\% with five AODs.
The larger improvement under limited AOD resources reflects the greater impact of movement scheduling when routing parallelism is constrained.

We next enable the alternating-column microarchitecture illustrated in \autoref{fig:strategy_opt_vs_unopt}, together with dynamic angle collection, routing-aware factory assignment, resource rematerialization, and asynchronous execution.
Relative to the routing-optimized configuration, these combined changes reduce execution time by an additional 60\% with one AOD and 44\% with five AODs.
A major contributor is the interleaving of factories and logical qubits, which reduces communication distance and allows the routing and assignment optimizations to use the available movement resources more efficiently.

Finally, replacing greedy asynchronous execution with coordinated high-parallelism synchronous execution yields a further 10\% improvement with one AOD and 1\% with five AODs under the evaluated timing and movement model. 
Although asynchronous execution exposes finer-grained overlap, 
synchronization provides larger optimization windows for resource assignment, movement batching, and rematerialization. 
Its benefit therefore decreases as additional movement controls reduce contention, 
illustrating a trade-off between fine-grained control and global coordination rather than an inherent advantage of synchronous execution.

For $T$-state cultivation, we use synchronous execution, 
which favors coordinated redistribution across the staged preparation pipeline, 
and evaluate the remaining optimizations cumulatively.
Movement and routing optimization reduces execution time by 15\% with one AOD and 7\% with five AODs. 
The optimized microarchitecture provides the largest additional improvement, reducing execution time by 54\% and 36\%, respectively. 
Check-stage patch redistribution provides a further 8\% improvement by allowing successful cultivation patches to proceed to available escape factories instead of waiting for their original factory pipelines.

Across both architectures, 
under the current hardware model, 
the largest cumulative improvements occur in configurations that introduce the optimized microarchitecture, although the corresponding STAR configuration also enables several runtime techniques.
Interleaving factories and logical qubits shortens communication distances and allows the assignment and routing techniques to operate more effectively.
Although microarchitecture design is not itself a real-time contribution of \ftname, 
this result highlights its importance when communication latency is significant. 
More generally, nominal factory throughput does not directly translate into application throughput: 
probabilistically generated resources must still be assigned, routed, corrected, and, for cultivation, redistributed at runtime. 
Without this coordination, idle factories, underutilized prepared states, and delayed consumption can substantially increase execution cost beyond factory-level space--time estimates. 
\ftname{} provides the runtime mechanisms for bridging this gap.

\subsection{Impact of AOD Parallelism}
The bottom row of \autoref{fig:star_t_runtime_comp} evaluates how execution time changes as the number of AODs increases. 
For STAR, increasing the number of AODs from one to two produces the largest gain, 
reducing execution time by 36\% for the baseline and 15\% for the optimized high-parallelism strategy. 
Additional AODs provide diminishing returns: 
for the optimized strategy, increasing from two to three AODs yields a further 5\% improvement, 
while increasing from three to five AODs provides approximately 1\%.

These results suggest that, 
under the evaluated model, 
a modest number of independent movement controls is sufficient to approach saturated performance for the evaluated logical-qubit arrays. 
Effective real-time coordination improves the utilization of this limited control budget: 
synchronization batches compatible movements and coordinates resource assignment, 
allowing the optimized strategy to saturate with fewer AODs. 
In contrast, the baseline relies more heavily on additional AODs to compensate for placement-induced routing conflicts. 
Although the precise saturation point depends on the movement and latency model, 
the broader hardware-design implication is that stronger runtime coordination can reduce the amount of independent movement control required to exploit available parallelism.

The $T$-state cultivation architecture follows a similar trend. 
Increasing the number of AODs from one to two reduces execution time by 21\%; increasing from two to three and from three to five AODs yields further reductions of 9\% and 6\%, respectively. 
The later saturation relative to optimized STAR reflects a different use of movement resources: 
cultivation requires transport not only to deliver completed magic states but also to redistribute successful intermediate patches across the preparation pipeline. 
Thus, the useful number of independent movement controls depends on how an architecture uses communication, 
while \ftname{} provides a common framework for quantifying this provisioning trade-off.

\subsection{Architecture-Dependent Roles of Real-Time Compilation}

\autoref{fig:runtime_profile} decomposes preparation and movement time to illustrate how STAR and the $T$-state cultivation architecture place different demands on the \ftname{} runtime under the evaluated timing model.
For STAR, communication accounts for an increasing share of execution time as the evaluated system scales. 
With one AOD, forward and return movement grow from approximately 43\% of runtime for $4\times4$ to more than 72\% for $10\times10$.
Even with five AODs, movement accounts for more than 51\% of runtime for the largest benchmark. 
Under these conditions, routing-aware assignment and movement batching coordinate limited AOD resources, 
while dynamic angle collection adapts preparation to evolving RUS demand.

For $T$-state cultivation architecture, \ftname{} instead focuses on preserving throughput through the structured cultivation-and-escape pipeline. 
Cultivation accounts for approximately 52--60\% of runtime for the smaller benchmarks, 
while movement becomes more prominent as the system scales or AOD resources are constrained. 
With one AOD, movement grows from approximately 31\% to 67\% of runtime; for the largest benchmark, increasing the number of AODs to five reduces this fraction to approximately 37\%. 
Check-stage redistribution allows successful intermediate states to continue through available escape factories, 
while routing and assignment coordinate their eventual delivery to the consuming logical qubits.

These profiles illustrate that different FTQC architectures rely on different forms of real-time coordination: 
STAR requires highly adaptive preparation and communication, 
whereas cultivation relies more heavily on maintaining throughput through a staged preparation pipeline.

Overall, although the quantitative improvements depend on the evaluated preparation and movement latency model, 
the results expose broader system-level interactions. 
Nominal factory space--time performance alone does not determine application throughput: 
across the evaluated compilation and microarchitectural strategies, improved system-level orchestration achieves a speedup of up to $3\times$ over the baseline.
When communication is costly, 
physical organization can strongly affect performance, 
while runtime coordination can reduce the number of independently controlled AODs needed to exploit available parallelism. 
Because these requirements differ across FTQC architectures, 
\ftname{} provides a common framework for jointly evaluating architecture, hardware provisioning, and real-time execution policy.
\section{Outlook}
\label{sec:conclusion}

The path toward scalable fault-tolerant quantum computing requires more than increasing hardware resources or optimizing individual fault-tolerant primitives.
As quantum systems scale,
resource preparation, communication, and logical execution become tightly coupled through stochastic outcomes, hardware constraints, and feedback-latency requirements.
This work establishes real-time compilation as a systems layer for coordinating these interactions.
We introduce \ftname,
a stage-aware compilation framework that combines offline optimization with runtime resource management, routing, and correction handling.
\ftname{} produces an execution timeline and a movement schedule that together specify the temporal and spatial execution of the program and can be lowered to architecture-specific hardware-control instructions.

We instantiate \ftname{} on two representative neutral-atom EFTQC architectures,
transversal STAR and a $T$-state cultivation architecture.
Under the evaluated timing and movement model,
\ftname{} achieves a speedup of up to $3\times$ over the baseline compilation flow.
The magnitude of this improvement is specific to the evaluated model and should not be interpreted as a universal performance gain; different hardware regimes may produce substantially larger or smaller benefits.
Nevertheless, the result demonstrates the potential importance of real-time coordination when stochastic resource generation, communication, and hardware contention are included in end-to-end execution.
By expressing architecture-specific preparation, assignment, routing, and execution policies through common stage interfaces,
\ftname{} provides a testbed for studying these interactions under different architectural assumptions.

More broadly, our evaluation highlights three system-level trade-offs relevant to FTQC design.
First, nominal factory throughput does not necessarily translate into application throughput because probabilistically generated resources must still be assigned, transported, corrected, and, when appropriate, redistributed.
Second, additional concurrency is not always beneficial.
Fine-grained asynchronous control can reduce local idle time, whereas synchronization provides larger optimization windows for resource assignment, movement batching, and rematerialization.
The preferred balance depends on the availability and latency of hardware controls.
Under the evaluated model, performance approaches saturation with a modest number of independently controlled AODs, particularly when coordinated scheduling reduces contention.
This suggests a hardware--software co-design opportunity: effective real-time coordination can reduce the amount of independent movement control needed to exploit available parallelism.
Finally, STAR and cultivation exercise different parts of \ftname{} because of their distinct resource-generation and communication structures, illustrating why architectures expressed through the same compilation abstraction may nevertheless require different runtime policies.

An important direction is to extend this abstraction across the expanding FTQC architecture and QEC design space.
Recent proposals combine reconfigurable atom arrays with high-rate quantum low-density parity-check (qLDPC) codes and transversal logical operations~\cite{ismail2026fastparallelhighratestar,yang2026spacetimeefficienthardwarecompatiblecomplexquantum,zhao2026ultrahighratequantumerrorcorrection,bhardwaj2026highrateqldpcprocessors,Xu2024ConstantOverhead,gu2026qgpuparallellogicquantum},
while others combine bivariate-bicycle and other qLDPC codes with modular or long-range-connected hardware~\cite{yoder2025tourgrossmodularquantum,liu2026assessingcapabilitiesbottlenecksearly,cain2026shorsalgorithmpossible10000,webster2026pinnaclearchitecturereducingcost}.
Although these systems retain the need for real-time coordination, they may change both its granularity and its constraints.
In particular, high-rate codes that encode more than one logical qubits per block introduce new static decisions about logical-qubit placement across blocks and new dynamic dependencies when operations or measurement outcomes couple multiple encoded qubits.
The available logical gadgets may also depend strongly on the code, hardware connectivity, and surrounding circuit context.
A central question for future work is therefore how \ftname{}'s stage interfaces should evolve when the unit of scheduling changes from individual logical patches to multi-logical-qubit code blocks.

A second direction is to model more realistic hardware organization.
The present implementation largely assumes a common computational region with reconfigurable movement, 
whereas neutral-atom processors are increasingly adopting zoned architectures with dedicated storage, entangling, readout, and reservoir regions~\cite{Bluvstein_2023, Bluvstein_2025, cain2026shorsalgorithmpossible10000}.
In such systems, routing depends not only on distance and AOD contention but also on when logical blocks should enter specialized zones, 
how limited zone capacity should be allocated, 
and how movement should be coordinated with measurement, reset, and atom replenishment.
Extending \ftname{} with explicit zone-aware resource-allocation and routing models would enable the runtime to jointly schedule computation, communication, measurement, and qubit reuse across heterogeneous hardware regions.


The framework also distinguishes the components of \ftname{} that are specific to reconfigurable neutral-atom hardware from those that apply more broadly.
The static--dynamic partition, stage-based execution model, feedback-driven resource assignment, correction handling, and comparison of execution policies under a common runtime are not inherently tied to atom movement.
In contrast, AOD scheduling, atom-transport optimization, and the movement schedules studied here directly exploit neutral-atom reconfigurability.
For architectures with fixed connectivity, modular interconnects or lattice-surgery-based communication, 
the routing stage would instead expose the corresponding communication primitives and constraints.
We therefore view the current implementation of \ftname{} primarily as a framework for reconfigurable FTQC architectures, while its stage-aware real-time compilation principles could support hardware-specific instantiations for a broader range of architectures.

Finally, future real-time compilers could adapt execution policies dynamically rather than selecting a fixed policy for an entire run.
Based on online feedback, a runtime could transition between synchronous and asynchronous scheduling, adjust factory allocation, or reconfigure logical layouts in response to hardware utilization, routing congestion, factory availability, and predicted resource demand.
Such adaptation must account for the latency and disruption introduced by policy changes, making the design of low-overhead decision mechanisms an important direction for future work.

This capability may be especially valuable for large, heterogeneous fault-tolerant workloads.
Applications such as quantum chemistry simulation and factoring contain subroutines with substantially different non-Clifford demand, available parallelism, communication patterns, and logical layouts~\cite{RevModPhys.68.733,proos2004shorsdiscretelogarithmquantum,10.1007/978-3-319-70697-9_9,harry2025resource_estimation,cain2026shorsalgorithmpossible10000,low2019hamiltonian,93zr-1ykb}.
Although our evaluation focuses on a relatively homogeneous TFIM workload, \ftname{} exposes policy parameters that could be adapted at subroutine boundaries or dynamically during execution.
Extending the evaluation to heterogeneous workloads would test when online policy selection improves end-to-end performance.
More broadly, these capabilities point toward a self-adaptive systems layer that continuously co-optimizes logical execution, resource generation, communication, and hardware control as FTQC architectures grow in scale and diversity.

\begin{acknowledgments}
W.-H. Lin and J. Cong are partially supported by NSF grants No.~2313083 and No.~2533041.
The authors thank Hanyu Wang, Adrian Liu, Dennis Liew and Jens Palsberg for valuable discussions.
The \ftname{} implementation and evaluation examples are publicly available at
\url{https://github.com/UCLA-VAST/FT-Weave}.
\end{acknowledgments}

\appendix
\section{Compilation Methods for EFTQC Architectures}
\label{appendix:compilation_method}

This section provides the detailed formulations and implementation procedures for the compilation techniques introduced in the main text. 
These techniques address runtime tasks shared by the EFTQC architectures considered in this work, 
including dynamic angle collection, factory assignment, teleportation routing, movement scheduling, and resource rematerialization.
We formulate each technique as a separate compilation module with a well-defined input, objective, and output. 
Although several modules interact during execution, 
this separation allows \ftname\ to replace individual heuristics without changing the overall runtime flow. 

\subsection{Implementation of Dynamic Angle Collection}
\label{appendix:angle_collection}

This section describes the heuristic used to distribute the available factory budget across current rotation demands and speculative RUS correction angles. 
The complete procedure is summarized in~\autoref{alg:angle_collection}.

\begin{algorithm}[t]
\caption{Angle Collection}
\label{alg:angle_collection}
\KwInput{
Active rotation requests $\mathcal{Q}$,
available factory budget $F$
}
\KwOutput{Angle-demand slots $\mathcal{A}$}

$L:=$maximum lookahead level \;
$L':=$coarse lookahead level with $L'\leq L$ \;
\ForEach{$q\in\mathcal{Q}$}{ \label{line:global_allocation}
    \ForEach{$\ell\in[0\ldots L]$}{ 
        Generate RUS angle
        $\theta_q^{(\ell)}$\;
        Estimate demand
        $\delta_{q,\ell}
        \leftarrow 2^{-\ell}/p_{\mathrm{succ}}(\theta_q^{(\ell)})$\;  \label{line:estimate_demand}
    }
    Coarse demand
    $\Delta_q\leftarrow0$\;
    \ForEach{$\ell\in[0\ldots L']$}{
        $\Delta_q\leftarrow\Delta_q+\delta_{q,\ell}$\; 
    }
}

Allocate $F$ across $q\in\mathcal{Q}$
proportionally to $\Delta_q$,
using largest-remainder rounding,
yielding $a_q$\; \label{line:global_allocation_end}

$\mathcal{A}\leftarrow\emptyset$\;

\ForEach{$q$ with $a_q>0$}{ \label{line:local_allocation}
    $B\leftarrow a_q$\;

    \For{$\ell\leftarrow0$ \KwTo $L$}{ \label{line:sub_local_allocation}
        $b\leftarrow
        \min(B,\lceil\delta{q,\ell}\rceil)$\;
        Add $b$ slots for $(q,\theta_q^{(\ell)},\ell)$ to $\mathcal{A}$\;
        $B\leftarrow B-b$\;
    
        \If{$B=0$}{\Break} \label{line:sub_local_allocation_end}
    }
    
    \If{$B>0$}{ \label{line:final_distribution}
        Distribute $B$ across levels proportionally to
        $\delta_{q,\ell}$ using largest-remainder rounding, and add the resulting slots to $\mathcal{A}$\; \label{line:local_allocation_end}
    }
}

\Return{$\mathcal{A}$}\;
\end{algorithm}

\autoref{alg:angle_collection} consists of two allocation stages. 
It first determines how much of the global factory budget should be assigned to each active rotation request (line~\ref{line:global_allocation}--line~\ref{line:global_allocation_end}), 
and then distributes each request's assigned budget along its RUS correction sequence (line~\ref{line:local_allocation}--line~\ref{line:local_allocation_end}).
The estimated demand $\delta_{q,\ell}$ combines the probability $2^{-\ell}$ of reaching lookahead level $\ell$ with the expected preparation cost $1/p_{\mathrm{succ}}(\theta_q^{(\ell)})$ at that level.
Thus, lower preparation-success probabilities increase estimated demand,
while the factor $2^{-\ell}$ discounts increasingly speculative correction levels.
With maximum lookahead level $L=3$ and coarse lookahead level $L'=1$,
the coarse demand $\Delta_q$ includes only the current rotation and first correction level, when present, so that the global allocation emphasizes near-term demand.
The available factories are distributed across active requests proportionally to $\Delta_q$, 
using largest-remainder rounding to obtain integer allocations.

In the second stage, each request's budget is distributed from shallow to deeper RUS levels. 
Each level receives up to its estimated demand before the next level is considered, 
giving priority to the current rotation and near-term corrections (line~\ref{line:sub_local_allocation}--line~\ref{line:sub_local_allocation_end}). 
If capacity remains after all considered levels are satisfied, 
the residual budget is redistributed proportionally across the levels using largest-remainder rounding (line~\ref{line:final_distribution}). 
The resulting angle-demand slots $\mathcal{A}$ are passed to the subsequent factory-assignment stage.

\subsection{Implementation of Routing-Aware Factory Assignment}
\label{appendix:factory_assignment_details}

As introduced in \autoref{sec:compilation_eftqc:factory_assignment},
\ftname{} uses the same bipartite-matching abstraction for both angle assignment before resource preparation and teleportation assignment after preparation.
The two cases differ in the candidate requests, compatibility conditions, and edge weights used to construct the graph.
Once the graph is constructed, 
both use the same minimum-cost maximum-cardinality assignment procedure, 
as summarized in \autoref{alg:factory_assignment}.

\begin{algorithm}[t]
\caption{Routing-Aware Factory Assignment}
\label{alg:factory_assignment}
\KwInput{
Assignment type $\mu\in\{\textsc{Prepare},\textsc{Teleport}\}$,
candidate factories $\mathcal{F}$,
candidate requests $\mathcal{R}$
}
\KwOutput{Assignment $\mathcal{M}$}

Initialize bipartite graph $G=(\mathcal{F}\cup\mathcal{R},E)$\;

\ForEach{$f\in\mathcal{F}$ and $r\in\mathcal{R}$}{
    \If{$f$ and $r$ are compatible under assignment type $\mu$}{
        Compute edge weight $w_\mu(f,r)$ 
        \Comment{\autoref{eq:prepare_assignment_cost},\autoref{eq:teleport_assignment_cost}}Add weighted edge $(f,r)$ to $E$\;
    }
}

$\mathcal{M}\leftarrow$
minimum-cost maximum matching on $G$,
implemented as a rectangular linear-assignment problem using the
Jonker--Volgenant algorithm~\cite{jonker1988shortest};

\Return{$\mathcal{M}$}\;
\end{algorithm}

\paragraph{Angle Assignment}

For preparation assignment, 
the candidate factory set $\mathcal{F}$ contains idle factories, 
while each request is an angle-demand slot $a=(q,\theta,\ell)$,
where $q$ is the target logical qubit, $\theta$ is the resource angle,
and $\ell$ is its RUS lookahead level.
Because any idle STAR factory can prepare any requested angle,
the corresponding bipartite graph is complete.

The edge weight jointly captures the expected routing cost and the priority of the demand:
\begin{equation}
\label{eq:prepare_assignment_cost}
w_{\textsc{Prepare}}(f,a)
=
\frac{d(f,q)}{\ell+1},
\end{equation}
where $d(f,q)$ is the estimated routing cost from factory $f$ to logical qubit $q$.
We use Manhattan distance in our implementation.

The lookahead factor $\frac{1}{\ell+1}$ assigns greater weight to routing distance at lower lookahead levels. 
Since all demand slots participate in the fixed-cardinality matching, 
this prioritizes minimizing movement for current requests while allowing deeper speculative requests to absorb larger routing costs.

\paragraph{Teleportation Assignment}

After preparation, the candidate factory set contains successfully prepared resource states, 
and each request is represented as $r=(q,\theta)$.
Unlike preparation assignment, 
an edge exists only when the resource held by factory $f$ matches the angle requested by $q$.
For each compatible pair, the edge weight is simply
\begin{equation}
\label{eq:teleport_assignment_cost}
w_{\textsc{Teleport}}(f,r)
=
d(f,q).
\end{equation}

The resulting graph decomposes into independent connected components corresponding to  different resource angles.
Solving the common assignment formulation therefore allows successfully prepared states to be shared among all compatible logical-qubit requests while minimizing their total routing cost.

\subsection{Movement and Routing Optimization}
\label{appendix:movement_optimization}

This section provides the detailed formulation of the movement and routing optimization introduced in \autoref{sec:compilation_eftqc:move_optimization}.
The overall procedure is summarized in \autoref{alg:forward_movement} and \autoref{alg:return_movement}.
Forward routing first partitions movements according to the AOD atom-transfer constraints and minimizes the number of sequential movement batches within each partition.
After teleportation, the lower occupancy of the array provides additional routing flexibility.
\ftname{} exploits this flexibility through destination reassignment, movement rebatching, and relay decomposition.

\begin{algorithm}[t]
\caption{Forward Movement Optimization}
\label{alg:forward_movement}
\KwInput{Factory-to-qubit movements $\mathcal{M}$}
\KwOutput{Forward batches $\mathcal{B}_{\mathrm{fwd}}$}

$\mathcal{P}_{\mathrm{fwd}}
\leftarrow
\textsc{PartitionByTransferSignature}(
    \mathcal{M})$\;
\label{alg:fwd_movement:partition}

$\mathcal{B}_{\mathrm{fwd}}\leftarrow\emptyset$\;

\ForEach{$\mathcal{P}\in\mathcal{P}_{\mathrm{fwd}}$}{
    $\mathcal{G}_{\mathcal{P}}
    \leftarrow
    \textsc{BuildPartialOrder}(
        \mathcal{P})$\;
    \Comment{\autoref{eq:movement_order}}
    \label{alg:fwd_movement:chain_start}

    $\mathcal{C}
    \leftarrow
    \textsc{MinChainDecomposition}(
        \mathcal{G}_{\mathcal{P}})$\;
    \label{alg:fwd_movement:chain_end}

    $\mathcal{B}_{\mathrm{fwd}}
    \leftarrow
    \mathcal{B}_{\mathrm{fwd}}\cup\mathcal{C}$\;
}

$\mathcal{B}_{\mathrm{fwd}}
\leftarrow
\textsc{MergeSameDisplacementBatches}(
    \mathcal{B}_{\mathrm{fwd}})$\;
\label{alg:fwd_movement:merge}

\Return{$\mathcal{B}_{\mathrm{fwd}}$}\;
\end{algorithm}

\begin{algorithm}[t]
\caption{Return Movement Optimization}
\label{alg:return_movement}
\KwInput{
Returning factories $\mathcal{F}_{\mathrm{ret}}$,
}
\KwOutput{Return batches $\mathcal{B}_{\mathrm{ret}}$}
$\mathcal{E}\leftarrow$ Extract available sites \; \label{alg:return_movement:available}
$\pi
\leftarrow
\textsc{MinCostReturnAssignment}(
    \mathcal{F}_{\mathrm{ret}},\mathcal{E})$\;

$\mathcal{M}_{\mathrm{ret}}
\leftarrow
\textsc{BuildReturnMoves}(\pi)$\;
\label{alg:return_movement:assignment}

$\mathcal{P}_{\mathrm{ret}}
\leftarrow
\textsc{PartitionByTransferSignature}(
    \mathcal{M}_{\mathrm{ret}})$\;

$\mathcal{B}_{\mathrm{ret}}\leftarrow\emptyset$\;

\ForEach{$\mathcal{P}\in\mathcal{P}_{\mathrm{ret}}$}{
    $\mathcal{G}_{\mathcal{P}}
    \leftarrow
    \textsc{BuildPartialOrder}(
        \mathcal{P})$\;

    $\mathcal{C}
    \leftarrow
    \textsc{MinChainDecomposition}(
        \mathcal{G}_{\mathcal{P}})$\;

    $\mathcal{B}_{\mathrm{ret}}
    \leftarrow
    \mathcal{B}_{\mathrm{ret}}\cup\mathcal{C}$\;
}

$\mathcal{B}_{\mathrm{ret}}
\leftarrow
\textsc{MergeAODCompatibleBatches}(
    \mathcal{B}_{\mathrm{ret}})$\;
\label{alg:return_movement:rebatch}

$\mathcal{B}_{\mathrm{ret}}
\leftarrow
\textsc{RelayCriticalMoves}(
    \mathcal{B}_{\mathrm{ret}},\mathcal{E})$\;
\label{alg:return_movement:relay}

\Return{$\mathcal{B}_{\mathrm{ret}}$}\;
\end{algorithm}

\subsubsection{AOD-Compatible Movement Batching}
Forward routing first partitions movements according to their source--destination transfer signatures (\autoref{alg:fwd_movement:partition}).
For row-based movement, transfers sharing the same source and destination rows
belong to the same partition. 
Column-based movement is handled analogously.
Within each partition, \ftname{} minimizes the number of sequential movement batches by formulating trajectory compatibility as a minimum chain-decomposition problem.

Consider two movements $i$ and $j$.
Let $\mathrm{pos}_{\mathrm{src}}(\cdot)$ and $\mathrm{pos}_{\mathrm{dst}}(\cdot)$ denote their positions along the
source and destination rows, respectively.
We define
\begin{equation} \label{eq:movement_order}
  \begin{aligned}
    i \prec j \iff &\ \mathrm{pos}_{\mathrm{src}}(i) < \mathrm{pos}_{\mathrm{src}}(j) \\
                  &\ \land\ \mathrm{pos}_{\mathrm{dst}}(i) < \mathrm{pos}_{\mathrm{dst}}(j).
  \end{aligned}
\end{equation}
Movements related by this partial order preserve their relative ordering and can therefore belong to the same movement batch.
A valid batch corresponds to a chain in the resulting partial order, 
so minimizing movement depth is equivalent to finding a minimum chain decomposition.

As shown in \autoref{alg:fwd_movement:chain_start}, 
\ftname{} constructs this partial order for each transfer partition and obtains its minimum chain decomposition using the standard reduction from Dilworth's theorem~\cite{Dilworth1990} to bipartite maximum matching (\autoref{alg:fwd_movement:chain_end}).
Specifically, we construct left and right copies of the movements and add an edge $(i_L,j_R)$ whenever $i\prec j$.
We compute the maximum matching using the Edmonds--Karp algorithm~\cite{edmonds1972}, 
and the resulting chains directly define the parallel movement batches.
Because the transfer partitions are independent, their matching problems can be solved in parallel.

After constructing the batches within individual partitions, 
\ftname{} merges batches across rows or columns when they have the same displacement vector and atom-transfer signature (\autoref{alg:fwd_movement:merge}).
This cross-partition merging preserves the transfer constraints while reducing the overall forward-routing depth.

\subsubsection{Return-Move Optimizations}

After teleportation, \ftname{} identifies the currently available empty sites as candidate return locations (\autoref{alg:return_movement:available}),
and exploits these sites through destination reassignment, movement rebatching, and relay decomposition.

\paragraph{Destination Reassignment.}
Factories are not required to return to their original locations.
Given the returning factories $\mathcal{F}_{\mathrm{ret}}$ and available sites $\mathcal{E}$, 
the destination assignment in \autoref{alg:return_movement:assignment} solves
\begin{equation}
\label{eq:return_assignment}
\min_{\pi}
\sum_{f\in\mathcal{F}_{\mathrm{ret}}}
d(f,\pi(f)),
\end{equation}
where $\pi(f)\in\mathcal{E}$ and $d(\cdot,\cdot)$ is the Manhattan distance.
We formulate this optimization as a minimum-cost bipartite matching between returning factories and available sites.
The resulting assignment minimizes aggregate return distance and determines the movements processed by the subsequent batching stage; it does not by itself minimize the latency of the complete return schedule.

\paragraph{Movement Rebatching.}
The induced return movements are first processed using the same transfer-signature partitioning and minimum-chain-decomposition procedure as forward routing.
Because return routing operates on a sparser array, 
the AOD activation constraints are generally easier to satisfy.
For example, even when the intended pickup locations do not form a complete Cartesian product, 
the activation is still valid as long as all other sites in the induced Cartesian product are empty.
Accordingly, \ftname{} greedily merges the initial return batches whenever their combined movement remains AOD-compatible (\autoref{alg:return_movement:rebatch}).
This second batching stage exploits the additional pickup opportunities created by the available empty sites without changing the underlying forward-batching formulation.

\paragraph{Relay Decomposition.}
The latency of a movement batch is determined by its longest transfer.
To reduce the return-routing critical path,
$\textsc{RelayCriticalMoves}$ in \autoref{alg:return_movement:relay} may replace a long movement $s\rightarrow d$ with
$s\rightarrow p, p\rightarrow d,$
where $p$ is an available intermediate site.
The decomposition is retained only when the resulting movements remain feasible and reduce the estimated latency of the complete return schedule.
This allows long return movements to be divided into shorter transfers
that can overlap with other movement batches.

\subsection{Resource Rematerialization Heuristic}
\label{appendix:operation_dropout}

This section details the resource-rematerialization heuristic introduced in
\autoref{sec:compilation_eftqc:operation_dropout}.
As summarized in \autoref{alg:rematerialization}, the runtime first evaluates
whether individual routing batches should be deferred and then determines
whether the remaining teleportation round should be deferred entirely.

\begin{algorithm}[t]
\caption{Resource Rematerialization}
\label{alg:rematerialization}
\KwInput{
Routing batches $\mathcal{B}$,
}
\KwOutput{Retained routing batches $\mathcal{B}'$}

$\mathcal{B}'\leftarrow\mathcal{B}$\;
ationsurfacecode
\tcp{Partial rematerialization}
\If{partial rematerialization is enabled}{
    \ForEach{$B\in\mathcal{B}$}{
        Compute $C_{\mathrm{batch}}(B)$
        \Comment{\autoref{eq:remat_batch_cost}}
        \label{alg:remat:batch_cost}
        \If{$C_{\mathrm{batch}}(B)\ge t_{\mathrm{prep}}$
            \textbf{and} $\textsc{CanRematerialize}(\mathcal{B},\mathcal{S})$\label{alg:remat:condition}
}{
            \If{\textsc{HasBetterIdleFactory}$(B)$
                \label{alg:remat:better_factory}}{
                $\mathcal{B}'\leftarrow
                \mathcal{B}'\setminus\{B\}$\;
            }
        }
    }
    \label{alg:remat:partial}
}
\tcp{Full-round rematerialization}
$T_{\mathrm{move}}
\leftarrow
\textsc{EstimateMovementLatency}(\mathcal{B}')$\;
\label{alg:remat:aod_schedule}

\If{$T_{\mathrm{move}}+t_{\mathrm{CNOT}}>t_{\mathrm{prep}}$
    \textbf{and} $\textsc{CanRematerialize}(\mathcal{B},\mathcal{S})$ 
}{
    \Return{$\emptyset$}\;
\label{alg:remat:full}
}
\Return{$\mathcal{B}'$}\;
\end{algorithm}

Here, the runtime state $\mathcal{S}$ contains the number of logical qubits, the available idle factories and AOD resources, and whether additional resource preparation is required.
The predicate $\textsc{CanRematerialize}(\mathcal{B}',\mathcal{S})$ is true when the requests represented by $\mathcal{B}'$ involve no more than 10\% of the logical qubits and additional preparation is required.

\subsubsection{Partial Rematerialization}

For each routing batch $B$, 
\ftname{} estimates the amortized cost of
executing its teleportations as
\begin{equation}
\label{eq:remat_batch_cost}
C_{\mathrm{batch}}(B)
=
\frac{2\tau(B)}{|B|}
+
t_{\mathrm{CNOT}},
\end{equation}
where $\tau(B)$ is the latency of the longest factory-to-qubit movement in the batch and $|B|$ is the number of teleportations in the batch.
The factor of two approximates the corresponding forward and return movement.

As shown in \autoref{alg:remat:partial}, 
partial rematerialization is considered when this cost is at least the latency of one additional preparation round, no more than 10\% of the logical qubits currently participate in teleportation, and additional preparation is required.
Before removing the batch, 
\textsc{HasBetterIdleFactory} verifies that every qubit in the batch has at least one idle factory whose movement latency is smaller than that of its current assignment (\autoref{alg:remat:better_factory}).
If so, the batch is deferred and its factory-to-qubit assignments are removed
from the current round.

If a candidate batch does not admit such an improving reassignment, 
\ftname{} retains that batch.
This conservative rule avoids delaying its teleportations when re-preparation is unlikely to improve their routing configuration.

\subsubsection{Full-Round Rematerialization}

After partial rematerialization, 
\ftname{} estimates the forward and return movement latency of
the retained routing schedule under the available AOD parallelism using the scheduling strategy from Ref.~\cite{zac}.
The runtime defers the entire remaining teleportation round when
\begin{equation}
\label{eq:remat_round_cost}
T_{\mathrm{move}}
+
t_{\mathrm{CNOT}}
>
t_{\mathrm{prep}},
\end{equation}
the remaining teleportation requests involve no more than $10\%$ of the logical
qubits, and additional resource preparation is required
(\autoref{alg:remat:full}).
In this case, the runtime begins another preparation round instead of executing the retained teleportations.
Because replacement preparation is stochastic, this comparison is a heuristic rather than a guarantee of lower realized latency.
\section{Experimental Settings and Fidelity Evaluation}
\label{sec:appendix:exp}

This appendix describes the execution-time and fidelity models used in our evaluation.
The runtime evaluation uses surface-code distance $d=9$.
The fidelity evaluation additionally considers STAR at $d=7$ and the $T$-state cultivation architecture at $d=7$ and $d=13$, alongside the common $d=9$ configuration.

\subsection{Execution Time Model}
\label{sec:appendix:exp_setting}

\subsubsection{Common Timing and Movement Model} 
\label{sec:exp_setting:common} 
All execution times are measured in QEC cycles, 
with one cycle corresponding to the duration of one syndrome-extraction round.
We adopt timing parameters from prior architecture studies: 
STAR follows Ref.~\cite{ismail2025transversal}, while 
$T$-state cultivation architecture follows Ref.~\cite{sahay2026foldtransversalsurfacecodecultivation}.
\autoref{tab:operation-timing} summarizes the operation latencies.
Following the correlated-decoding scheme in Ref.~\cite{correlated_decoding}, 
the one-cycle latency assigned to each logical Clifford operation includes the additional syndrome-extraction round performed after that operation.

\begin{table}[htbp] 
\centering 
\begin{tabular}{l|c} 
\hline\hline 
Operation & QEC cycles \\ \hline 
$H$, $S$, CNOT & 1 \\ 
TMR & 6 \\ 
$T$-state cultivation check stage & 12.5 \\ 
$T$-state cultivation escape stage, $d=7,9$ & 0.5 \\ 
$T$-state cultivation escape stage, $d=13$ & 6 \\ \hline\hline \end{tabular} 
\caption{Operation latencies used throughout the evaluation. 
Durations are normalized to QEC cycles; the preparation-stage entries give the duration of one attempt.}
\label{tab:operation-timing} 
\end{table} 

For logical-patch movement, we use horizontal-then-vertical routing as the evaluated heuristic.
Moving a logical patch by one row or column requires 0.5 QEC cycles, giving the latency
$t_{\mathrm{move}}
=
0.5\left(|x'-x|+|y'-y|\right)$
between patch locations $(x,y)$ and $(x',y')$.
Idle logical qubits continue to accumulate physical errors.
Whenever a logical qubit remains idle for ten QEC cycles, we schedule an additional syndrome-extraction round.

\subsubsection{STAR Execution Model} 
\label{sec:exp_setting:star} 

The success probability of preparing a logical rotation state depends on the target logical angle $\theta$ and code distance $d$.
Because this probability is symmetric under $\theta\mapsto-\theta$, we model it as a function of $|\theta|$.

For distance $d=7$, we fit the simulation results from~\cite{ismail2025transversal} using
\begin{equation}
p_{\mathrm{succ}}(\theta,7)
=
a|\theta|^b+c,
\end{equation}
where $(a,b,c)=(-0.47,\,0.47,\,0.49)$. This empirical fit is used over the range of rotation angles considered in the evaluation.

For distance $d=9$, we adopt the analytical model from~\cite{ismail2025transversal},
\begin{equation}
p_{\mathrm{succ}}(\theta,9)
=
\epsilon_{\mathrm{init}}
\left(
\sin^{2k}(\theta_{\mathrm{phys}})
+
\cos^{2k}(\theta_{\mathrm{phys}})
\right),
\end{equation}
where $\theta_{\mathrm{phys}}$ is the physical rotation angle corresponding
to the target logical rotation,
$k=4$, and
$\epsilon_{\mathrm{init}}=0.7872$ is the logical-qubit initialization
success probability obtained by fitting the simulation results in
Ref.~\cite{ismail2025transversal}.
and $\theta_{\mathrm{phys}}$ is the physical rotation angle corresponding to the target logical rotation.
We adopt $k=\lfloor d/2\rfloor$ following the model of
Ref.~\cite{ismail2025transversal}; this choice is specific to that model
rather than an intrinsic property of TMR.


\subsubsection{T-State Cultivation Architecture Execution Model} 
\label{sec:exp_setting:t_cultivation} 
For the fault-distance-5 magic-state cultivation protocol (MSC-5), we model preparation as a check stage followed by an escape stage.
Each check-stage attempt succeeds with probability
$p_{\mathrm{check}}=0.25$.
A state that passes the check stage proceeds to escape, whose acceptance probability depends on the postselection criterion and target logical error rate.
To produce $|T\rangle$ states with an assigned output error of $10^{-8}$, comparable to the STAR resource-state target used in our fidelity model, we set
$p_{\mathrm{escape}}=0.90$.
These probabilities are parameters of the evaluated execution model rather than universal properties of MSC-5.

As summarized in \autoref{tab:operation-timing}, each check-stage attempt requires 12.5 QEC cycles.
Each escape-stage attempt requires 0.5 QEC cycles for final code distances $d=7,9$ and 6 QEC cycles for $d=13$ according to Ref.~\cite{sahay2026foldtransversalsurfacecodecultivation}.

\subsection{Fidelity Model}
\label{sec:appendix:fidelity_model}
We next describe the physical and logical error models used to estimate circuit fidelity.

\subsubsection{Physical Error Model}
We adopt the neutral-atom physical error model from Ref.~\cite{ismail2025transversal}, 
summarized in \autoref{tab:physical-error-model}.
The \emph{basic} model represents current neutral-atom operation error rates.
For the fidelity evaluation, 
we use a \emph{projected} model in which each operation error rate is reduced by a factor of 10.
The coherence time is similarly increased from $T_2=1.5\,\mathrm{s}$ to $15\,\mathrm{s}$.


\begin{table}[htbp]
\centering
\begin{tabular}{lc}
    \hline
    Operation & Basic error rate\\
    \hline
    Initialization $p_{\mathrm{init}}$ & $3.0 \times 10^{-3}$ \\
    Single-qubit gate $p_1$ & $3.0 \times 10^{-3}$ \\
    Move $p_\mathrm{move}$ & $2.8 \times 10^{-4}$ \\
    Move loss $p_{\mathrm{move},\mathrm{loss}}$  & $2.0 \times 10^{-5}$ \\
    CZ $p_\mathrm{CZ}$ & $2.5 \times 10^{-3}$ \\
    CZ spectator $p_\mathrm{CZ,spec}$ & $2.3 \times 10^{-3}$ \\
    CZ loss $p_\mathrm{CZ,loss}$ & $1.3 \times 10^{-3}$ \\
    Measurement $p_\mathrm{meas}$ & $4.0 \times 10^{-3}$\\
    Effective physical rate $p_{\mathrm{ph}}$ & $7.3 \times 10^{-3}$\\
    \hline
\end{tabular}
\caption{Neutral-atom physical error parameters used in this work. 
The projected model is obtained by dividing every error rate in the basic model by 10.}
\label{tab:physical-error-model}
\end{table}

Because the evaluated TFIM circuits are too large for direct state-vector simulation, 
we estimate the fidelity of the unencoded physical implementation using a multiplicative error model similar to Ref.~\cite{zac}:
\begin{equation}
    f_{\mathrm{phys}}
    =
    f_{\mathrm{init}} \cdot
    f_{\mathrm{1q}} \cdot
    f_{\mathrm{CZ}} \cdot
    f_{\mathrm{move}} \cdot
    f_{\mathrm{meas}} \cdot
    f_{\mathrm{idle}}.
\end{equation}

The individual contributions are
\begin{align*}
    f_{\mathrm{init}}
    &=
    (1-p_{\mathrm{init}})^{n_{\mathrm{init}}},
    &
    f_{\mathrm{1q}}
    &=
    (1-p_1)^{n_{\mathrm{1q}}},
    \\
    f_{\mathrm{meas}}
    &=
    (1-p_{\mathrm{meas}})^{n_{\mathrm{meas}}},
    &
    f_{\mathrm{idle}}
    &=
    \prod_{q\in Q}
    \exp\left(-\frac{t_q}{T_2}\right),
\end{align*}

\begin{equation*}
    f_{\mathrm{CZ}}
    =
    (1-p_{\mathrm{CZ}})^{n_{\mathrm{CZ}}}
    (1-p_{\mathrm{CZ,spec}})^{n_{\mathrm{spec}}}
    (1-p_{\mathrm{CZ,loss}})^{n_{\mathrm{CZ,loss}}},
\end{equation*}
and
\begin{equation*}
    f_{\mathrm{move}}
    =
    (1-p_{\mathrm{move}})^{n_{\mathrm{move}}}
    (1-p_{\mathrm{move,loss}})^{n_{\mathrm{move}}}.
\end{equation*}
Here, $Q$ is the set of physical qubits, 
$t_q$ is the cumulative idle time of qubit $q$, 
and the model assumes one movement-error and one movement-loss opportunity per counted movement.
The operation counts are derived from the compiled or analytically constructed execution schedule; the movement count is estimated separately as described below.

For each second-order Trotter step, we estimate the number of movement operations as
\begin{equation*}
n_\mathrm{move}
=
2\times2\times3
\left\lfloor
\frac{L}{2}
\right\rfloor
L.
\end{equation*}
The first factor of two accounts for the two lattice directions, 
the second accounts for the forward and reverse halves of the second-order Trotter sequence, 
and the factor of three accounts for displacement through the two CZ layers followed by return movement.
We assume $12\,\mu\mathrm{m}$ spacing between entanglement sites and an atom-transport acceleration of $5500\,\mathrm{m}/\mathrm{s}^2$.
Idle errors are computed from the cumulative waiting time of each physical qubit, 
including movement and atom-transfer latency. For the coherence factor, $t_q$ and $T_2$ are both expressed in seconds.

\subsubsection{Logical Error Model}
For surface-code distance $d=7$ under the lookahead physical error model, 
we use the logical error rates reported in Ref.~\cite{ismail2025transversal}.
A logical identity $I$ corresponds to one syndrome-extraction round.
Each entry in \autoref{tab:logical-error-model} is the total logical error probability of the indicated operation together with its subsequent syndrome-extraction round.

\begin{table}[t]
\centering
\begin{tabular}{l|cccc}
    \hline\hline
    Logical operation & $I$ & $H$ & $S$ & $\mathrm{CNOT}$ \\ \hline
    Error rate & $2 \times 10^{-7}$ & $4 \times 10^{-7}$ & $1.46 \times 10^{-6}$ & $3 \times 10^{-6}$ \\
    \hline\hline
\end{tabular}
\caption{
Logical error rates at surface-code distance $d=7$ under the projected physical error rate $p_{\mathrm{ph}}=7.3\times10^{-4}$. Each logical gadget includes one subsequent syndrome-extraction round.
}
\label{tab:logical-error-model}
\end{table}

For other code distances, we extrapolate the logical CNOT error rate using
\begin{equation}
\epsilon_{\mathrm{CNOT}}(d) = A\left(\frac{p_{\mathrm{ph}}}{p_c}\right)^{\lceil (d+1)/2 \rceil},
\end{equation}
where $A=0.0579$ and $p_c=8.5\times 10^{-3}$.
We compute the scaling relative to the distance-$7$ CNOT error rate and apply the same factor to the $I$, $H$, and $S$ error rates.

For STAR, we fit the logical $R_Z(\theta)$ fidelity from the simulation data of Ref.~\cite{ismail2025transversal}:
\begin{equation}
f_{R_Z(\theta)} = 1 - a|\theta|^b + c,
\end{equation}
where $a=3.14\times 10^{-4}$, $b=1.46$, and $c=0$.
For the evaluated TFIM circuit, the chosen Trotter step produces
the logical rotation angle is approximately $1.46\times10^{-3}$, 
corresponding to a fitted rotation error of approximately $2.3 \times 10^{-8}$.

\subsubsection{STAR Architecture Fidelity}
Under an independent-logical-gadget approximation, we estimate the STAR logical-circuit fidelity as
\begin{equation}
\begin{split}
f_{\mathrm{STAR}}
={}
(f_{\mathrm{CNOT}})^{n_{\mathrm{CNOT}}}
&(f_H)^{n_H}
(f_S)^{n_S} 
\\
& \times (f_I)^{n_I}
\prod_{\theta\in\Theta} f_{R_Z}(\theta),
\end{split}
\label{eq:logical_circuit_fidelity}
\end{equation}
where $n_{\mathrm{CNOT}}$, $n_H$, $n_S$, and $n_I$ are the corresponding logical-gadget counts, and $\Theta$ is the multiset of all rotation angles teleported during the RUS protocol.
The logical-operation counts are obtained from the runtime trace.
In particular, RUS outcomes that require correction generate additional rotation requests and teleportation CNOTs, so $n_{\mathrm{CNOT}}$ and $\Theta$ depend on the realized execution.
To model idle error correction, an additional syndrome-extraction round is inserted whenever a logical qubit accumulates ten QEC cycles of idle time; these rounds contribute to $n_I$.
Initialization, measurement, and movement errors are already captured by the logical-gadget error rates and are therefore not included again as separate logical-level factors

\subsubsection{T-State Cultivation Architecture Fidelity}
For $T$-state cultivation architecture, each arbitrary-angle $R_Z$ rotation is synthesized into $T$, $H$, and $S$ gates using Gridsynth~\cite{rzsynthesis} with synthesis accuracy $\epsilon$.
The resulting $T$-gate count per rotation is approximately
$n_T^\theta \approx
-3\log_2\left(\sqrt{\epsilon}\right)$.
Under a first-order independent-error approximation, the error of a synthesized rotation is
\begin{equation}
\epsilon_{R_Z}
=
\epsilon
+
n_T^\theta\epsilon_T
+
n_{\mathrm{Clifford}}^{(R_Z)}
\epsilon_{\mathrm{Clifford}},
\end{equation}
where the terms capture synthesis error, logical $T$-gate error, and logical Clifford error, respectively.
We set $\delta=10^{-8}$ to provide a target comparable to STAR, 
yielding approximately 40 $T$ gates per logical $R_Z$ rotation.
For MSC-5 cultivation, each logical $|T\rangle$ state is assigned an output error of $10^{-8}$, such that $f_T=1-\epsilon_T$. This factor captures the cultivated-state error; the logical Clifford errors associated with gate teleportation are counted separately.

Under the same independent-gadget approximation, the logical-circuit fidelity is estimated as
\begin{equation}
\begin{split}
f_{\mathrm{cult}}
={}
&(f_{\mathrm{CNOT}})^{n_{\mathrm{CNOT}}}
(f_H)^{n_H}
\\
&\times
(f_S)^{n_S}
(f_I)^{n_I}
(1-\delta)^{n_{R_Z}}
(f_T)^{n_T},
\end{split}
\end{equation}
where $n_T$ is the total number of logical $T$ gates after synthesis and $n_{R_Z}$ is the number of synthesized rotations.
Unlike in STAR, the number of teleportation CNOTs is fixed once the synthesized sequences are known: each $T$-gate teleportation consumes one $|T\rangle$ state regardless of the measurement outcome.
An outcome may require an $S$ correction, but it does not recursively generate additional resource-state demand; consequently, the realized $S$-gate count can depend on the teleportation outcomes.

\subsection{Fidelity Evaluation}
\label{sec:appendix:fidelity_eval}

\begin{figure}[t]
    \centering
    \includegraphics[width=\linewidth]{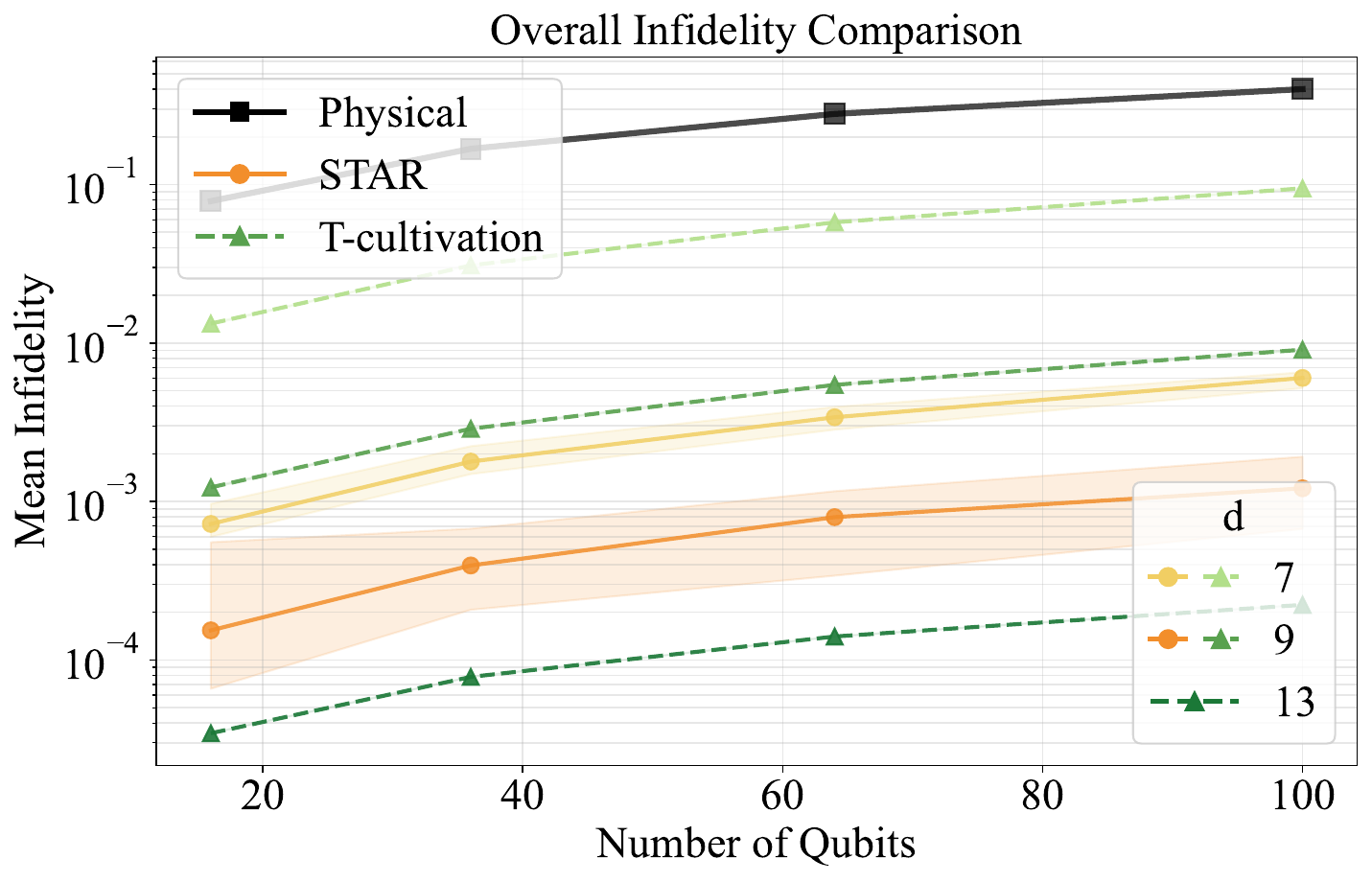}
    \caption{
    \textbf{Estimated execution infidelity across TFIM system sizes.}
    We compare unencoded physical execution, STAR at surface-code distances
    $d=7$ and $d=9$, and the $T$-state cultivation architecture at
    $d=7$, $d=9$, and $d=13$.
    Markers show the mean estimated execution infidelity across 10 trials,
    while the shaded regions indicate the minimum-to-maximum range.
   }
    \label{fig:fidelity}
\end{figure}

\autoref{fig:fidelity} compares the estimated execution infidelity of unencoded physical execution, STAR, and the $T$-state cultivation architecture.
We evaluate STAR at $d=7,9$ and cultivation at $d=7,9,13$.
STAR at $d=13$ is omitted because the success probability of arbitrary-angle resource-state preparation decreases with code distance, producing prohibitively long preparation latency under the adopted preparation model.

The two fault-tolerant approaches reduce infidelity by up to several orders of magnitude relative to unencoded physical execution, depending on code distance.
At the same evaluated distance, STAR achieves lower infidelity under our model.
Its native arbitrary-angle rotations avoid decomposing each $R_Z$ rotation into a long sequence of $T$ and Clifford gates, reducing the number of teleportations, circuit duration, and accumulated logical error.
The dominant modeled contributions to STAR infidelity arise from logical $R_Z(\theta)$ rotations and the CNOT gates used for teleportation.

In contrast, the $T$-state cultivation architecture incurs additional error from repeated $T$-gate teleportation, Clifford corrections, and longer preparation-induced idle intervals.
Increasing the code distance suppresses logical Clifford errors; at $d=13$, cultivation therefore achieves the lowest infidelity among the evaluated configurations despite its synthesis overhead.

\bibliography{main}

\begin{thebibliography}{78}%
\makeatletter
\providecommand \@ifxundefined [1]{%
 \@ifx{#1\undefined}
}%
\providecommand \@ifnum [1]{%
 \ifnum #1\expandafter \@firstoftwo
 \else \expandafter \@secondoftwo
 \fi
}%
\providecommand \@ifx [1]{%
 \ifx #1\expandafter \@firstoftwo
 \else \expandafter \@secondoftwo
 \fi
}%
\providecommand \natexlab [1]{#1}%
\providecommand \enquote  [1]{``#1''}%
\providecommand \bibnamefont  [1]{#1}%
\providecommand \bibfnamefont [1]{#1}%
\providecommand \citenamefont [1]{#1}%
\providecommand \href@noop [0]{\@secondoftwo}%
\providecommand \href [0]{\begingroup \@sanitize@url \@href}%
\providecommand \@href[1]{\@@startlink{#1}\@@href}%
\providecommand \@@href[1]{\endgroup#1\@@endlink}%
\providecommand \@sanitize@url [0]{\catcode `\\12\catcode `\$12\catcode `\&12\catcode `\#12\catcode `\^12\catcode `\_12\catcode `\%12\relax}%
\providecommand \@@startlink[1]{}%
\providecommand \@@endlink[0]{}%
\providecommand \url  [0]{\begingroup\@sanitize@url \@url }%
\providecommand \@url [1]{\endgroup\@href {#1}{\urlprefix }}%
\providecommand \urlprefix  [0]{URL }%
\providecommand \Eprint [0]{\href }%
\providecommand \doibase [0]{https://doi.org/}%
\providecommand \selectlanguage [0]{\@gobble}%
\providecommand \bibinfo  [0]{\@secondoftwo}%
\providecommand \bibfield  [0]{\@secondoftwo}%
\providecommand \translation [1]{[#1]}%
\providecommand \BibitemOpen [0]{}%
\providecommand \bibitemStop [0]{}%
\providecommand \bibitemNoStop [0]{.\EOS\space}%
\providecommand \EOS [0]{\spacefactor3000\relax}%
\providecommand \BibitemShut  [1]{\csname bibitem#1\endcsname}%
\let\auto@bib@innerbib\@empty
\bibitem [{\citenamefont {Acharya}\ \emph {et~al.}(2024)\citenamefont {Acharya}, \citenamefont {Abanin}, \citenamefont {Aghababaie-Beni}, \citenamefont {Aleiner}, \citenamefont {Andersen}, \citenamefont {Ansmann}, \citenamefont {Arute}, \citenamefont {Arya}, \citenamefont {Asfaw}, \citenamefont {Astrakhantsev}, \citenamefont {Atalaya}, \citenamefont {Babbush}, \citenamefont {Bacon}, \citenamefont {Ballard}, \citenamefont {Bardin}, \citenamefont {Bausch}, \citenamefont {Bengtsson}, \citenamefont {Bilmes}, \citenamefont {Blackwell}, \citenamefont {Boixo}, \citenamefont {Bortoli}, \citenamefont {Bourassa}, \citenamefont {Bovaird}, \citenamefont {Brill}, \citenamefont {Broughton}, \citenamefont {Browne}, \citenamefont {Buchea}, \citenamefont {Buckley}, \citenamefont {Buell}, \citenamefont {Burger}, \citenamefont {Burkett}, \citenamefont {Bushnell}, \citenamefont {Cabrera}, \citenamefont {Campero}, \citenamefont {Chang}, \citenamefont {Chen}, \citenamefont {Chen}, \citenamefont {Chiaro}, \citenamefont {Chik},
  \citenamefont {Chou}, \citenamefont {Claes}, \citenamefont {Cleland}, \citenamefont {Cogan}, \citenamefont {Collins}, \citenamefont {Conner}, \citenamefont {Courtney}, \citenamefont {Crook}, \citenamefont {Curtin}, \citenamefont {Das}, \citenamefont {Davies}, \citenamefont {De~Lorenzo}, \citenamefont {Debroy}, \citenamefont {Demura}, \citenamefont {Devoret}, \citenamefont {Di~Paolo}, \citenamefont {Donohoe}, \citenamefont {Drozdov}, \citenamefont {Dunsworth}, \citenamefont {Earle}, \citenamefont {Edlich}, \citenamefont {Eickbusch}, \citenamefont {Elbag}, \citenamefont {Elzouka}, \citenamefont {Erickson}, \citenamefont {Faoro}, \citenamefont {Farhi}, \citenamefont {Ferreira}, \citenamefont {Burgos}, \citenamefont {Forati}, \citenamefont {Fowler}, \citenamefont {Foxen}, \citenamefont {Ganjam}, \citenamefont {Garcia}, \citenamefont {Gasca}, \citenamefont {Genois}, \citenamefont {Giang}, \citenamefont {Gidney}, \citenamefont {Gilboa}, \citenamefont {Gosula}, \citenamefont {Dau}, \citenamefont {Graumann},
  \citenamefont {Greene}, \citenamefont {Gross}, \citenamefont {Habegger}, \citenamefont {Hall}, \citenamefont {Hamilton}, \citenamefont {Hansen}, \citenamefont {Harrigan}, \citenamefont {Harrington}, \citenamefont {Heras}, \citenamefont {Heslin}, \citenamefont {Heu}, \citenamefont {Higgott}, \citenamefont {Hill}, \citenamefont {Hilton}, \citenamefont {Holland}, \citenamefont {Hong}, \citenamefont {Huang}, \citenamefont {Huff}, \citenamefont {Huggins}, \citenamefont {Ioffe}, \citenamefont {Isakov}, \citenamefont {Iveland}, \citenamefont {Jeffrey}, \citenamefont {Jiang}, \citenamefont {Jones}, \citenamefont {Jordan}, \citenamefont {Joshi}, \citenamefont {Juhas}, \citenamefont {Kafri}, \citenamefont {Kang}, \citenamefont {Karamlou}, \citenamefont {Kechedzhi}, \citenamefont {Kelly}, \citenamefont {Khaire}, \citenamefont {Khattar}, \citenamefont {Khezri}, \citenamefont {Kim}, \citenamefont {Klimov}, \citenamefont {Klots}, \citenamefont {Kobrin}, \citenamefont {Kohli}, \citenamefont {Korotkov}, \citenamefont
  {Kostritsa}, \citenamefont {Kothari}, \citenamefont {Kozlovskii}, \citenamefont {Kreikebaum}, \citenamefont {Kurilovich}, \citenamefont {Lacroix}, \citenamefont {Landhuis}, \citenamefont {Lange-Dei}, \citenamefont {Langley}, \citenamefont {Laptev}, \citenamefont {Lau}, \citenamefont {Le~Guevel}, \citenamefont {Ledford}, \citenamefont {Lee}, \citenamefont {Lee}, \citenamefont {Lensky}, \citenamefont {Leon}, \citenamefont {Lester}, \citenamefont {Li}, \citenamefont {Li}, \citenamefont {Lill}, \citenamefont {Liu}, \citenamefont {Livingston}, \citenamefont {Locharla}, \citenamefont {Lucero}, \citenamefont {Lundahl}, \citenamefont {Lunt}, \citenamefont {Madhuk}, \citenamefont {Malone}, \citenamefont {Maloney}, \citenamefont {Mandrà}, \citenamefont {Manyika}, \citenamefont {Martin}, \citenamefont {Martin}, \citenamefont {Martin}, \citenamefont {Maxfield}, \citenamefont {McClean}, \citenamefont {McEwen}, \citenamefont {Meeks}, \citenamefont {Megrant}, \citenamefont {Mi}, \citenamefont {Miao}, \citenamefont
  {Mieszala}, \citenamefont {Molavi}, \citenamefont {Molina}, \citenamefont {Montazeri}, \citenamefont {Morvan}, \citenamefont {Movassagh}, \citenamefont {Mruczkiewicz}, \citenamefont {Naaman}, \citenamefont {Neeley}, \citenamefont {Neill}, \citenamefont {Nersisyan}, \citenamefont {Neven}, \citenamefont {Newman}, \citenamefont {Ng}, \citenamefont {Nguyen}, \citenamefont {Nguyen}, \citenamefont {Ni}, \citenamefont {Niu}, \citenamefont {O’Brien}, \citenamefont {Oliver}, \citenamefont {Opremcak}, \citenamefont {Ottosson}, \citenamefont {Petukhov}, \citenamefont {Pizzuto}, \citenamefont {Platt}, \citenamefont {Potter}, \citenamefont {Pritchard}, \citenamefont {Pryadko}, \citenamefont {Quintana}, \citenamefont {Ramachandran}, \citenamefont {Reagor}, \citenamefont {Redding}, \citenamefont {Rhodes}, \citenamefont {Roberts}, \citenamefont {Rosenberg}, \citenamefont {Rosenfeld}, \citenamefont {Roushan}, \citenamefont {Rubin}, \citenamefont {Saei}, \citenamefont {Sank}, \citenamefont {Sankaragomathi}, \citenamefont
  {Satzinger}, \citenamefont {Schurkus}, \citenamefont {Schuster}, \citenamefont {Senior}, \citenamefont {Shearn}, \citenamefont {Shorter}, \citenamefont {Shutty}, \citenamefont {Shvarts}, \citenamefont {Singh}, \citenamefont {Sivak}, \citenamefont {Skruzny}, \citenamefont {Small}, \citenamefont {Smelyanskiy}, \citenamefont {Smith}, \citenamefont {Somma}, \citenamefont {Springer}, \citenamefont {Sterling}, \citenamefont {Strain}, \citenamefont {Suchard}, \citenamefont {Szasz}, \citenamefont {Sztein}, \citenamefont {Thor}, \citenamefont {Torres}, \citenamefont {Torunbalci}, \citenamefont {Vaishnav}, \citenamefont {Vargas}, \citenamefont {Vdovichev}, \citenamefont {Vidal}, \citenamefont {Villalonga}, \citenamefont {Heidweiller}, \citenamefont {Waltman}, \citenamefont {Wang}, \citenamefont {Ware}, \citenamefont {Weber}, \citenamefont {Weidel}, \citenamefont {White}, \citenamefont {Wong}, \citenamefont {Woo}, \citenamefont {Xing}, \citenamefont {Yao}, \citenamefont {Yeh}, \citenamefont {Ying}, \citenamefont
  {Yoo}, \citenamefont {Yosri}, \citenamefont {Young}, \citenamefont {Zalcman}, \citenamefont {Zhang}, \citenamefont {Zhu},\ and\ \citenamefont {Zobrist}}]{google2024}%
  \BibitemOpen
  \bibfield  {author} {\bibinfo {author} {\bibfnamefont {R.}~\bibnamefont {Acharya}}, \bibinfo {author} {\bibfnamefont {D.~A.}\ \bibnamefont {Abanin}}, \bibinfo {author} {\bibfnamefont {L.}~\bibnamefont {Aghababaie-Beni}}, \bibinfo {author} {\bibfnamefont {I.}~\bibnamefont {Aleiner}}, \bibinfo {author} {\bibfnamefont {T.~I.}\ \bibnamefont {Andersen}}, \bibinfo {author} {\bibfnamefont {M.}~\bibnamefont {Ansmann}}, \bibinfo {author} {\bibfnamefont {F.}~\bibnamefont {Arute}}, \bibinfo {author} {\bibfnamefont {K.}~\bibnamefont {Arya}}, \bibinfo {author} {\bibfnamefont {A.}~\bibnamefont {Asfaw}}, \bibinfo {author} {\bibfnamefont {N.}~\bibnamefont {Astrakhantsev}}, \bibinfo {author} {\bibfnamefont {J.}~\bibnamefont {Atalaya}}, \bibinfo {author} {\bibfnamefont {R.}~\bibnamefont {Babbush}}, \bibinfo {author} {\bibfnamefont {D.}~\bibnamefont {Bacon}}, \bibinfo {author} {\bibfnamefont {B.}~\bibnamefont {Ballard}}, \bibinfo {author} {\bibfnamefont {J.~C.}\ \bibnamefont {Bardin}}, \bibinfo {author} {\bibfnamefont
  {J.}~\bibnamefont {Bausch}}, \bibinfo {author} {\bibfnamefont {A.}~\bibnamefont {Bengtsson}}, \bibinfo {author} {\bibfnamefont {A.}~\bibnamefont {Bilmes}}, \bibinfo {author} {\bibfnamefont {S.}~\bibnamefont {Blackwell}}, \bibinfo {author} {\bibfnamefont {S.}~\bibnamefont {Boixo}}, \bibinfo {author} {\bibfnamefont {G.}~\bibnamefont {Bortoli}}, \bibinfo {author} {\bibfnamefont {A.}~\bibnamefont {Bourassa}}, \bibinfo {author} {\bibfnamefont {J.}~\bibnamefont {Bovaird}}, \bibinfo {author} {\bibfnamefont {L.}~\bibnamefont {Brill}}, \bibinfo {author} {\bibfnamefont {M.}~\bibnamefont {Broughton}}, \bibinfo {author} {\bibfnamefont {D.~A.}\ \bibnamefont {Browne}}, \bibinfo {author} {\bibfnamefont {B.}~\bibnamefont {Buchea}}, \bibinfo {author} {\bibfnamefont {B.~B.}\ \bibnamefont {Buckley}}, \bibinfo {author} {\bibfnamefont {D.~A.}\ \bibnamefont {Buell}}, \bibinfo {author} {\bibfnamefont {T.}~\bibnamefont {Burger}}, \bibinfo {author} {\bibfnamefont {B.}~\bibnamefont {Burkett}}, \bibinfo {author} {\bibfnamefont
  {N.}~\bibnamefont {Bushnell}}, \bibinfo {author} {\bibfnamefont {A.}~\bibnamefont {Cabrera}}, \bibinfo {author} {\bibfnamefont {J.}~\bibnamefont {Campero}}, \bibinfo {author} {\bibfnamefont {H.-S.}\ \bibnamefont {Chang}}, \bibinfo {author} {\bibfnamefont {Y.}~\bibnamefont {Chen}}, \bibinfo {author} {\bibfnamefont {Z.}~\bibnamefont {Chen}}, \bibinfo {author} {\bibfnamefont {B.}~\bibnamefont {Chiaro}}, \bibinfo {author} {\bibfnamefont {D.}~\bibnamefont {Chik}}, \bibinfo {author} {\bibfnamefont {C.}~\bibnamefont {Chou}}, \bibinfo {author} {\bibfnamefont {J.}~\bibnamefont {Claes}}, \bibinfo {author} {\bibfnamefont {A.~Y.}\ \bibnamefont {Cleland}}, \bibinfo {author} {\bibfnamefont {J.}~\bibnamefont {Cogan}}, \bibinfo {author} {\bibfnamefont {R.}~\bibnamefont {Collins}}, \bibinfo {author} {\bibfnamefont {P.}~\bibnamefont {Conner}}, \bibinfo {author} {\bibfnamefont {W.}~\bibnamefont {Courtney}}, \bibinfo {author} {\bibfnamefont {A.~L.}\ \bibnamefont {Crook}}, \bibinfo {author} {\bibfnamefont {B.}~\bibnamefont
  {Curtin}}, \bibinfo {author} {\bibfnamefont {S.}~\bibnamefont {Das}}, \bibinfo {author} {\bibfnamefont {A.}~\bibnamefont {Davies}}, \bibinfo {author} {\bibfnamefont {L.}~\bibnamefont {De~Lorenzo}}, \bibinfo {author} {\bibfnamefont {D.~M.}\ \bibnamefont {Debroy}}, \bibinfo {author} {\bibfnamefont {S.}~\bibnamefont {Demura}}, \bibinfo {author} {\bibfnamefont {M.}~\bibnamefont {Devoret}}, \bibinfo {author} {\bibfnamefont {A.}~\bibnamefont {Di~Paolo}}, \bibinfo {author} {\bibfnamefont {P.}~\bibnamefont {Donohoe}}, \bibinfo {author} {\bibfnamefont {I.}~\bibnamefont {Drozdov}}, \bibinfo {author} {\bibfnamefont {A.}~\bibnamefont {Dunsworth}}, \bibinfo {author} {\bibfnamefont {C.}~\bibnamefont {Earle}}, \bibinfo {author} {\bibfnamefont {T.}~\bibnamefont {Edlich}}, \bibinfo {author} {\bibfnamefont {A.}~\bibnamefont {Eickbusch}}, \bibinfo {author} {\bibfnamefont {A.~M.}\ \bibnamefont {Elbag}}, \bibinfo {author} {\bibfnamefont {M.}~\bibnamefont {Elzouka}}, \bibinfo {author} {\bibfnamefont {C.}~\bibnamefont
  {Erickson}}, \bibinfo {author} {\bibfnamefont {L.}~\bibnamefont {Faoro}}, \bibinfo {author} {\bibfnamefont {E.}~\bibnamefont {Farhi}}, \bibinfo {author} {\bibfnamefont {V.~S.}\ \bibnamefont {Ferreira}}, \bibinfo {author} {\bibfnamefont {L.~F.}\ \bibnamefont {Burgos}}, \bibinfo {author} {\bibfnamefont {E.}~\bibnamefont {Forati}}, \bibinfo {author} {\bibfnamefont {A.~G.}\ \bibnamefont {Fowler}}, \bibinfo {author} {\bibfnamefont {B.}~\bibnamefont {Foxen}}, \bibinfo {author} {\bibfnamefont {S.}~\bibnamefont {Ganjam}}, \bibinfo {author} {\bibfnamefont {G.}~\bibnamefont {Garcia}}, \bibinfo {author} {\bibfnamefont {R.}~\bibnamefont {Gasca}}, \bibinfo {author} {\bibfnamefont {{\'E}.}~\bibnamefont {Genois}}, \bibinfo {author} {\bibfnamefont {W.}~\bibnamefont {Giang}}, \bibinfo {author} {\bibfnamefont {C.}~\bibnamefont {Gidney}}, \bibinfo {author} {\bibfnamefont {D.}~\bibnamefont {Gilboa}}, \bibinfo {author} {\bibfnamefont {R.}~\bibnamefont {Gosula}}, \bibinfo {author} {\bibfnamefont {A.~G.}\ \bibnamefont {Dau}},
  \bibinfo {author} {\bibfnamefont {D.}~\bibnamefont {Graumann}}, \bibinfo {author} {\bibfnamefont {A.}~\bibnamefont {Greene}}, \bibinfo {author} {\bibfnamefont {J.~A.}\ \bibnamefont {Gross}}, \bibinfo {author} {\bibfnamefont {S.}~\bibnamefont {Habegger}}, \bibinfo {author} {\bibfnamefont {J.}~\bibnamefont {Hall}}, \bibinfo {author} {\bibfnamefont {M.~C.}\ \bibnamefont {Hamilton}}, \bibinfo {author} {\bibfnamefont {M.}~\bibnamefont {Hansen}}, \bibinfo {author} {\bibfnamefont {M.~P.}\ \bibnamefont {Harrigan}}, \bibinfo {author} {\bibfnamefont {S.~D.}\ \bibnamefont {Harrington}}, \bibinfo {author} {\bibfnamefont {F.~J.~H.}\ \bibnamefont {Heras}}, \bibinfo {author} {\bibfnamefont {S.}~\bibnamefont {Heslin}}, \bibinfo {author} {\bibfnamefont {P.}~\bibnamefont {Heu}}, \bibinfo {author} {\bibfnamefont {O.}~\bibnamefont {Higgott}}, \bibinfo {author} {\bibfnamefont {G.}~\bibnamefont {Hill}}, \bibinfo {author} {\bibfnamefont {J.}~\bibnamefont {Hilton}}, \bibinfo {author} {\bibfnamefont {G.}~\bibnamefont {Holland}},
  \bibinfo {author} {\bibfnamefont {S.}~\bibnamefont {Hong}}, \bibinfo {author} {\bibfnamefont {H.-Y.}\ \bibnamefont {Huang}}, \bibinfo {author} {\bibfnamefont {A.}~\bibnamefont {Huff}}, \bibinfo {author} {\bibfnamefont {W.~J.}\ \bibnamefont {Huggins}}, \bibinfo {author} {\bibfnamefont {L.~B.}\ \bibnamefont {Ioffe}}, \bibinfo {author} {\bibfnamefont {S.~V.}\ \bibnamefont {Isakov}}, \bibinfo {author} {\bibfnamefont {J.}~\bibnamefont {Iveland}}, \bibinfo {author} {\bibfnamefont {E.}~\bibnamefont {Jeffrey}}, \bibinfo {author} {\bibfnamefont {Z.}~\bibnamefont {Jiang}}, \bibinfo {author} {\bibfnamefont {C.}~\bibnamefont {Jones}}, \bibinfo {author} {\bibfnamefont {S.}~\bibnamefont {Jordan}}, \bibinfo {author} {\bibfnamefont {C.}~\bibnamefont {Joshi}}, \bibinfo {author} {\bibfnamefont {P.}~\bibnamefont {Juhas}}, \bibinfo {author} {\bibfnamefont {D.}~\bibnamefont {Kafri}}, \bibinfo {author} {\bibfnamefont {H.}~\bibnamefont {Kang}}, \bibinfo {author} {\bibfnamefont {A.~H.}\ \bibnamefont {Karamlou}}, \bibinfo {author}
  {\bibfnamefont {K.}~\bibnamefont {Kechedzhi}}, \bibinfo {author} {\bibfnamefont {J.}~\bibnamefont {Kelly}}, \bibinfo {author} {\bibfnamefont {T.}~\bibnamefont {Khaire}}, \bibinfo {author} {\bibfnamefont {T.}~\bibnamefont {Khattar}}, \bibinfo {author} {\bibfnamefont {M.}~\bibnamefont {Khezri}}, \bibinfo {author} {\bibfnamefont {S.}~\bibnamefont {Kim}}, \bibinfo {author} {\bibfnamefont {P.~V.}\ \bibnamefont {Klimov}}, \bibinfo {author} {\bibfnamefont {A.~R.}\ \bibnamefont {Klots}}, \bibinfo {author} {\bibfnamefont {B.}~\bibnamefont {Kobrin}}, \bibinfo {author} {\bibfnamefont {P.}~\bibnamefont {Kohli}}, \bibinfo {author} {\bibfnamefont {A.~N.}\ \bibnamefont {Korotkov}}, \bibinfo {author} {\bibfnamefont {F.}~\bibnamefont {Kostritsa}}, \bibinfo {author} {\bibfnamefont {R.}~\bibnamefont {Kothari}}, \bibinfo {author} {\bibfnamefont {B.}~\bibnamefont {Kozlovskii}}, \bibinfo {author} {\bibfnamefont {J.~M.}\ \bibnamefont {Kreikebaum}}, \bibinfo {author} {\bibfnamefont {V.~D.}\ \bibnamefont {Kurilovich}}, \bibinfo
  {author} {\bibfnamefont {N.}~\bibnamefont {Lacroix}}, \bibinfo {author} {\bibfnamefont {D.}~\bibnamefont {Landhuis}}, \bibinfo {author} {\bibfnamefont {T.}~\bibnamefont {Lange-Dei}}, \bibinfo {author} {\bibfnamefont {B.~W.}\ \bibnamefont {Langley}}, \bibinfo {author} {\bibfnamefont {P.}~\bibnamefont {Laptev}}, \bibinfo {author} {\bibfnamefont {K.-M.}\ \bibnamefont {Lau}}, \bibinfo {author} {\bibfnamefont {L.}~\bibnamefont {Le~Guevel}}, \bibinfo {author} {\bibfnamefont {J.}~\bibnamefont {Ledford}}, \bibinfo {author} {\bibfnamefont {J.}~\bibnamefont {Lee}}, \bibinfo {author} {\bibfnamefont {K.}~\bibnamefont {Lee}}, \bibinfo {author} {\bibfnamefont {Y.~D.}\ \bibnamefont {Lensky}}, \bibinfo {author} {\bibfnamefont {S.}~\bibnamefont {Leon}}, \bibinfo {author} {\bibfnamefont {B.~J.}\ \bibnamefont {Lester}}, \bibinfo {author} {\bibfnamefont {W.~Y.}\ \bibnamefont {Li}}, \bibinfo {author} {\bibfnamefont {Y.}~\bibnamefont {Li}}, \bibinfo {author} {\bibfnamefont {A.~T.}\ \bibnamefont {Lill}}, \bibinfo {author}
  {\bibfnamefont {W.}~\bibnamefont {Liu}}, \bibinfo {author} {\bibfnamefont {W.~P.}\ \bibnamefont {Livingston}}, \bibinfo {author} {\bibfnamefont {A.}~\bibnamefont {Locharla}}, \bibinfo {author} {\bibfnamefont {E.}~\bibnamefont {Lucero}}, \bibinfo {author} {\bibfnamefont {D.}~\bibnamefont {Lundahl}}, \bibinfo {author} {\bibfnamefont {A.}~\bibnamefont {Lunt}}, \bibinfo {author} {\bibfnamefont {S.}~\bibnamefont {Madhuk}}, \bibinfo {author} {\bibfnamefont {F.~D.}\ \bibnamefont {Malone}}, \bibinfo {author} {\bibfnamefont {A.}~\bibnamefont {Maloney}}, \bibinfo {author} {\bibfnamefont {S.}~\bibnamefont {Mandrà}}, \bibinfo {author} {\bibfnamefont {J.}~\bibnamefont {Manyika}}, \bibinfo {author} {\bibfnamefont {L.~S.}\ \bibnamefont {Martin}}, \bibinfo {author} {\bibfnamefont {O.}~\bibnamefont {Martin}}, \bibinfo {author} {\bibfnamefont {S.}~\bibnamefont {Martin}}, \bibinfo {author} {\bibfnamefont {C.}~\bibnamefont {Maxfield}}, \bibinfo {author} {\bibfnamefont {J.~R.}\ \bibnamefont {McClean}}, \bibinfo {author}
  {\bibfnamefont {M.}~\bibnamefont {McEwen}}, \bibinfo {author} {\bibfnamefont {S.}~\bibnamefont {Meeks}}, \bibinfo {author} {\bibfnamefont {A.}~\bibnamefont {Megrant}}, \bibinfo {author} {\bibfnamefont {X.}~\bibnamefont {Mi}}, \bibinfo {author} {\bibfnamefont {K.~C.}\ \bibnamefont {Miao}}, \bibinfo {author} {\bibfnamefont {A.}~\bibnamefont {Mieszala}}, \bibinfo {author} {\bibfnamefont {R.}~\bibnamefont {Molavi}}, \bibinfo {author} {\bibfnamefont {S.}~\bibnamefont {Molina}}, \bibinfo {author} {\bibfnamefont {S.}~\bibnamefont {Montazeri}}, \bibinfo {author} {\bibfnamefont {A.}~\bibnamefont {Morvan}}, \bibinfo {author} {\bibfnamefont {R.}~\bibnamefont {Movassagh}}, \bibinfo {author} {\bibfnamefont {W.}~\bibnamefont {Mruczkiewicz}}, \bibinfo {author} {\bibfnamefont {O.}~\bibnamefont {Naaman}}, \bibinfo {author} {\bibfnamefont {M.}~\bibnamefont {Neeley}}, \bibinfo {author} {\bibfnamefont {C.}~\bibnamefont {Neill}}, \bibinfo {author} {\bibfnamefont {A.}~\bibnamefont {Nersisyan}}, \bibinfo {author} {\bibfnamefont
  {H.}~\bibnamefont {Neven}}, \bibinfo {author} {\bibfnamefont {M.}~\bibnamefont {Newman}}, \bibinfo {author} {\bibfnamefont {J.~H.}\ \bibnamefont {Ng}}, \bibinfo {author} {\bibfnamefont {A.}~\bibnamefont {Nguyen}}, \bibinfo {author} {\bibfnamefont {M.}~\bibnamefont {Nguyen}}, \bibinfo {author} {\bibfnamefont {C.-H.}\ \bibnamefont {Ni}}, \bibinfo {author} {\bibfnamefont {M.~Y.}\ \bibnamefont {Niu}}, \bibinfo {author} {\bibfnamefont {T.~E.}\ \bibnamefont {O’Brien}}, \bibinfo {author} {\bibfnamefont {W.~D.}\ \bibnamefont {Oliver}}, \bibinfo {author} {\bibfnamefont {A.}~\bibnamefont {Opremcak}}, \bibinfo {author} {\bibfnamefont {K.}~\bibnamefont {Ottosson}}, \bibinfo {author} {\bibfnamefont {A.}~\bibnamefont {Petukhov}}, \bibinfo {author} {\bibfnamefont {A.}~\bibnamefont {Pizzuto}}, \bibinfo {author} {\bibfnamefont {J.}~\bibnamefont {Platt}}, \bibinfo {author} {\bibfnamefont {R.}~\bibnamefont {Potter}}, \bibinfo {author} {\bibfnamefont {O.}~\bibnamefont {Pritchard}}, \bibinfo {author} {\bibfnamefont {L.~P.}\
  \bibnamefont {Pryadko}}, \bibinfo {author} {\bibfnamefont {C.}~\bibnamefont {Quintana}}, \bibinfo {author} {\bibfnamefont {G.}~\bibnamefont {Ramachandran}}, \bibinfo {author} {\bibfnamefont {M.~J.}\ \bibnamefont {Reagor}}, \bibinfo {author} {\bibfnamefont {J.}~\bibnamefont {Redding}}, \bibinfo {author} {\bibfnamefont {D.~M.}\ \bibnamefont {Rhodes}}, \bibinfo {author} {\bibfnamefont {G.}~\bibnamefont {Roberts}}, \bibinfo {author} {\bibfnamefont {E.}~\bibnamefont {Rosenberg}}, \bibinfo {author} {\bibfnamefont {E.}~\bibnamefont {Rosenfeld}}, \bibinfo {author} {\bibfnamefont {P.}~\bibnamefont {Roushan}}, \bibinfo {author} {\bibfnamefont {N.~C.}\ \bibnamefont {Rubin}}, \bibinfo {author} {\bibfnamefont {N.}~\bibnamefont {Saei}}, \bibinfo {author} {\bibfnamefont {D.}~\bibnamefont {Sank}}, \bibinfo {author} {\bibfnamefont {K.}~\bibnamefont {Sankaragomathi}}, \bibinfo {author} {\bibfnamefont {K.~J.}\ \bibnamefont {Satzinger}}, \bibinfo {author} {\bibfnamefont {H.~F.}\ \bibnamefont {Schurkus}}, \bibinfo {author}
  {\bibfnamefont {C.}~\bibnamefont {Schuster}}, \bibinfo {author} {\bibfnamefont {A.~W.}\ \bibnamefont {Senior}}, \bibinfo {author} {\bibfnamefont {M.~J.}\ \bibnamefont {Shearn}}, \bibinfo {author} {\bibfnamefont {A.}~\bibnamefont {Shorter}}, \bibinfo {author} {\bibfnamefont {N.}~\bibnamefont {Shutty}}, \bibinfo {author} {\bibfnamefont {V.}~\bibnamefont {Shvarts}}, \bibinfo {author} {\bibfnamefont {S.}~\bibnamefont {Singh}}, \bibinfo {author} {\bibfnamefont {V.}~\bibnamefont {Sivak}}, \bibinfo {author} {\bibfnamefont {J.}~\bibnamefont {Skruzny}}, \bibinfo {author} {\bibfnamefont {S.}~\bibnamefont {Small}}, \bibinfo {author} {\bibfnamefont {V.}~\bibnamefont {Smelyanskiy}}, \bibinfo {author} {\bibfnamefont {W.~C.}\ \bibnamefont {Smith}}, \bibinfo {author} {\bibfnamefont {R.~D.}\ \bibnamefont {Somma}}, \bibinfo {author} {\bibfnamefont {S.}~\bibnamefont {Springer}}, \bibinfo {author} {\bibfnamefont {G.}~\bibnamefont {Sterling}}, \bibinfo {author} {\bibfnamefont {D.}~\bibnamefont {Strain}}, \bibinfo {author}
  {\bibfnamefont {J.}~\bibnamefont {Suchard}}, \bibinfo {author} {\bibfnamefont {A.}~\bibnamefont {Szasz}}, \bibinfo {author} {\bibfnamefont {A.}~\bibnamefont {Sztein}}, \bibinfo {author} {\bibfnamefont {D.}~\bibnamefont {Thor}}, \bibinfo {author} {\bibfnamefont {A.}~\bibnamefont {Torres}}, \bibinfo {author} {\bibfnamefont {M.~M.}\ \bibnamefont {Torunbalci}}, \bibinfo {author} {\bibfnamefont {A.}~\bibnamefont {Vaishnav}}, \bibinfo {author} {\bibfnamefont {J.}~\bibnamefont {Vargas}}, \bibinfo {author} {\bibfnamefont {S.}~\bibnamefont {Vdovichev}}, \bibinfo {author} {\bibfnamefont {G.}~\bibnamefont {Vidal}}, \bibinfo {author} {\bibfnamefont {B.}~\bibnamefont {Villalonga}}, \bibinfo {author} {\bibfnamefont {C.~V.}\ \bibnamefont {Heidweiller}}, \bibinfo {author} {\bibfnamefont {S.}~\bibnamefont {Waltman}}, \bibinfo {author} {\bibfnamefont {S.~X.}\ \bibnamefont {Wang}}, \bibinfo {author} {\bibfnamefont {B.}~\bibnamefont {Ware}}, \bibinfo {author} {\bibfnamefont {K.}~\bibnamefont {Weber}}, \bibinfo {author}
  {\bibfnamefont {T.}~\bibnamefont {Weidel}}, \bibinfo {author} {\bibfnamefont {T.}~\bibnamefont {White}}, \bibinfo {author} {\bibfnamefont {K.}~\bibnamefont {Wong}}, \bibinfo {author} {\bibfnamefont {B.~W.~K.}\ \bibnamefont {Woo}}, \bibinfo {author} {\bibfnamefont {C.}~\bibnamefont {Xing}}, \bibinfo {author} {\bibfnamefont {Z.~J.}\ \bibnamefont {Yao}}, \bibinfo {author} {\bibfnamefont {P.}~\bibnamefont {Yeh}}, \bibinfo {author} {\bibfnamefont {B.}~\bibnamefont {Ying}}, \bibinfo {author} {\bibfnamefont {J.}~\bibnamefont {Yoo}}, \bibinfo {author} {\bibfnamefont {N.}~\bibnamefont {Yosri}}, \bibinfo {author} {\bibfnamefont {G.}~\bibnamefont {Young}}, \bibinfo {author} {\bibfnamefont {A.}~\bibnamefont {Zalcman}}, \bibinfo {author} {\bibfnamefont {Y.}~\bibnamefont {Zhang}}, \bibinfo {author} {\bibfnamefont {N.}~\bibnamefont {Zhu}},\ and\ \bibinfo {author} {\bibfnamefont {N.}~\bibnamefont {Zobrist}},\ }\bibfield  {title} {\bibinfo {title} {Quantum error correction below the surface code threshold},\ }\href
  {https://doi.org/10.1038/s41586-024-08449-y} {\bibfield  {journal} {\bibinfo  {journal} {Nature}\ }\textbf {\bibinfo {volume} {638}},\ \bibinfo {pages} {920–926} (\bibinfo {year} {2024})}\BibitemShut {NoStop}%
\bibitem [{\citenamefont {Manetsch}\ \emph {et~al.}(2025)\citenamefont {Manetsch}, \citenamefont {Nomura}, \citenamefont {Bataille}, \citenamefont {Lv}, \citenamefont {Leung},\ and\ \citenamefont {Endres}}]{Manetsch_2025}%
  \BibitemOpen
  \bibfield  {author} {\bibinfo {author} {\bibfnamefont {H.~J.}\ \bibnamefont {Manetsch}}, \bibinfo {author} {\bibfnamefont {G.}~\bibnamefont {Nomura}}, \bibinfo {author} {\bibfnamefont {E.}~\bibnamefont {Bataille}}, \bibinfo {author} {\bibfnamefont {X.}~\bibnamefont {Lv}}, \bibinfo {author} {\bibfnamefont {K.~H.}\ \bibnamefont {Leung}},\ and\ \bibinfo {author} {\bibfnamefont {M.}~\bibnamefont {Endres}},\ }\bibfield  {title} {\bibinfo {title} {A tweezer array with 6,100 highly coherent atomic qubits},\ }\href {https://doi.org/10.1038/s41586-025-09641-4} {\bibfield  {journal} {\bibinfo  {journal} {Nature}\ }\textbf {\bibinfo {volume} {647}},\ \bibinfo {pages} {60} (\bibinfo {year} {2025})}\BibitemShut {NoStop}%
\bibitem [{\citenamefont {Ransford}\ \emph {et~al.}(2025)\citenamefont {Ransford}, \citenamefont {Allman}, \citenamefont {Arkinstall}, \citenamefont {III}, \citenamefont {Cooper}, \citenamefont {Delaney}, \citenamefont {Dreiling}, \citenamefont {Estey}, \citenamefont {Figgatt}, \citenamefont {Hall}, \citenamefont {Husain}, \citenamefont {Isanaka}, \citenamefont {Kennedy}, \citenamefont {Kotibhaskar}, \citenamefont {Madjarov}, \citenamefont {Mayer}, \citenamefont {Milne}, \citenamefont {Park}, \citenamefont {Reed}, \citenamefont {Ancona}, \citenamefont {Andersen}, \citenamefont {Andres-Martinez}, \citenamefont {Angenent}, \citenamefont {Argueta}, \citenamefont {Arkin}, \citenamefont {Ascarrunz}, \citenamefont {Baker}, \citenamefont {Barnes}, \citenamefont {Bartolotta}, \citenamefont {Berg}, \citenamefont {Besand}, \citenamefont {Bjork}, \citenamefont {Blain}, \citenamefont {Blanchard}, \citenamefont {Blume-Kohout}, \citenamefont {Bohn}, \citenamefont {Borgna}, \citenamefont {Botamanenko}, \citenamefont
  {Boutelle}, \citenamefont {Brown}, \citenamefont {Buckingham}, \citenamefont {Burdick}, \citenamefont {Burton}, \citenamefont {Carey}, \citenamefont {Carron}, \citenamefont {Chambers}, \citenamefont {Children}, \citenamefont {Colussi}, \citenamefont {Crepinsek}, \citenamefont {Cureton}, \citenamefont {Davies}, \citenamefont {Davis}, \citenamefont {DeCross}, \citenamefont {Deen}, \citenamefont {Delaney}, \citenamefont {DelVento}, \citenamefont {DeSalvo}, \citenamefont {Dominy}, \citenamefont {Duncan}, \citenamefont {Eccles}, \citenamefont {Edgington}, \citenamefont {Erickson}, \citenamefont {Erickson}, \citenamefont {Ertsgaard}, \citenamefont {Evans}, \citenamefont {Evans}, \citenamefont {Fabrikant}, \citenamefont {Fischer}, \citenamefont {Foltz}, \citenamefont {Foss-Feig}, \citenamefont {Francois}, \citenamefont {Freyberg}, \citenamefont {Gao}, \citenamefont {Garay}, \citenamefont {Garvin}, \citenamefont {Gaudiosi}, \citenamefont {Gilbreth}, \citenamefont {Giles}, \citenamefont {Glynn}, \citenamefont
  {Graves}, \citenamefont {Hansen}, \citenamefont {Hayes}, \citenamefont {Heidemann}, \citenamefont {Higashi}, \citenamefont {Hilbun}, \citenamefont {Hines}, \citenamefont {Hlavaty}, \citenamefont {Hoffman}, \citenamefont {Hoffman}, \citenamefont {Holliman}, \citenamefont {Hooper}, \citenamefont {Horning}, \citenamefont {Hostetter}, \citenamefont {Hothem}, \citenamefont {Houlton}, \citenamefont {Hout}, \citenamefont {Hutson}, \citenamefont {Jacobs}, \citenamefont {Jacobs}, \citenamefont {Johannsen}, \citenamefont {Johansen}, \citenamefont {Jones}, \citenamefont {Julian}, \citenamefont {Jung}, \citenamefont {Keay}, \citenamefont {Klein}, \citenamefont {Koch}, \citenamefont {Kondo}, \citenamefont {Kong}, \citenamefont {Kosto}, \citenamefont {Lawrence}, \citenamefont {Liefer}, \citenamefont {Lollie}, \citenamefont {Lucchetti}, \citenamefont {Lysne}, \citenamefont {Lytle}, \citenamefont {MacPherson}, \citenamefont {Malm}, \citenamefont {Mather}, \citenamefont {Mathewson}, \citenamefont {Maxwell}, \citenamefont
  {McCaffrey}, \citenamefont {McDougall}, \citenamefont {Mendoza}, \citenamefont {Mills}, \citenamefont {Morrison}, \citenamefont {Narmour}, \citenamefont {Nguyen}, \citenamefont {Nugent}, \citenamefont {Olson}, \citenamefont {Ouellette}, \citenamefont {Parks}, \citenamefont {Peters}, \citenamefont {Petricka}, \citenamefont {Pino}, \citenamefont {Polito}, \citenamefont {Preidl}, \citenamefont {Price}, \citenamefont {Proctor}, \citenamefont {Pugh}, \citenamefont {Ratcliff}, \citenamefont {Raymondson}, \citenamefont {Rhodes}, \citenamefont {Roman}, \citenamefont {Roy}, \citenamefont {Ryan-Anderson}, \citenamefont {Sanchez}, \citenamefont {Sangiolo}, \citenamefont {Sawadski}, \citenamefont {Schaffer}, \citenamefont {Schow}, \citenamefont {Sedlacek}, \citenamefont {Semenenko}, \citenamefont {Shevchuk}, \citenamefont {Shore}, \citenamefont {Siegfried}, \citenamefont {Singhal}, \citenamefont {Sivarajah}, \citenamefont {Skripka}, \citenamefont {Sletten}, \citenamefont {Spaun}, \citenamefont {Sprenkle}, \citenamefont
  {Stoufer}, \citenamefont {Tader}, \citenamefont {Taylor}, \citenamefont {Thompson}, \citenamefont {Tobey}, \citenamefont {Tran}, \citenamefont {Tran}, \citenamefont {Vittorini}, \citenamefont {Volin}, \citenamefont {Walker}, \citenamefont {White}, \citenamefont {Wilson}, \citenamefont {Wolf}, \citenamefont {Wringe}, \citenamefont {Young}, \citenamefont {Zheng}, \citenamefont {Zuraski}, \citenamefont {Baldwin}, \citenamefont {Chernoguzov}, \citenamefont {Gaebler}, \citenamefont {Sanders}, \citenamefont {Neyenhuis}, \citenamefont {Stutz},\ and\ \citenamefont {Bohnet}}]{ransford2025helios98qubittrappedionquantum}%
  \BibitemOpen
  \bibfield  {author} {\bibinfo {author} {\bibfnamefont {A.}~\bibnamefont {Ransford}}, \bibinfo {author} {\bibfnamefont {M.~S.}\ \bibnamefont {Allman}}, \bibinfo {author} {\bibfnamefont {J.}~\bibnamefont {Arkinstall}}, \bibinfo {author} {\bibfnamefont {J.~P.~C.}\ \bibnamefont {III}}, \bibinfo {author} {\bibfnamefont {S.~F.}\ \bibnamefont {Cooper}}, \bibinfo {author} {\bibfnamefont {R.~D.}\ \bibnamefont {Delaney}}, \bibinfo {author} {\bibfnamefont {J.~M.}\ \bibnamefont {Dreiling}}, \bibinfo {author} {\bibfnamefont {B.}~\bibnamefont {Estey}}, \bibinfo {author} {\bibfnamefont {C.}~\bibnamefont {Figgatt}}, \bibinfo {author} {\bibfnamefont {A.}~\bibnamefont {Hall}}, \bibinfo {author} {\bibfnamefont {A.~A.}\ \bibnamefont {Husain}}, \bibinfo {author} {\bibfnamefont {A.}~\bibnamefont {Isanaka}}, \bibinfo {author} {\bibfnamefont {C.~J.}\ \bibnamefont {Kennedy}}, \bibinfo {author} {\bibfnamefont {N.}~\bibnamefont {Kotibhaskar}}, \bibinfo {author} {\bibfnamefont {I.~S.}\ \bibnamefont {Madjarov}}, \bibinfo {author}
  {\bibfnamefont {K.}~\bibnamefont {Mayer}}, \bibinfo {author} {\bibfnamefont {A.~R.}\ \bibnamefont {Milne}}, \bibinfo {author} {\bibfnamefont {A.~J.}\ \bibnamefont {Park}}, \bibinfo {author} {\bibfnamefont {A.~P.}\ \bibnamefont {Reed}}, \bibinfo {author} {\bibfnamefont {R.}~\bibnamefont {Ancona}}, \bibinfo {author} {\bibfnamefont {M.~P.}\ \bibnamefont {Andersen}}, \bibinfo {author} {\bibfnamefont {P.}~\bibnamefont {Andres-Martinez}}, \bibinfo {author} {\bibfnamefont {W.}~\bibnamefont {Angenent}}, \bibinfo {author} {\bibfnamefont {L.}~\bibnamefont {Argueta}}, \bibinfo {author} {\bibfnamefont {B.}~\bibnamefont {Arkin}}, \bibinfo {author} {\bibfnamefont {L.}~\bibnamefont {Ascarrunz}}, \bibinfo {author} {\bibfnamefont {W.}~\bibnamefont {Baker}}, \bibinfo {author} {\bibfnamefont {C.}~\bibnamefont {Barnes}}, \bibinfo {author} {\bibfnamefont {J.}~\bibnamefont {Bartolotta}}, \bibinfo {author} {\bibfnamefont {J.}~\bibnamefont {Berg}}, \bibinfo {author} {\bibfnamefont {R.}~\bibnamefont {Besand}}, \bibinfo {author}
  {\bibfnamefont {B.}~\bibnamefont {Bjork}}, \bibinfo {author} {\bibfnamefont {M.}~\bibnamefont {Blain}}, \bibinfo {author} {\bibfnamefont {P.}~\bibnamefont {Blanchard}}, \bibinfo {author} {\bibfnamefont {R.}~\bibnamefont {Blume-Kohout}}, \bibinfo {author} {\bibfnamefont {M.}~\bibnamefont {Bohn}}, \bibinfo {author} {\bibfnamefont {A.}~\bibnamefont {Borgna}}, \bibinfo {author} {\bibfnamefont {D.~Y.}\ \bibnamefont {Botamanenko}}, \bibinfo {author} {\bibfnamefont {R.}~\bibnamefont {Boutelle}}, \bibinfo {author} {\bibfnamefont {N.}~\bibnamefont {Brown}}, \bibinfo {author} {\bibfnamefont {G.~T.}\ \bibnamefont {Buckingham}}, \bibinfo {author} {\bibfnamefont {N.~Q.}\ \bibnamefont {Burdick}}, \bibinfo {author} {\bibfnamefont {W.~C.}\ \bibnamefont {Burton}}, \bibinfo {author} {\bibfnamefont {V.}~\bibnamefont {Carey}}, \bibinfo {author} {\bibfnamefont {C.~J.}\ \bibnamefont {Carron}}, \bibinfo {author} {\bibfnamefont {J.}~\bibnamefont {Chambers}}, \bibinfo {author} {\bibfnamefont {J.}~\bibnamefont {Children}}, \bibinfo
  {author} {\bibfnamefont {V.~E.}\ \bibnamefont {Colussi}}, \bibinfo {author} {\bibfnamefont {S.}~\bibnamefont {Crepinsek}}, \bibinfo {author} {\bibfnamefont {A.}~\bibnamefont {Cureton}}, \bibinfo {author} {\bibfnamefont {J.}~\bibnamefont {Davies}}, \bibinfo {author} {\bibfnamefont {D.}~\bibnamefont {Davis}}, \bibinfo {author} {\bibfnamefont {M.}~\bibnamefont {DeCross}}, \bibinfo {author} {\bibfnamefont {D.}~\bibnamefont {Deen}}, \bibinfo {author} {\bibfnamefont {C.}~\bibnamefont {Delaney}}, \bibinfo {author} {\bibfnamefont {D.}~\bibnamefont {DelVento}}, \bibinfo {author} {\bibfnamefont {B.~J.}\ \bibnamefont {DeSalvo}}, \bibinfo {author} {\bibfnamefont {J.}~\bibnamefont {Dominy}}, \bibinfo {author} {\bibfnamefont {R.}~\bibnamefont {Duncan}}, \bibinfo {author} {\bibfnamefont {V.}~\bibnamefont {Eccles}}, \bibinfo {author} {\bibfnamefont {A.}~\bibnamefont {Edgington}}, \bibinfo {author} {\bibfnamefont {N.}~\bibnamefont {Erickson}}, \bibinfo {author} {\bibfnamefont {S.}~\bibnamefont {Erickson}}, \bibinfo {author}
  {\bibfnamefont {C.~T.}\ \bibnamefont {Ertsgaard}}, \bibinfo {author} {\bibfnamefont {B.}~\bibnamefont {Evans}}, \bibinfo {author} {\bibfnamefont {T.}~\bibnamefont {Evans}}, \bibinfo {author} {\bibfnamefont {M.~I.}\ \bibnamefont {Fabrikant}}, \bibinfo {author} {\bibfnamefont {A.}~\bibnamefont {Fischer}}, \bibinfo {author} {\bibfnamefont {C.}~\bibnamefont {Foltz}}, \bibinfo {author} {\bibfnamefont {M.}~\bibnamefont {Foss-Feig}}, \bibinfo {author} {\bibfnamefont {D.}~\bibnamefont {Francois}}, \bibinfo {author} {\bibfnamefont {B.}~\bibnamefont {Freyberg}}, \bibinfo {author} {\bibfnamefont {C.}~\bibnamefont {Gao}}, \bibinfo {author} {\bibfnamefont {R.}~\bibnamefont {Garay}}, \bibinfo {author} {\bibfnamefont {J.}~\bibnamefont {Garvin}}, \bibinfo {author} {\bibfnamefont {D.~M.}\ \bibnamefont {Gaudiosi}}, \bibinfo {author} {\bibfnamefont {C.~N.}\ \bibnamefont {Gilbreth}}, \bibinfo {author} {\bibfnamefont {J.}~\bibnamefont {Giles}}, \bibinfo {author} {\bibfnamefont {E.}~\bibnamefont {Glynn}}, \bibinfo {author}
  {\bibfnamefont {J.}~\bibnamefont {Graves}}, \bibinfo {author} {\bibfnamefont {A.}~\bibnamefont {Hansen}}, \bibinfo {author} {\bibfnamefont {D.}~\bibnamefont {Hayes}}, \bibinfo {author} {\bibfnamefont {L.}~\bibnamefont {Heidemann}}, \bibinfo {author} {\bibfnamefont {B.}~\bibnamefont {Higashi}}, \bibinfo {author} {\bibfnamefont {T.}~\bibnamefont {Hilbun}}, \bibinfo {author} {\bibfnamefont {J.}~\bibnamefont {Hines}}, \bibinfo {author} {\bibfnamefont {A.}~\bibnamefont {Hlavaty}}, \bibinfo {author} {\bibfnamefont {K.}~\bibnamefont {Hoffman}}, \bibinfo {author} {\bibfnamefont {I.~M.}\ \bibnamefont {Hoffman}}, \bibinfo {author} {\bibfnamefont {C.}~\bibnamefont {Holliman}}, \bibinfo {author} {\bibfnamefont {I.}~\bibnamefont {Hooper}}, \bibinfo {author} {\bibfnamefont {B.}~\bibnamefont {Horning}}, \bibinfo {author} {\bibfnamefont {J.}~\bibnamefont {Hostetter}}, \bibinfo {author} {\bibfnamefont {D.}~\bibnamefont {Hothem}}, \bibinfo {author} {\bibfnamefont {J.}~\bibnamefont {Houlton}}, \bibinfo {author} {\bibfnamefont
  {J.}~\bibnamefont {Hout}}, \bibinfo {author} {\bibfnamefont {R.}~\bibnamefont {Hutson}}, \bibinfo {author} {\bibfnamefont {R.~T.}\ \bibnamefont {Jacobs}}, \bibinfo {author} {\bibfnamefont {T.}~\bibnamefont {Jacobs}}, \bibinfo {author} {\bibfnamefont {M.}~\bibnamefont {Johannsen}}, \bibinfo {author} {\bibfnamefont {J.}~\bibnamefont {Johansen}}, \bibinfo {author} {\bibfnamefont {L.}~\bibnamefont {Jones}}, \bibinfo {author} {\bibfnamefont {S.}~\bibnamefont {Julian}}, \bibinfo {author} {\bibfnamefont {R.}~\bibnamefont {Jung}}, \bibinfo {author} {\bibfnamefont {A.}~\bibnamefont {Keay}}, \bibinfo {author} {\bibfnamefont {T.}~\bibnamefont {Klein}}, \bibinfo {author} {\bibfnamefont {M.}~\bibnamefont {Koch}}, \bibinfo {author} {\bibfnamefont {R.}~\bibnamefont {Kondo}}, \bibinfo {author} {\bibfnamefont {C.}~\bibnamefont {Kong}}, \bibinfo {author} {\bibfnamefont {A.}~\bibnamefont {Kosto}}, \bibinfo {author} {\bibfnamefont {A.}~\bibnamefont {Lawrence}}, \bibinfo {author} {\bibfnamefont {D.}~\bibnamefont {Liefer}},
  \bibinfo {author} {\bibfnamefont {M.}~\bibnamefont {Lollie}}, \bibinfo {author} {\bibfnamefont {D.}~\bibnamefont {Lucchetti}}, \bibinfo {author} {\bibfnamefont {N.~K.}\ \bibnamefont {Lysne}}, \bibinfo {author} {\bibfnamefont {C.}~\bibnamefont {Lytle}}, \bibinfo {author} {\bibfnamefont {C.}~\bibnamefont {MacPherson}}, \bibinfo {author} {\bibfnamefont {A.}~\bibnamefont {Malm}}, \bibinfo {author} {\bibfnamefont {S.}~\bibnamefont {Mather}}, \bibinfo {author} {\bibfnamefont {B.}~\bibnamefont {Mathewson}}, \bibinfo {author} {\bibfnamefont {D.}~\bibnamefont {Maxwell}}, \bibinfo {author} {\bibfnamefont {L.}~\bibnamefont {McCaffrey}}, \bibinfo {author} {\bibfnamefont {H.}~\bibnamefont {McDougall}}, \bibinfo {author} {\bibfnamefont {R.}~\bibnamefont {Mendoza}}, \bibinfo {author} {\bibfnamefont {M.}~\bibnamefont {Mills}}, \bibinfo {author} {\bibfnamefont {R.}~\bibnamefont {Morrison}}, \bibinfo {author} {\bibfnamefont {L.}~\bibnamefont {Narmour}}, \bibinfo {author} {\bibfnamefont {N.}~\bibnamefont {Nguyen}}, \bibinfo
  {author} {\bibfnamefont {L.}~\bibnamefont {Nugent}}, \bibinfo {author} {\bibfnamefont {S.}~\bibnamefont {Olson}}, \bibinfo {author} {\bibfnamefont {D.}~\bibnamefont {Ouellette}}, \bibinfo {author} {\bibfnamefont {J.}~\bibnamefont {Parks}}, \bibinfo {author} {\bibfnamefont {Z.}~\bibnamefont {Peters}}, \bibinfo {author} {\bibfnamefont {J.}~\bibnamefont {Petricka}}, \bibinfo {author} {\bibfnamefont {J.~M.}\ \bibnamefont {Pino}}, \bibinfo {author} {\bibfnamefont {F.}~\bibnamefont {Polito}}, \bibinfo {author} {\bibfnamefont {M.}~\bibnamefont {Preidl}}, \bibinfo {author} {\bibfnamefont {G.}~\bibnamefont {Price}}, \bibinfo {author} {\bibfnamefont {T.}~\bibnamefont {Proctor}}, \bibinfo {author} {\bibfnamefont {M.}~\bibnamefont {Pugh}}, \bibinfo {author} {\bibfnamefont {N.}~\bibnamefont {Ratcliff}}, \bibinfo {author} {\bibfnamefont {D.}~\bibnamefont {Raymondson}}, \bibinfo {author} {\bibfnamefont {P.}~\bibnamefont {Rhodes}}, \bibinfo {author} {\bibfnamefont {C.}~\bibnamefont {Roman}}, \bibinfo {author}
  {\bibfnamefont {C.}~\bibnamefont {Roy}}, \bibinfo {author} {\bibfnamefont {C.}~\bibnamefont {Ryan-Anderson}}, \bibinfo {author} {\bibfnamefont {F.~B.}\ \bibnamefont {Sanchez}}, \bibinfo {author} {\bibfnamefont {G.}~\bibnamefont {Sangiolo}}, \bibinfo {author} {\bibfnamefont {T.}~\bibnamefont {Sawadski}}, \bibinfo {author} {\bibfnamefont {A.}~\bibnamefont {Schaffer}}, \bibinfo {author} {\bibfnamefont {P.}~\bibnamefont {Schow}}, \bibinfo {author} {\bibfnamefont {J.}~\bibnamefont {Sedlacek}}, \bibinfo {author} {\bibfnamefont {H.}~\bibnamefont {Semenenko}}, \bibinfo {author} {\bibfnamefont {P.}~\bibnamefont {Shevchuk}}, \bibinfo {author} {\bibfnamefont {S.}~\bibnamefont {Shore}}, \bibinfo {author} {\bibfnamefont {P.}~\bibnamefont {Siegfried}}, \bibinfo {author} {\bibfnamefont {K.}~\bibnamefont {Singhal}}, \bibinfo {author} {\bibfnamefont {S.}~\bibnamefont {Sivarajah}}, \bibinfo {author} {\bibfnamefont {T.}~\bibnamefont {Skripka}}, \bibinfo {author} {\bibfnamefont {L.}~\bibnamefont {Sletten}}, \bibinfo {author}
  {\bibfnamefont {B.}~\bibnamefont {Spaun}}, \bibinfo {author} {\bibfnamefont {R.~T.}\ \bibnamefont {Sprenkle}}, \bibinfo {author} {\bibfnamefont {P.}~\bibnamefont {Stoufer}}, \bibinfo {author} {\bibfnamefont {M.}~\bibnamefont {Tader}}, \bibinfo {author} {\bibfnamefont {S.~F.}\ \bibnamefont {Taylor}}, \bibinfo {author} {\bibfnamefont {T.~H.}\ \bibnamefont {Thompson}}, \bibinfo {author} {\bibfnamefont {R.}~\bibnamefont {Tobey}}, \bibinfo {author} {\bibfnamefont {A.}~\bibnamefont {Tran}}, \bibinfo {author} {\bibfnamefont {T.}~\bibnamefont {Tran}}, \bibinfo {author} {\bibfnamefont {G.}~\bibnamefont {Vittorini}}, \bibinfo {author} {\bibfnamefont {C.}~\bibnamefont {Volin}}, \bibinfo {author} {\bibfnamefont {J.}~\bibnamefont {Walker}}, \bibinfo {author} {\bibfnamefont {S.}~\bibnamefont {White}}, \bibinfo {author} {\bibfnamefont {D.}~\bibnamefont {Wilson}}, \bibinfo {author} {\bibfnamefont {Q.}~\bibnamefont {Wolf}}, \bibinfo {author} {\bibfnamefont {C.}~\bibnamefont {Wringe}}, \bibinfo {author} {\bibfnamefont
  {K.}~\bibnamefont {Young}}, \bibinfo {author} {\bibfnamefont {J.}~\bibnamefont {Zheng}}, \bibinfo {author} {\bibfnamefont {K.}~\bibnamefont {Zuraski}}, \bibinfo {author} {\bibfnamefont {C.~H.}\ \bibnamefont {Baldwin}}, \bibinfo {author} {\bibfnamefont {A.}~\bibnamefont {Chernoguzov}}, \bibinfo {author} {\bibfnamefont {J.~P.}\ \bibnamefont {Gaebler}}, \bibinfo {author} {\bibfnamefont {S.~J.}\ \bibnamefont {Sanders}}, \bibinfo {author} {\bibfnamefont {B.}~\bibnamefont {Neyenhuis}}, \bibinfo {author} {\bibfnamefont {R.}~\bibnamefont {Stutz}},\ and\ \bibinfo {author} {\bibfnamefont {J.~G.}\ \bibnamefont {Bohnet}},\ }\href {https://arxiv.org/abs/2511.05465} {\bibinfo {title} {Helios: A 98-qubit trapped-ion quantum computer}} (\bibinfo {year} {2025}),\ \Eprint {https://arxiv.org/abs/2511.05465} {arXiv:2511.05465 [quant-ph]} \BibitemShut {NoStop}%
\bibitem [{\citenamefont {Evered}\ \emph {et~al.}(2026)\citenamefont {Evered}, \citenamefont {Xu}, \citenamefont {Li}, \citenamefont {Geim}, \citenamefont {Ataides}, \citenamefont {Kalinowski}, \citenamefont {Bluvstein}, \citenamefont {Maskara}, \citenamefont {Kokail}, \citenamefont {Greiner}, \citenamefont {Vuletić},\ and\ \citenamefont {Lukin}}]{evered2026highfidelityentanglinggatesnonlocal}%
  \BibitemOpen
  \bibfield  {author} {\bibinfo {author} {\bibfnamefont {S.~J.}\ \bibnamefont {Evered}}, \bibinfo {author} {\bibfnamefont {M.}~\bibnamefont {Xu}}, \bibinfo {author} {\bibfnamefont {S.~H.}\ \bibnamefont {Li}}, \bibinfo {author} {\bibfnamefont {A.~A.}\ \bibnamefont {Geim}}, \bibinfo {author} {\bibfnamefont {J.~P.~B.}\ \bibnamefont {Ataides}}, \bibinfo {author} {\bibfnamefont {M.}~\bibnamefont {Kalinowski}}, \bibinfo {author} {\bibfnamefont {D.}~\bibnamefont {Bluvstein}}, \bibinfo {author} {\bibfnamefont {N.}~\bibnamefont {Maskara}}, \bibinfo {author} {\bibfnamefont {C.}~\bibnamefont {Kokail}}, \bibinfo {author} {\bibfnamefont {M.}~\bibnamefont {Greiner}}, \bibinfo {author} {\bibfnamefont {V.}~\bibnamefont {Vuletić}},\ and\ \bibinfo {author} {\bibfnamefont {M.~D.}\ \bibnamefont {Lukin}},\ }\href {https://arxiv.org/abs/2604.25987} {\bibinfo {title} {High-fidelity entangling gates and nonlocal circuits with neutral atoms}} (\bibinfo {year} {2026}),\ \Eprint {https://arxiv.org/abs/2604.25987} {arXiv:2604.25987
  [quant-ph]} \BibitemShut {NoStop}%
\bibitem [{\citenamefont {Williamson}\ and\ \citenamefont {Yoder}(2026)}]{Williamson_2026}%
  \BibitemOpen
  \bibfield  {author} {\bibinfo {author} {\bibfnamefont {D.~J.}\ \bibnamefont {Williamson}}\ and\ \bibinfo {author} {\bibfnamefont {T.~J.}\ \bibnamefont {Yoder}},\ }\bibfield  {title} {\bibinfo {title} {Low-overhead fault-tolerant quantum computation by gauging logical operators},\ }\href {https://doi.org/10.1038/s41567-026-03220-8} {\bibfield  {journal} {\bibinfo  {journal} {Nature Physics}\ }\textbf {\bibinfo {volume} {22}},\ \bibinfo {pages} {598} (\bibinfo {year} {2026})}\BibitemShut {NoStop}%
\bibitem [{\citenamefont {Gidney}\ \emph {et~al.}(2024)\citenamefont {Gidney}, \citenamefont {Shutty},\ and\ \citenamefont {Jones}}]{gidney2024magicstatecultivationgrowing}%
  \BibitemOpen
  \bibfield  {author} {\bibinfo {author} {\bibfnamefont {C.}~\bibnamefont {Gidney}}, \bibinfo {author} {\bibfnamefont {N.}~\bibnamefont {Shutty}},\ and\ \bibinfo {author} {\bibfnamefont {C.}~\bibnamefont {Jones}},\ }\href {https://arxiv.org/abs/2409.17595} {\bibinfo {title} {Magic state cultivation: growing t states as cheap as cnot gates}} (\bibinfo {year} {2024}),\ \Eprint {https://arxiv.org/abs/2409.17595} {arXiv:2409.17595 [quant-ph]} \BibitemShut {NoStop}%
\bibitem [{\citenamefont {Claes}(2025)}]{claes2025cultivatingtstatessurface}%
  \BibitemOpen
  \bibfield  {author} {\bibinfo {author} {\bibfnamefont {J.}~\bibnamefont {Claes}},\ }\href {https://arxiv.org/abs/2509.05232} {\bibinfo {title} {Cultivating t states on the surface code with only two-qubit gates}} (\bibinfo {year} {2025}),\ \Eprint {https://arxiv.org/abs/2509.05232} {arXiv:2509.05232 [quant-ph]} \BibitemShut {NoStop}%
\bibitem [{\citenamefont {Chen}\ \emph {et~al.}(2026)\citenamefont {Chen}, \citenamefont {Chen}, \citenamefont {Lu},\ and\ \citenamefont {Pan}}]{chen2025efficientmagicstatecultivation}%
  \BibitemOpen
  \bibfield  {author} {\bibinfo {author} {\bibfnamefont {Z.-H.}\ \bibnamefont {Chen}}, \bibinfo {author} {\bibfnamefont {M.-C.}\ \bibnamefont {Chen}}, \bibinfo {author} {\bibfnamefont {C.-Y.}\ \bibnamefont {Lu}},\ and\ \bibinfo {author} {\bibfnamefont {J.-W.}\ \bibnamefont {Pan}},\ }\bibfield  {title} {\bibinfo {title} {Efficient magic state cultivation on ${\mathbb{r}\mathbb{p}}^{2}$},\ }\href {https://doi.org/10.1103/9kys-3whh} {\bibfield  {journal} {\bibinfo  {journal} {PRX Quantum}\ }\textbf {\bibinfo {volume} {7}},\ \bibinfo {pages} {010315} (\bibinfo {year} {2026})}\BibitemShut {NoStop}%
\bibitem [{\citenamefont {Sahay}\ \emph {et~al.}(2026)\citenamefont {Sahay}, \citenamefont {Tsai}, \citenamefont {Chang}, \citenamefont {Su}, \citenamefont {Smith}, \citenamefont {Singh},\ and\ \citenamefont {Puri}}]{sahay2026foldtransversalsurfacecodecultivation}%
  \BibitemOpen
  \bibfield  {author} {\bibinfo {author} {\bibfnamefont {K.}~\bibnamefont {Sahay}}, \bibinfo {author} {\bibfnamefont {P.-K.}\ \bibnamefont {Tsai}}, \bibinfo {author} {\bibfnamefont {K.}~\bibnamefont {Chang}}, \bibinfo {author} {\bibfnamefont {Q.}~\bibnamefont {Su}}, \bibinfo {author} {\bibfnamefont {T.~B.}\ \bibnamefont {Smith}}, \bibinfo {author} {\bibfnamefont {S.}~\bibnamefont {Singh}},\ and\ \bibinfo {author} {\bibfnamefont {S.}~\bibnamefont {Puri}},\ }\href {https://arxiv.org/abs/2509.05212} {\bibinfo {title} {Fold-transversal surface code cultivation}} (\bibinfo {year} {2026}),\ \Eprint {https://arxiv.org/abs/2509.05212} {arXiv:2509.05212 [quant-ph]} \BibitemShut {NoStop}%
\bibitem [{\citenamefont {Delfosse}\ and\ \citenamefont {Tham}(2025)}]{PhysRevLett.134.090603}%
  \BibitemOpen
  \bibfield  {author} {\bibinfo {author} {\bibfnamefont {N.}~\bibnamefont {Delfosse}}\ and\ \bibinfo {author} {\bibfnamefont {E.}~\bibnamefont {Tham}},\ }\bibfield  {title} {\bibinfo {title} {Low-cost noise reduction for clifford circuits},\ }\href {https://doi.org/10.1103/PhysRevLett.134.090603} {\bibfield  {journal} {\bibinfo  {journal} {Phys. Rev. Lett.}\ }\textbf {\bibinfo {volume} {134}},\ \bibinfo {pages} {090603} (\bibinfo {year} {2025})}\BibitemShut {NoStop}%
\bibitem [{\citenamefont {Eickbusch}\ \emph {et~al.}(2025)\citenamefont {Eickbusch}, \citenamefont {McEwen}, \citenamefont {Sivak}, \citenamefont {Bourassa}, \citenamefont {Atalaya}, \citenamefont {Claes}, \citenamefont {Kafri}, \citenamefont {Gidney}, \citenamefont {Warren}, \citenamefont {Gross}, \citenamefont {Opremcak}, \citenamefont {Zobrist}, \citenamefont {Miao}, \citenamefont {Roberts}, \citenamefont {Satzinger}, \citenamefont {Bengtsson}, \citenamefont {Neeley}, \citenamefont {Livingston}, \citenamefont {Greene}, \citenamefont {Acharya}, \citenamefont {Beni}, \citenamefont {Aigeldinger}, \citenamefont {Alcaraz}, \citenamefont {Andersen}, \citenamefont {Ansmann}, \citenamefont {Arute}, \citenamefont {Arya}, \citenamefont {Asfaw}, \citenamefont {Babbush}, \citenamefont {Ballard}, \citenamefont {Bardin}, \citenamefont {Bilmes}, \citenamefont {Bovaird}, \citenamefont {Bowers}, \citenamefont {Brill}, \citenamefont {Broughton}, \citenamefont {Browne}, \citenamefont {Buchea}, \citenamefont {Buckley},
  \citenamefont {Burger}, \citenamefont {Burkett}, \citenamefont {Bushnell}, \citenamefont {Cabrera}, \citenamefont {Campero}, \citenamefont {Chang}, \citenamefont {Chiaro}, \citenamefont {Chih}, \citenamefont {Cleland}, \citenamefont {Cogan}, \citenamefont {Collins}, \citenamefont {Conner}, \citenamefont {Courtney}, \citenamefont {Crook}, \citenamefont {Curtin}, \citenamefont {Das}, \citenamefont {Barba}, \citenamefont {Demura}, \citenamefont {Lorenzo}, \citenamefont {Paolo}, \citenamefont {Donohoe}, \citenamefont {Drozdov}, \citenamefont {Dunsworth}, \citenamefont {Elbag}, \citenamefont {Elzouka}, \citenamefont {Erickson}, \citenamefont {Ferreira}, \citenamefont {Burgos}, \citenamefont {Forati}, \citenamefont {Fowler}, \citenamefont {Foxen}, \citenamefont {Ganjam}, \citenamefont {Garcia}, \citenamefont {Gasca}, \citenamefont {Élie Genois}, \citenamefont {Giang}, \citenamefont {Gilboa}, \citenamefont {Gosula}, \citenamefont {Dau}, \citenamefont {Graumann}, \citenamefont {Ha}, \citenamefont {Habegger},
  \citenamefont {Hamilton}, \citenamefont {Hansen}, \citenamefont {Harrigan}, \citenamefont {Harrington}, \citenamefont {Heslin}, \citenamefont {Heu}, \citenamefont {Higgott}, \citenamefont {Hiltermann}, \citenamefont {Hilton}, \citenamefont {Huang}, \citenamefont {Huff}, \citenamefont {Huggins}, \citenamefont {Jeffrey}, \citenamefont {Jiang}, \citenamefont {Jin}, \citenamefont {Jones}, \citenamefont {Joshi}, \citenamefont {Juhas}, \citenamefont {Kabel}, \citenamefont {Kang}, \citenamefont {Karamlou}, \citenamefont {Kechedzhi}, \citenamefont {Khaire}, \citenamefont {Khattar}, \citenamefont {Khezri}, \citenamefont {Kim}, \citenamefont {Kobrin}, \citenamefont {Korotkov}, \citenamefont {Kostritsa}, \citenamefont {Kreikebaum}, \citenamefont {Kurilovich}, \citenamefont {Landhuis}, \citenamefont {Lange-Dei}, \citenamefont {Langley}, \citenamefont {Lau}, \citenamefont {Ledford}, \citenamefont {Lee}, \citenamefont {Lester}, \citenamefont {Guevel}, \citenamefont {Li}, \citenamefont {Lill}, \citenamefont {Locharla},
  \citenamefont {Lucero}, \citenamefont {Lundahl}, \citenamefont {Lunt}, \citenamefont {Madhuk}, \citenamefont {Maloney}, \citenamefont {Mandrà}, \citenamefont {Martin}, \citenamefont {Martin}, \citenamefont {Maxfield}, \citenamefont {McClean}, \citenamefont {Meeks}, \citenamefont {Megrant}, \citenamefont {Molavi}, \citenamefont {Molina}, \citenamefont {Montazeri}, \citenamefont {Movassagh}, \citenamefont {Newman}, \citenamefont {Nguyen}, \citenamefont {Nguyen}, \citenamefont {Ni}, \citenamefont {Oas}, \citenamefont {Orosco}, \citenamefont {Ottosson}, \citenamefont {Pizzuto}, \citenamefont {Potter}, \citenamefont {Pritchard}, \citenamefont {Quintana}, \citenamefont {Ramachandran}, \citenamefont {Reagor}, \citenamefont {Rhodes}, \citenamefont {Rosenberg}, \citenamefont {Rossi}, \citenamefont {Sankaragomathi}, \citenamefont {Schurkus}, \citenamefont {Shearn}, \citenamefont {Shorter}, \citenamefont {Shutty}, \citenamefont {Shvarts}, \citenamefont {Small}, \citenamefont {Smith}, \citenamefont {Springer},
  \citenamefont {Sterling}, \citenamefont {Suchard}, \citenamefont {Szasz}, \citenamefont {Sztein}, \citenamefont {Thor}, \citenamefont {Tomita}, \citenamefont {Torres}, \citenamefont {Torunbalci}, \citenamefont {Vaishnav}, \citenamefont {Vargas}, \citenamefont {Vdovichev}, \citenamefont {Vidal}, \citenamefont {Heidweiller}, \citenamefont {Waltman}, \citenamefont {Waltz}, \citenamefont {Wang}, \citenamefont {Ware}, \citenamefont {Weidel}, \citenamefont {White}, \citenamefont {Wong}, \citenamefont {Woo}, \citenamefont {Woodson}, \citenamefont {Xing}, \citenamefont {Yao}, \citenamefont {Yeh}, \citenamefont {Ying}, \citenamefont {Yoo}, \citenamefont {Yosri}, \citenamefont {Young}, \citenamefont {Zalcman}, \citenamefont {Zhang}, \citenamefont {Zhu}, \citenamefont {Boixo}, \citenamefont {Kelly}, \citenamefont {Smelyanskiy}, \citenamefont {Neven}, \citenamefont {Bacon}, \citenamefont {Chen}, \citenamefont {Klimov}, \citenamefont {Roushan}, \citenamefont {Neill}, \citenamefont {Chen},\ and\ \citenamefont
  {Morvan}}]{eickbusch2025demonstratingdynamicsurfacecodes}%
  \BibitemOpen
  \bibfield  {author} {\bibinfo {author} {\bibfnamefont {A.}~\bibnamefont {Eickbusch}}, \bibinfo {author} {\bibfnamefont {M.}~\bibnamefont {McEwen}}, \bibinfo {author} {\bibfnamefont {V.}~\bibnamefont {Sivak}}, \bibinfo {author} {\bibfnamefont {A.}~\bibnamefont {Bourassa}}, \bibinfo {author} {\bibfnamefont {J.}~\bibnamefont {Atalaya}}, \bibinfo {author} {\bibfnamefont {J.}~\bibnamefont {Claes}}, \bibinfo {author} {\bibfnamefont {D.}~\bibnamefont {Kafri}}, \bibinfo {author} {\bibfnamefont {C.}~\bibnamefont {Gidney}}, \bibinfo {author} {\bibfnamefont {C.~W.}\ \bibnamefont {Warren}}, \bibinfo {author} {\bibfnamefont {J.}~\bibnamefont {Gross}}, \bibinfo {author} {\bibfnamefont {A.}~\bibnamefont {Opremcak}}, \bibinfo {author} {\bibfnamefont {N.}~\bibnamefont {Zobrist}}, \bibinfo {author} {\bibfnamefont {K.~C.}\ \bibnamefont {Miao}}, \bibinfo {author} {\bibfnamefont {G.}~\bibnamefont {Roberts}}, \bibinfo {author} {\bibfnamefont {K.~J.}\ \bibnamefont {Satzinger}}, \bibinfo {author} {\bibfnamefont {A.}~\bibnamefont
  {Bengtsson}}, \bibinfo {author} {\bibfnamefont {M.}~\bibnamefont {Neeley}}, \bibinfo {author} {\bibfnamefont {W.~P.}\ \bibnamefont {Livingston}}, \bibinfo {author} {\bibfnamefont {A.}~\bibnamefont {Greene}}, \bibinfo {author} {\bibfnamefont {R.}~\bibnamefont {Acharya}}, \bibinfo {author} {\bibfnamefont {L.~A.}\ \bibnamefont {Beni}}, \bibinfo {author} {\bibfnamefont {G.}~\bibnamefont {Aigeldinger}}, \bibinfo {author} {\bibfnamefont {R.}~\bibnamefont {Alcaraz}}, \bibinfo {author} {\bibfnamefont {T.~I.}\ \bibnamefont {Andersen}}, \bibinfo {author} {\bibfnamefont {M.}~\bibnamefont {Ansmann}}, \bibinfo {author} {\bibfnamefont {F.}~\bibnamefont {Arute}}, \bibinfo {author} {\bibfnamefont {K.}~\bibnamefont {Arya}}, \bibinfo {author} {\bibfnamefont {A.}~\bibnamefont {Asfaw}}, \bibinfo {author} {\bibfnamefont {R.}~\bibnamefont {Babbush}}, \bibinfo {author} {\bibfnamefont {B.}~\bibnamefont {Ballard}}, \bibinfo {author} {\bibfnamefont {J.~C.}\ \bibnamefont {Bardin}}, \bibinfo {author} {\bibfnamefont {A.}~\bibnamefont
  {Bilmes}}, \bibinfo {author} {\bibfnamefont {J.}~\bibnamefont {Bovaird}}, \bibinfo {author} {\bibfnamefont {D.}~\bibnamefont {Bowers}}, \bibinfo {author} {\bibfnamefont {L.}~\bibnamefont {Brill}}, \bibinfo {author} {\bibfnamefont {M.}~\bibnamefont {Broughton}}, \bibinfo {author} {\bibfnamefont {D.~A.}\ \bibnamefont {Browne}}, \bibinfo {author} {\bibfnamefont {B.}~\bibnamefont {Buchea}}, \bibinfo {author} {\bibfnamefont {B.~B.}\ \bibnamefont {Buckley}}, \bibinfo {author} {\bibfnamefont {T.}~\bibnamefont {Burger}}, \bibinfo {author} {\bibfnamefont {B.}~\bibnamefont {Burkett}}, \bibinfo {author} {\bibfnamefont {N.}~\bibnamefont {Bushnell}}, \bibinfo {author} {\bibfnamefont {A.}~\bibnamefont {Cabrera}}, \bibinfo {author} {\bibfnamefont {J.}~\bibnamefont {Campero}}, \bibinfo {author} {\bibfnamefont {H.-S.}\ \bibnamefont {Chang}}, \bibinfo {author} {\bibfnamefont {B.}~\bibnamefont {Chiaro}}, \bibinfo {author} {\bibfnamefont {L.-Y.}\ \bibnamefont {Chih}}, \bibinfo {author} {\bibfnamefont {A.~Y.}\ \bibnamefont
  {Cleland}}, \bibinfo {author} {\bibfnamefont {J.}~\bibnamefont {Cogan}}, \bibinfo {author} {\bibfnamefont {R.}~\bibnamefont {Collins}}, \bibinfo {author} {\bibfnamefont {P.}~\bibnamefont {Conner}}, \bibinfo {author} {\bibfnamefont {W.}~\bibnamefont {Courtney}}, \bibinfo {author} {\bibfnamefont {A.~L.}\ \bibnamefont {Crook}}, \bibinfo {author} {\bibfnamefont {B.}~\bibnamefont {Curtin}}, \bibinfo {author} {\bibfnamefont {S.}~\bibnamefont {Das}}, \bibinfo {author} {\bibfnamefont {A.~D.~T.}\ \bibnamefont {Barba}}, \bibinfo {author} {\bibfnamefont {S.}~\bibnamefont {Demura}}, \bibinfo {author} {\bibfnamefont {L.~D.}\ \bibnamefont {Lorenzo}}, \bibinfo {author} {\bibfnamefont {A.~D.}\ \bibnamefont {Paolo}}, \bibinfo {author} {\bibfnamefont {P.}~\bibnamefont {Donohoe}}, \bibinfo {author} {\bibfnamefont {I.~K.}\ \bibnamefont {Drozdov}}, \bibinfo {author} {\bibfnamefont {A.}~\bibnamefont {Dunsworth}}, \bibinfo {author} {\bibfnamefont {A.~M.}\ \bibnamefont {Elbag}}, \bibinfo {author} {\bibfnamefont {M.}~\bibnamefont
  {Elzouka}}, \bibinfo {author} {\bibfnamefont {C.}~\bibnamefont {Erickson}}, \bibinfo {author} {\bibfnamefont {V.~S.}\ \bibnamefont {Ferreira}}, \bibinfo {author} {\bibfnamefont {L.~F.}\ \bibnamefont {Burgos}}, \bibinfo {author} {\bibfnamefont {E.}~\bibnamefont {Forati}}, \bibinfo {author} {\bibfnamefont {A.~G.}\ \bibnamefont {Fowler}}, \bibinfo {author} {\bibfnamefont {B.}~\bibnamefont {Foxen}}, \bibinfo {author} {\bibfnamefont {S.}~\bibnamefont {Ganjam}}, \bibinfo {author} {\bibfnamefont {G.}~\bibnamefont {Garcia}}, \bibinfo {author} {\bibfnamefont {R.}~\bibnamefont {Gasca}}, \bibinfo {author} {\bibnamefont {Élie Genois}}, \bibinfo {author} {\bibfnamefont {W.}~\bibnamefont {Giang}}, \bibinfo {author} {\bibfnamefont {D.}~\bibnamefont {Gilboa}}, \bibinfo {author} {\bibfnamefont {R.}~\bibnamefont {Gosula}}, \bibinfo {author} {\bibfnamefont {A.~G.}\ \bibnamefont {Dau}}, \bibinfo {author} {\bibfnamefont {D.}~\bibnamefont {Graumann}}, \bibinfo {author} {\bibfnamefont {T.}~\bibnamefont {Ha}}, \bibinfo {author}
  {\bibfnamefont {S.}~\bibnamefont {Habegger}}, \bibinfo {author} {\bibfnamefont {M.~C.}\ \bibnamefont {Hamilton}}, \bibinfo {author} {\bibfnamefont {M.}~\bibnamefont {Hansen}}, \bibinfo {author} {\bibfnamefont {M.~P.}\ \bibnamefont {Harrigan}}, \bibinfo {author} {\bibfnamefont {S.~D.}\ \bibnamefont {Harrington}}, \bibinfo {author} {\bibfnamefont {S.}~\bibnamefont {Heslin}}, \bibinfo {author} {\bibfnamefont {P.}~\bibnamefont {Heu}}, \bibinfo {author} {\bibfnamefont {O.}~\bibnamefont {Higgott}}, \bibinfo {author} {\bibfnamefont {R.}~\bibnamefont {Hiltermann}}, \bibinfo {author} {\bibfnamefont {J.}~\bibnamefont {Hilton}}, \bibinfo {author} {\bibfnamefont {H.-Y.}\ \bibnamefont {Huang}}, \bibinfo {author} {\bibfnamefont {A.}~\bibnamefont {Huff}}, \bibinfo {author} {\bibfnamefont {W.~J.}\ \bibnamefont {Huggins}}, \bibinfo {author} {\bibfnamefont {E.}~\bibnamefont {Jeffrey}}, \bibinfo {author} {\bibfnamefont {Z.}~\bibnamefont {Jiang}}, \bibinfo {author} {\bibfnamefont {X.}~\bibnamefont {Jin}}, \bibinfo {author}
  {\bibfnamefont {C.}~\bibnamefont {Jones}}, \bibinfo {author} {\bibfnamefont {C.}~\bibnamefont {Joshi}}, \bibinfo {author} {\bibfnamefont {P.}~\bibnamefont {Juhas}}, \bibinfo {author} {\bibfnamefont {A.}~\bibnamefont {Kabel}}, \bibinfo {author} {\bibfnamefont {H.}~\bibnamefont {Kang}}, \bibinfo {author} {\bibfnamefont {A.~H.}\ \bibnamefont {Karamlou}}, \bibinfo {author} {\bibfnamefont {K.}~\bibnamefont {Kechedzhi}}, \bibinfo {author} {\bibfnamefont {T.}~\bibnamefont {Khaire}}, \bibinfo {author} {\bibfnamefont {T.}~\bibnamefont {Khattar}}, \bibinfo {author} {\bibfnamefont {M.}~\bibnamefont {Khezri}}, \bibinfo {author} {\bibfnamefont {S.}~\bibnamefont {Kim}}, \bibinfo {author} {\bibfnamefont {B.}~\bibnamefont {Kobrin}}, \bibinfo {author} {\bibfnamefont {A.~N.}\ \bibnamefont {Korotkov}}, \bibinfo {author} {\bibfnamefont {F.}~\bibnamefont {Kostritsa}}, \bibinfo {author} {\bibfnamefont {J.~M.}\ \bibnamefont {Kreikebaum}}, \bibinfo {author} {\bibfnamefont {V.~D.}\ \bibnamefont {Kurilovich}}, \bibinfo {author}
  {\bibfnamefont {D.}~\bibnamefont {Landhuis}}, \bibinfo {author} {\bibfnamefont {T.}~\bibnamefont {Lange-Dei}}, \bibinfo {author} {\bibfnamefont {B.~W.}\ \bibnamefont {Langley}}, \bibinfo {author} {\bibfnamefont {K.-M.}\ \bibnamefont {Lau}}, \bibinfo {author} {\bibfnamefont {J.}~\bibnamefont {Ledford}}, \bibinfo {author} {\bibfnamefont {K.}~\bibnamefont {Lee}}, \bibinfo {author} {\bibfnamefont {B.~J.}\ \bibnamefont {Lester}}, \bibinfo {author} {\bibfnamefont {L.~L.}\ \bibnamefont {Guevel}}, \bibinfo {author} {\bibfnamefont {W.~Y.}\ \bibnamefont {Li}}, \bibinfo {author} {\bibfnamefont {A.~T.}\ \bibnamefont {Lill}}, \bibinfo {author} {\bibfnamefont {A.}~\bibnamefont {Locharla}}, \bibinfo {author} {\bibfnamefont {E.}~\bibnamefont {Lucero}}, \bibinfo {author} {\bibfnamefont {D.}~\bibnamefont {Lundahl}}, \bibinfo {author} {\bibfnamefont {A.}~\bibnamefont {Lunt}}, \bibinfo {author} {\bibfnamefont {S.}~\bibnamefont {Madhuk}}, \bibinfo {author} {\bibfnamefont {A.}~\bibnamefont {Maloney}}, \bibinfo {author}
  {\bibfnamefont {S.}~\bibnamefont {Mandrà}}, \bibinfo {author} {\bibfnamefont {L.~S.}\ \bibnamefont {Martin}}, \bibinfo {author} {\bibfnamefont {O.}~\bibnamefont {Martin}}, \bibinfo {author} {\bibfnamefont {C.}~\bibnamefont {Maxfield}}, \bibinfo {author} {\bibfnamefont {J.~R.}\ \bibnamefont {McClean}}, \bibinfo {author} {\bibfnamefont {S.}~\bibnamefont {Meeks}}, \bibinfo {author} {\bibfnamefont {A.}~\bibnamefont {Megrant}}, \bibinfo {author} {\bibfnamefont {R.}~\bibnamefont {Molavi}}, \bibinfo {author} {\bibfnamefont {S.}~\bibnamefont {Molina}}, \bibinfo {author} {\bibfnamefont {S.}~\bibnamefont {Montazeri}}, \bibinfo {author} {\bibfnamefont {R.}~\bibnamefont {Movassagh}}, \bibinfo {author} {\bibfnamefont {M.}~\bibnamefont {Newman}}, \bibinfo {author} {\bibfnamefont {A.}~\bibnamefont {Nguyen}}, \bibinfo {author} {\bibfnamefont {M.}~\bibnamefont {Nguyen}}, \bibinfo {author} {\bibfnamefont {C.-H.}\ \bibnamefont {Ni}}, \bibinfo {author} {\bibfnamefont {L.}~\bibnamefont {Oas}}, \bibinfo {author} {\bibfnamefont
  {R.}~\bibnamefont {Orosco}}, \bibinfo {author} {\bibfnamefont {K.}~\bibnamefont {Ottosson}}, \bibinfo {author} {\bibfnamefont {A.}~\bibnamefont {Pizzuto}}, \bibinfo {author} {\bibfnamefont {R.}~\bibnamefont {Potter}}, \bibinfo {author} {\bibfnamefont {O.}~\bibnamefont {Pritchard}}, \bibinfo {author} {\bibfnamefont {C.}~\bibnamefont {Quintana}}, \bibinfo {author} {\bibfnamefont {G.}~\bibnamefont {Ramachandran}}, \bibinfo {author} {\bibfnamefont {M.~J.}\ \bibnamefont {Reagor}}, \bibinfo {author} {\bibfnamefont {D.~M.}\ \bibnamefont {Rhodes}}, \bibinfo {author} {\bibfnamefont {E.}~\bibnamefont {Rosenberg}}, \bibinfo {author} {\bibfnamefont {E.}~\bibnamefont {Rossi}}, \bibinfo {author} {\bibfnamefont {K.}~\bibnamefont {Sankaragomathi}}, \bibinfo {author} {\bibfnamefont {H.~F.}\ \bibnamefont {Schurkus}}, \bibinfo {author} {\bibfnamefont {M.~J.}\ \bibnamefont {Shearn}}, \bibinfo {author} {\bibfnamefont {A.}~\bibnamefont {Shorter}}, \bibinfo {author} {\bibfnamefont {N.}~\bibnamefont {Shutty}}, \bibinfo {author}
  {\bibfnamefont {V.}~\bibnamefont {Shvarts}}, \bibinfo {author} {\bibfnamefont {S.}~\bibnamefont {Small}}, \bibinfo {author} {\bibfnamefont {W.~C.}\ \bibnamefont {Smith}}, \bibinfo {author} {\bibfnamefont {S.}~\bibnamefont {Springer}}, \bibinfo {author} {\bibfnamefont {G.}~\bibnamefont {Sterling}}, \bibinfo {author} {\bibfnamefont {J.}~\bibnamefont {Suchard}}, \bibinfo {author} {\bibfnamefont {A.}~\bibnamefont {Szasz}}, \bibinfo {author} {\bibfnamefont {A.}~\bibnamefont {Sztein}}, \bibinfo {author} {\bibfnamefont {D.}~\bibnamefont {Thor}}, \bibinfo {author} {\bibfnamefont {E.}~\bibnamefont {Tomita}}, \bibinfo {author} {\bibfnamefont {A.}~\bibnamefont {Torres}}, \bibinfo {author} {\bibfnamefont {M.~M.}\ \bibnamefont {Torunbalci}}, \bibinfo {author} {\bibfnamefont {A.}~\bibnamefont {Vaishnav}}, \bibinfo {author} {\bibfnamefont {J.}~\bibnamefont {Vargas}}, \bibinfo {author} {\bibfnamefont {S.}~\bibnamefont {Vdovichev}}, \bibinfo {author} {\bibfnamefont {G.}~\bibnamefont {Vidal}}, \bibinfo {author}
  {\bibfnamefont {C.~V.}\ \bibnamefont {Heidweiller}}, \bibinfo {author} {\bibfnamefont {S.}~\bibnamefont {Waltman}}, \bibinfo {author} {\bibfnamefont {J.}~\bibnamefont {Waltz}}, \bibinfo {author} {\bibfnamefont {S.~X.}\ \bibnamefont {Wang}}, \bibinfo {author} {\bibfnamefont {B.}~\bibnamefont {Ware}}, \bibinfo {author} {\bibfnamefont {T.}~\bibnamefont {Weidel}}, \bibinfo {author} {\bibfnamefont {T.}~\bibnamefont {White}}, \bibinfo {author} {\bibfnamefont {K.}~\bibnamefont {Wong}}, \bibinfo {author} {\bibfnamefont {B.~W.~K.}\ \bibnamefont {Woo}}, \bibinfo {author} {\bibfnamefont {M.}~\bibnamefont {Woodson}}, \bibinfo {author} {\bibfnamefont {C.}~\bibnamefont {Xing}}, \bibinfo {author} {\bibfnamefont {Z.~J.}\ \bibnamefont {Yao}}, \bibinfo {author} {\bibfnamefont {P.}~\bibnamefont {Yeh}}, \bibinfo {author} {\bibfnamefont {B.}~\bibnamefont {Ying}}, \bibinfo {author} {\bibfnamefont {J.}~\bibnamefont {Yoo}}, \bibinfo {author} {\bibfnamefont {N.}~\bibnamefont {Yosri}}, \bibinfo {author} {\bibfnamefont
  {G.}~\bibnamefont {Young}}, \bibinfo {author} {\bibfnamefont {A.}~\bibnamefont {Zalcman}}, \bibinfo {author} {\bibfnamefont {Y.}~\bibnamefont {Zhang}}, \bibinfo {author} {\bibfnamefont {N.}~\bibnamefont {Zhu}}, \bibinfo {author} {\bibfnamefont {S.}~\bibnamefont {Boixo}}, \bibinfo {author} {\bibfnamefont {J.}~\bibnamefont {Kelly}}, \bibinfo {author} {\bibfnamefont {V.}~\bibnamefont {Smelyanskiy}}, \bibinfo {author} {\bibfnamefont {H.}~\bibnamefont {Neven}}, \bibinfo {author} {\bibfnamefont {D.}~\bibnamefont {Bacon}}, \bibinfo {author} {\bibfnamefont {Z.}~\bibnamefont {Chen}}, \bibinfo {author} {\bibfnamefont {P.~V.}\ \bibnamefont {Klimov}}, \bibinfo {author} {\bibfnamefont {P.}~\bibnamefont {Roushan}}, \bibinfo {author} {\bibfnamefont {C.}~\bibnamefont {Neill}}, \bibinfo {author} {\bibfnamefont {Y.}~\bibnamefont {Chen}},\ and\ \bibinfo {author} {\bibfnamefont {A.}~\bibnamefont {Morvan}},\ }\href {https://arxiv.org/abs/2412.14360} {\bibinfo {title} {Demonstrating dynamic surface codes}} (\bibinfo {year}
  {2025}),\ \Eprint {https://arxiv.org/abs/2412.14360} {arXiv:2412.14360 [quant-ph]} \BibitemShut {NoStop}%
\bibitem [{\citenamefont {Zhao}\ \emph {et~al.}(2026)\citenamefont {Zhao}, \citenamefont {Duckering}, \citenamefont {Gu}, \citenamefont {Maskara},\ and\ \citenamefont {Zhou}}]{zhao2026ultrahighratequantumerrorcorrection}%
  \BibitemOpen
  \bibfield  {author} {\bibinfo {author} {\bibfnamefont {C.}~\bibnamefont {Zhao}}, \bibinfo {author} {\bibfnamefont {C.}~\bibnamefont {Duckering}}, \bibinfo {author} {\bibfnamefont {A.}~\bibnamefont {Gu}}, \bibinfo {author} {\bibfnamefont {N.}~\bibnamefont {Maskara}},\ and\ \bibinfo {author} {\bibfnamefont {H.}~\bibnamefont {Zhou}},\ }\href {https://arxiv.org/abs/2604.16209} {\bibinfo {title} {Towards ultra-high-rate quantum error correction with reconfigurable atom arrays}} (\bibinfo {year} {2026}),\ \Eprint {https://arxiv.org/abs/2604.16209} {arXiv:2604.16209 [quant-ph]} \BibitemShut {NoStop}%
\bibitem [{\citenamefont {Cain}\ \emph {et~al.}(2024)\citenamefont {Cain}, \citenamefont {Zhao}, \citenamefont {Zhou}, \citenamefont {Meister}, \citenamefont {Ataides}, \citenamefont {Jaffe}, \citenamefont {Bluvstein},\ and\ \citenamefont {Lukin}}]{correlated_decoding}%
  \BibitemOpen
  \bibfield  {author} {\bibinfo {author} {\bibfnamefont {M.}~\bibnamefont {Cain}}, \bibinfo {author} {\bibfnamefont {C.}~\bibnamefont {Zhao}}, \bibinfo {author} {\bibfnamefont {H.}~\bibnamefont {Zhou}}, \bibinfo {author} {\bibfnamefont {N.}~\bibnamefont {Meister}}, \bibinfo {author} {\bibfnamefont {J.~P.~B.}\ \bibnamefont {Ataides}}, \bibinfo {author} {\bibfnamefont {A.}~\bibnamefont {Jaffe}}, \bibinfo {author} {\bibfnamefont {D.}~\bibnamefont {Bluvstein}},\ and\ \bibinfo {author} {\bibfnamefont {M.~D.}\ \bibnamefont {Lukin}},\ }\bibfield  {title} {\bibinfo {title} {Correlated decoding of logical algorithms with transversal gates},\ }\href {https://doi.org/10.1103/PhysRevLett.133.240602} {\bibfield  {journal} {\bibinfo  {journal} {Phys. Rev. Lett.}\ }\textbf {\bibinfo {volume} {133}},\ \bibinfo {pages} {240602} (\bibinfo {year} {2024})}\BibitemShut {NoStop}%
\bibitem [{\citenamefont {Bluvstein}\ \emph {et~al.}(2023)\citenamefont {Bluvstein}, \citenamefont {Evered}, \citenamefont {Geim}, \citenamefont {Li}, \citenamefont {Zhou}, \citenamefont {Manovitz}, \citenamefont {Ebadi}, \citenamefont {Cain}, \citenamefont {Kalinowski}, \citenamefont {Hangleiter}, \citenamefont {Bonilla~Ataides}, \citenamefont {Maskara}, \citenamefont {Cong}, \citenamefont {Gao}, \citenamefont {Sales~Rodriguez}, \citenamefont {Karolyshyn}, \citenamefont {Semeghini}, \citenamefont {Gullans}, \citenamefont {Greiner}, \citenamefont {Vuletić},\ and\ \citenamefont {Lukin}}]{Bluvstein_2023}%
  \BibitemOpen
  \bibfield  {author} {\bibinfo {author} {\bibfnamefont {D.}~\bibnamefont {Bluvstein}}, \bibinfo {author} {\bibfnamefont {S.~J.}\ \bibnamefont {Evered}}, \bibinfo {author} {\bibfnamefont {A.~A.}\ \bibnamefont {Geim}}, \bibinfo {author} {\bibfnamefont {S.~H.}\ \bibnamefont {Li}}, \bibinfo {author} {\bibfnamefont {H.}~\bibnamefont {Zhou}}, \bibinfo {author} {\bibfnamefont {T.}~\bibnamefont {Manovitz}}, \bibinfo {author} {\bibfnamefont {S.}~\bibnamefont {Ebadi}}, \bibinfo {author} {\bibfnamefont {M.}~\bibnamefont {Cain}}, \bibinfo {author} {\bibfnamefont {M.}~\bibnamefont {Kalinowski}}, \bibinfo {author} {\bibfnamefont {D.}~\bibnamefont {Hangleiter}}, \bibinfo {author} {\bibfnamefont {J.~P.}\ \bibnamefont {Bonilla~Ataides}}, \bibinfo {author} {\bibfnamefont {N.}~\bibnamefont {Maskara}}, \bibinfo {author} {\bibfnamefont {I.}~\bibnamefont {Cong}}, \bibinfo {author} {\bibfnamefont {X.}~\bibnamefont {Gao}}, \bibinfo {author} {\bibfnamefont {P.}~\bibnamefont {Sales~Rodriguez}}, \bibinfo {author} {\bibfnamefont
  {T.}~\bibnamefont {Karolyshyn}}, \bibinfo {author} {\bibfnamefont {G.}~\bibnamefont {Semeghini}}, \bibinfo {author} {\bibfnamefont {M.~J.}\ \bibnamefont {Gullans}}, \bibinfo {author} {\bibfnamefont {M.}~\bibnamefont {Greiner}}, \bibinfo {author} {\bibfnamefont {V.}~\bibnamefont {Vuletić}},\ and\ \bibinfo {author} {\bibfnamefont {M.~D.}\ \bibnamefont {Lukin}},\ }\bibfield  {title} {\bibinfo {title} {Logical quantum processor based on reconfigurable atom arrays},\ }\href {https://doi.org/10.1038/s41586-023-06927-3} {\bibfield  {journal} {\bibinfo  {journal} {Nature}\ }\textbf {\bibinfo {volume} {626}},\ \bibinfo {pages} {58} (\bibinfo {year} {2023})}\BibitemShut {NoStop}%
\bibitem [{\citenamefont {Paetznick}\ \emph {et~al.}(2024)\citenamefont {Paetznick}, \citenamefont {da~Silva}, \citenamefont {Ryan-Anderson}, \citenamefont {Bello-Rivas}, \citenamefont {III}, \citenamefont {Chernoguzov}, \citenamefont {Dreiling}, \citenamefont {Foltz}, \citenamefont {Frachon}, \citenamefont {Gaebler}, \citenamefont {Gatterman}, \citenamefont {Grans-Samuelsson}, \citenamefont {Gresh}, \citenamefont {Hayes}, \citenamefont {Hewitt}, \citenamefont {Holliman}, \citenamefont {Horst}, \citenamefont {Johansen}, \citenamefont {Lucchetti}, \citenamefont {Matsuoka}, \citenamefont {Mills}, \citenamefont {Moses}, \citenamefont {Neyenhuis}, \citenamefont {Paz}, \citenamefont {Pino}, \citenamefont {Siegfried}, \citenamefont {Sundaram}, \citenamefont {Tom}, \citenamefont {Wernli}, \citenamefont {Zanner}, \citenamefont {Stutz},\ and\ \citenamefont {Svore}}]{paetznick2024demonstrationlogicalqubitsrepeated}%
  \BibitemOpen
  \bibfield  {author} {\bibinfo {author} {\bibfnamefont {A.}~\bibnamefont {Paetznick}}, \bibinfo {author} {\bibfnamefont {M.~P.}\ \bibnamefont {da~Silva}}, \bibinfo {author} {\bibfnamefont {C.}~\bibnamefont {Ryan-Anderson}}, \bibinfo {author} {\bibfnamefont {J.~M.}\ \bibnamefont {Bello-Rivas}}, \bibinfo {author} {\bibfnamefont {J.~P.~C.}\ \bibnamefont {III}}, \bibinfo {author} {\bibfnamefont {A.}~\bibnamefont {Chernoguzov}}, \bibinfo {author} {\bibfnamefont {J.~M.}\ \bibnamefont {Dreiling}}, \bibinfo {author} {\bibfnamefont {C.}~\bibnamefont {Foltz}}, \bibinfo {author} {\bibfnamefont {F.}~\bibnamefont {Frachon}}, \bibinfo {author} {\bibfnamefont {J.~P.}\ \bibnamefont {Gaebler}}, \bibinfo {author} {\bibfnamefont {T.~M.}\ \bibnamefont {Gatterman}}, \bibinfo {author} {\bibfnamefont {L.}~\bibnamefont {Grans-Samuelsson}}, \bibinfo {author} {\bibfnamefont {D.}~\bibnamefont {Gresh}}, \bibinfo {author} {\bibfnamefont {D.}~\bibnamefont {Hayes}}, \bibinfo {author} {\bibfnamefont {N.}~\bibnamefont {Hewitt}}, \bibinfo
  {author} {\bibfnamefont {C.}~\bibnamefont {Holliman}}, \bibinfo {author} {\bibfnamefont {C.~V.}\ \bibnamefont {Horst}}, \bibinfo {author} {\bibfnamefont {J.}~\bibnamefont {Johansen}}, \bibinfo {author} {\bibfnamefont {D.}~\bibnamefont {Lucchetti}}, \bibinfo {author} {\bibfnamefont {Y.}~\bibnamefont {Matsuoka}}, \bibinfo {author} {\bibfnamefont {M.}~\bibnamefont {Mills}}, \bibinfo {author} {\bibfnamefont {S.~A.}\ \bibnamefont {Moses}}, \bibinfo {author} {\bibfnamefont {B.}~\bibnamefont {Neyenhuis}}, \bibinfo {author} {\bibfnamefont {A.}~\bibnamefont {Paz}}, \bibinfo {author} {\bibfnamefont {J.}~\bibnamefont {Pino}}, \bibinfo {author} {\bibfnamefont {P.}~\bibnamefont {Siegfried}}, \bibinfo {author} {\bibfnamefont {A.}~\bibnamefont {Sundaram}}, \bibinfo {author} {\bibfnamefont {D.}~\bibnamefont {Tom}}, \bibinfo {author} {\bibfnamefont {S.~J.}\ \bibnamefont {Wernli}}, \bibinfo {author} {\bibfnamefont {M.}~\bibnamefont {Zanner}}, \bibinfo {author} {\bibfnamefont {R.~P.}\ \bibnamefont {Stutz}},\ and\ \bibinfo
  {author} {\bibfnamefont {K.~M.}\ \bibnamefont {Svore}},\ }\href {https://arxiv.org/abs/2404.02280} {\bibinfo {title} {Demonstration of logical qubits and repeated error correction with better-than-physical error rates}} (\bibinfo {year} {2024}),\ \Eprint {https://arxiv.org/abs/2404.02280} {arXiv:2404.02280 [quant-ph]} \BibitemShut {NoStop}%
\bibitem [{\citenamefont {Bluvstein}\ \emph {et~al.}(2025)\citenamefont {Bluvstein}, \citenamefont {Geim}, \citenamefont {Li}, \citenamefont {Evered}, \citenamefont {Bonilla~Ataides}, \citenamefont {Baranes}, \citenamefont {Gu}, \citenamefont {Manovitz}, \citenamefont {Xu}, \citenamefont {Kalinowski}, \citenamefont {Majidy}, \citenamefont {Kokail}, \citenamefont {Maskara}, \citenamefont {Trapp}, \citenamefont {Stewart}, \citenamefont {Hollerith}, \citenamefont {Zhou}, \citenamefont {Gullans}, \citenamefont {Yelin}, \citenamefont {Greiner}, \citenamefont {Vuletić}, \citenamefont {Cain},\ and\ \citenamefont {Lukin}}]{Bluvstein_2025}%
  \BibitemOpen
  \bibfield  {author} {\bibinfo {author} {\bibfnamefont {D.}~\bibnamefont {Bluvstein}}, \bibinfo {author} {\bibfnamefont {A.~A.}\ \bibnamefont {Geim}}, \bibinfo {author} {\bibfnamefont {S.~H.}\ \bibnamefont {Li}}, \bibinfo {author} {\bibfnamefont {S.~J.}\ \bibnamefont {Evered}}, \bibinfo {author} {\bibfnamefont {J.~P.}\ \bibnamefont {Bonilla~Ataides}}, \bibinfo {author} {\bibfnamefont {G.}~\bibnamefont {Baranes}}, \bibinfo {author} {\bibfnamefont {A.}~\bibnamefont {Gu}}, \bibinfo {author} {\bibfnamefont {T.}~\bibnamefont {Manovitz}}, \bibinfo {author} {\bibfnamefont {M.}~\bibnamefont {Xu}}, \bibinfo {author} {\bibfnamefont {M.}~\bibnamefont {Kalinowski}}, \bibinfo {author} {\bibfnamefont {S.}~\bibnamefont {Majidy}}, \bibinfo {author} {\bibfnamefont {C.}~\bibnamefont {Kokail}}, \bibinfo {author} {\bibfnamefont {N.}~\bibnamefont {Maskara}}, \bibinfo {author} {\bibfnamefont {E.~C.}\ \bibnamefont {Trapp}}, \bibinfo {author} {\bibfnamefont {L.~M.}\ \bibnamefont {Stewart}}, \bibinfo {author} {\bibfnamefont
  {S.}~\bibnamefont {Hollerith}}, \bibinfo {author} {\bibfnamefont {H.}~\bibnamefont {Zhou}}, \bibinfo {author} {\bibfnamefont {M.~J.}\ \bibnamefont {Gullans}}, \bibinfo {author} {\bibfnamefont {S.~F.}\ \bibnamefont {Yelin}}, \bibinfo {author} {\bibfnamefont {M.}~\bibnamefont {Greiner}}, \bibinfo {author} {\bibfnamefont {V.}~\bibnamefont {Vuletić}}, \bibinfo {author} {\bibfnamefont {M.}~\bibnamefont {Cain}},\ and\ \bibinfo {author} {\bibfnamefont {M.~D.}\ \bibnamefont {Lukin}},\ }\bibfield  {title} {\bibinfo {title} {A fault-tolerant neutral-atom architecture for universal quantum computation},\ }\href {https://doi.org/10.1038/s41586-025-09848-5} {\bibfield  {journal} {\bibinfo  {journal} {Nature}\ }\textbf {\bibinfo {volume} {649}},\ \bibinfo {pages} {39} (\bibinfo {year} {2025})}\BibitemShut {NoStop}%
\bibitem [{\citenamefont {Zhang}\ \emph {et~al.}(2025)\citenamefont {Zhang}, \citenamefont {Chen}, \citenamefont {Wang}, \citenamefont {Lu}, \citenamefont {Zhang}, \citenamefont {Li}, \citenamefont {Duan}, \citenamefont {Wu},\ and\ \citenamefont {Guo}}]{zhang2025demonstrating}%
  \BibitemOpen
  \bibfield  {author} {\bibinfo {author} {\bibfnamefont {J.}~\bibnamefont {Zhang}}, \bibinfo {author} {\bibfnamefont {Z.-Y.}\ \bibnamefont {Chen}}, \bibinfo {author} {\bibfnamefont {Y.-J.}\ \bibnamefont {Wang}}, \bibinfo {author} {\bibfnamefont {B.-H.}\ \bibnamefont {Lu}}, \bibinfo {author} {\bibfnamefont {H.-F.}\ \bibnamefont {Zhang}}, \bibinfo {author} {\bibfnamefont {J.-N.}\ \bibnamefont {Li}}, \bibinfo {author} {\bibfnamefont {P.}~\bibnamefont {Duan}}, \bibinfo {author} {\bibfnamefont {Y.-C.}\ \bibnamefont {Wu}},\ and\ \bibinfo {author} {\bibfnamefont {G.-P.}\ \bibnamefont {Guo}},\ }\bibfield  {title} {\bibinfo {title} {Demonstrating a universal logical gate set in error-detecting surface codes on a superconducting quantum processor},\ }\href {https://doi.org/10.1038/s41534-025-01118-6} {\bibfield  {journal} {\bibinfo  {journal} {npj Quantum Information}\ }\textbf {\bibinfo {volume} {11}},\ \bibinfo {pages} {177} (\bibinfo {year} {2025})}\BibitemShut {NoStop}%
\bibitem [{\citenamefont {Dasu}\ \emph {et~al.}(2026)\citenamefont {Dasu}, \citenamefont {DeCross}, \citenamefont {Guo}, \citenamefont {Lavasani}, \citenamefont {Behrends}, \citenamefont {Benhemou}, \citenamefont {Chen}, \citenamefont {Mayer}, \citenamefont {Self}, \citenamefont {Simsek}, \citenamefont {Srivastava}, \citenamefont {Allman}, \citenamefont {Arkinstall}, \citenamefont {Bohnet}, \citenamefont {Burdick}, \citenamefont {III}, \citenamefont {Chernoguzov}, \citenamefont {Cooper}, \citenamefont {Delaney}, \citenamefont {Dreiling}, \citenamefont {Estey}, \citenamefont {Figgatt}, \citenamefont {Foltz}, \citenamefont {Gaebler}, \citenamefont {Hall}, \citenamefont {Holliman}, \citenamefont {Husain}, \citenamefont {Isanaka}, \citenamefont {Kennedy}, \citenamefont {Kodama}, \citenamefont {Kotibhaskar}, \citenamefont {Lysne}, \citenamefont {Madjarov}, \citenamefont {Mills}, \citenamefont {Milne}, \citenamefont {Neyenhuis}, \citenamefont {Park}, \citenamefont {Ransford}, \citenamefont {Reed}, \citenamefont
  {Sanders}, \citenamefont {Baldwin}, \citenamefont {Hayes}, \citenamefont {Criger}, \citenamefont {Potter},\ and\ \citenamefont {Amaro}}]{dasu2026computingencodedlogicalqubits}%
  \BibitemOpen
  \bibfield  {author} {\bibinfo {author} {\bibfnamefont {S.}~\bibnamefont {Dasu}}, \bibinfo {author} {\bibfnamefont {M.}~\bibnamefont {DeCross}}, \bibinfo {author} {\bibfnamefont {A.~Y.}\ \bibnamefont {Guo}}, \bibinfo {author} {\bibfnamefont {A.}~\bibnamefont {Lavasani}}, \bibinfo {author} {\bibfnamefont {J.}~\bibnamefont {Behrends}}, \bibinfo {author} {\bibfnamefont {A.}~\bibnamefont {Benhemou}}, \bibinfo {author} {\bibfnamefont {Y.-H.}\ \bibnamefont {Chen}}, \bibinfo {author} {\bibfnamefont {K.}~\bibnamefont {Mayer}}, \bibinfo {author} {\bibfnamefont {C.~N.}\ \bibnamefont {Self}}, \bibinfo {author} {\bibfnamefont {S.}~\bibnamefont {Simsek}}, \bibinfo {author} {\bibfnamefont {B.}~\bibnamefont {Srivastava}}, \bibinfo {author} {\bibfnamefont {M.~S.}\ \bibnamefont {Allman}}, \bibinfo {author} {\bibfnamefont {J.}~\bibnamefont {Arkinstall}}, \bibinfo {author} {\bibfnamefont {J.~G.}\ \bibnamefont {Bohnet}}, \bibinfo {author} {\bibfnamefont {N.~Q.}\ \bibnamefont {Burdick}}, \bibinfo {author} {\bibfnamefont
  {J.~P.~C.}\ \bibnamefont {III}}, \bibinfo {author} {\bibfnamefont {A.}~\bibnamefont {Chernoguzov}}, \bibinfo {author} {\bibfnamefont {S.~F.}\ \bibnamefont {Cooper}}, \bibinfo {author} {\bibfnamefont {R.~D.}\ \bibnamefont {Delaney}}, \bibinfo {author} {\bibfnamefont {J.~M.}\ \bibnamefont {Dreiling}}, \bibinfo {author} {\bibfnamefont {B.}~\bibnamefont {Estey}}, \bibinfo {author} {\bibfnamefont {C.}~\bibnamefont {Figgatt}}, \bibinfo {author} {\bibfnamefont {C.}~\bibnamefont {Foltz}}, \bibinfo {author} {\bibfnamefont {J.~P.}\ \bibnamefont {Gaebler}}, \bibinfo {author} {\bibfnamefont {A.}~\bibnamefont {Hall}}, \bibinfo {author} {\bibfnamefont {C.~A.}\ \bibnamefont {Holliman}}, \bibinfo {author} {\bibfnamefont {A.~A.}\ \bibnamefont {Husain}}, \bibinfo {author} {\bibfnamefont {A.}~\bibnamefont {Isanaka}}, \bibinfo {author} {\bibfnamefont {C.~J.}\ \bibnamefont {Kennedy}}, \bibinfo {author} {\bibfnamefont {Y.}~\bibnamefont {Kodama}}, \bibinfo {author} {\bibfnamefont {N.}~\bibnamefont {Kotibhaskar}}, \bibinfo
  {author} {\bibfnamefont {N.~K.}\ \bibnamefont {Lysne}}, \bibinfo {author} {\bibfnamefont {I.~S.}\ \bibnamefont {Madjarov}}, \bibinfo {author} {\bibfnamefont {M.}~\bibnamefont {Mills}}, \bibinfo {author} {\bibfnamefont {A.~R.}\ \bibnamefont {Milne}}, \bibinfo {author} {\bibfnamefont {B.}~\bibnamefont {Neyenhuis}}, \bibinfo {author} {\bibfnamefont {A.~J.}\ \bibnamefont {Park}}, \bibinfo {author} {\bibfnamefont {A.}~\bibnamefont {Ransford}}, \bibinfo {author} {\bibfnamefont {A.~P.}\ \bibnamefont {Reed}}, \bibinfo {author} {\bibfnamefont {S.~J.}\ \bibnamefont {Sanders}}, \bibinfo {author} {\bibfnamefont {C.~H.}\ \bibnamefont {Baldwin}}, \bibinfo {author} {\bibfnamefont {D.}~\bibnamefont {Hayes}}, \bibinfo {author} {\bibfnamefont {B.}~\bibnamefont {Criger}}, \bibinfo {author} {\bibfnamefont {A.~C.}\ \bibnamefont {Potter}},\ and\ \bibinfo {author} {\bibfnamefont {D.}~\bibnamefont {Amaro}},\ }\href {https://arxiv.org/abs/2602.22211} {\bibinfo {title} {Computing with many encoded logical qubits beyond break-even}}
  (\bibinfo {year} {2026}),\ \Eprint {https://arxiv.org/abs/2602.22211} {arXiv:2602.22211 [quant-ph]} \BibitemShut {NoStop}%
\bibitem [{\citenamefont {Bluvstein}\ \emph {et~al.}(2022)\citenamefont {Bluvstein}, \citenamefont {Levine}, \citenamefont {Semeghini}, \citenamefont {Wang}, \citenamefont {Ebadi}, \citenamefont {Kalinowski}, \citenamefont {Keesling}, \citenamefont {Maskara}, \citenamefont {Pichler}, \citenamefont {Greiner}, \citenamefont {Vuletić},\ and\ \citenamefont {Lukin}}]{Bluvstein_2022}%
  \BibitemOpen
  \bibfield  {author} {\bibinfo {author} {\bibfnamefont {D.}~\bibnamefont {Bluvstein}}, \bibinfo {author} {\bibfnamefont {H.}~\bibnamefont {Levine}}, \bibinfo {author} {\bibfnamefont {G.}~\bibnamefont {Semeghini}}, \bibinfo {author} {\bibfnamefont {T.~T.}\ \bibnamefont {Wang}}, \bibinfo {author} {\bibfnamefont {S.}~\bibnamefont {Ebadi}}, \bibinfo {author} {\bibfnamefont {M.}~\bibnamefont {Kalinowski}}, \bibinfo {author} {\bibfnamefont {A.}~\bibnamefont {Keesling}}, \bibinfo {author} {\bibfnamefont {N.}~\bibnamefont {Maskara}}, \bibinfo {author} {\bibfnamefont {H.}~\bibnamefont {Pichler}}, \bibinfo {author} {\bibfnamefont {M.}~\bibnamefont {Greiner}}, \bibinfo {author} {\bibfnamefont {V.}~\bibnamefont {Vuletić}},\ and\ \bibinfo {author} {\bibfnamefont {M.~D.}\ \bibnamefont {Lukin}},\ }\bibfield  {title} {\bibinfo {title} {A quantum processor based on coherent transport of entangled atom arrays},\ }\href {https://doi.org/10.1038/s41586-022-04592-6} {\bibfield  {journal} {\bibinfo  {journal} {Nature}\ }\textbf
  {\bibinfo {volume} {604}},\ \bibinfo {pages} {451} (\bibinfo {year} {2022})}\BibitemShut {NoStop}%
\bibitem [{\citenamefont {Evered}\ \emph {et~al.}(2023)\citenamefont {Evered}, \citenamefont {Bluvstein}, \citenamefont {Kalinowski}, \citenamefont {Ebadi}, \citenamefont {Manovitz}, \citenamefont {Zhou}, \citenamefont {Li}, \citenamefont {Geim}, \citenamefont {Wang}, \citenamefont {Maskara}, \citenamefont {Levine}, \citenamefont {Semeghini}, \citenamefont {Greiner}, \citenamefont {Vuletić},\ and\ \citenamefont {Lukin}}]{Evered_2023}%
  \BibitemOpen
  \bibfield  {author} {\bibinfo {author} {\bibfnamefont {S.~J.}\ \bibnamefont {Evered}}, \bibinfo {author} {\bibfnamefont {D.}~\bibnamefont {Bluvstein}}, \bibinfo {author} {\bibfnamefont {M.}~\bibnamefont {Kalinowski}}, \bibinfo {author} {\bibfnamefont {S.}~\bibnamefont {Ebadi}}, \bibinfo {author} {\bibfnamefont {T.}~\bibnamefont {Manovitz}}, \bibinfo {author} {\bibfnamefont {H.}~\bibnamefont {Zhou}}, \bibinfo {author} {\bibfnamefont {S.~H.}\ \bibnamefont {Li}}, \bibinfo {author} {\bibfnamefont {A.~A.}\ \bibnamefont {Geim}}, \bibinfo {author} {\bibfnamefont {T.~T.}\ \bibnamefont {Wang}}, \bibinfo {author} {\bibfnamefont {N.}~\bibnamefont {Maskara}}, \bibinfo {author} {\bibfnamefont {H.}~\bibnamefont {Levine}}, \bibinfo {author} {\bibfnamefont {G.}~\bibnamefont {Semeghini}}, \bibinfo {author} {\bibfnamefont {M.}~\bibnamefont {Greiner}}, \bibinfo {author} {\bibfnamefont {V.}~\bibnamefont {Vuletić}},\ and\ \bibinfo {author} {\bibfnamefont {M.~D.}\ \bibnamefont {Lukin}},\ }\bibfield  {title} {\bibinfo {title}
  {High-fidelity parallel entangling gates on a neutral-atom quantum computer},\ }\href {https://doi.org/10.1038/s41586-023-06481-y} {\bibfield  {journal} {\bibinfo  {journal} {Nature}\ }\textbf {\bibinfo {volume} {622}},\ \bibinfo {pages} {268} (\bibinfo {year} {2023})}\BibitemShut {NoStop}%
\bibitem [{\citenamefont {Cain}\ \emph {et~al.}(2026)\citenamefont {Cain}, \citenamefont {Xu}, \citenamefont {King}, \citenamefont {Picard}, \citenamefont {Levine}, \citenamefont {Endres}, \citenamefont {Preskill}, \citenamefont {Huang},\ and\ \citenamefont {Bluvstein}}]{cain2026shorsalgorithmpossible10000}%
  \BibitemOpen
  \bibfield  {author} {\bibinfo {author} {\bibfnamefont {M.}~\bibnamefont {Cain}}, \bibinfo {author} {\bibfnamefont {Q.}~\bibnamefont {Xu}}, \bibinfo {author} {\bibfnamefont {R.}~\bibnamefont {King}}, \bibinfo {author} {\bibfnamefont {L.~R.~B.}\ \bibnamefont {Picard}}, \bibinfo {author} {\bibfnamefont {H.}~\bibnamefont {Levine}}, \bibinfo {author} {\bibfnamefont {M.}~\bibnamefont {Endres}}, \bibinfo {author} {\bibfnamefont {J.}~\bibnamefont {Preskill}}, \bibinfo {author} {\bibfnamefont {H.-Y.}\ \bibnamefont {Huang}},\ and\ \bibinfo {author} {\bibfnamefont {D.}~\bibnamefont {Bluvstein}},\ }\href {https://arxiv.org/abs/2603.28627} {\bibinfo {title} {Shor's algorithm is possible with as few as 10,000 reconfigurable atomic qubits}} (\bibinfo {year} {2026}),\ \Eprint {https://arxiv.org/abs/2603.28627} {arXiv:2603.28627 [quant-ph]} \BibitemShut {NoStop}%
\bibitem [{\citenamefont {Ismail}\ \emph {et~al.}(2026{\natexlab{a}})\citenamefont {Ismail}, \citenamefont {Chen}, \citenamefont {Zhao}, \citenamefont {Weiss}, \citenamefont {Liu}, \citenamefont {Zhou}, \citenamefont {Wang}, \citenamefont {Sornborger},\ and\ \citenamefont {Kornjača}}]{ismail2025transversal}%
  \BibitemOpen
  \bibfield  {author} {\bibinfo {author} {\bibfnamefont {R.}~\bibnamefont {Ismail}}, \bibinfo {author} {\bibfnamefont {I.-C.}\ \bibnamefont {Chen}}, \bibinfo {author} {\bibfnamefont {C.}~\bibnamefont {Zhao}}, \bibinfo {author} {\bibfnamefont {R.}~\bibnamefont {Weiss}}, \bibinfo {author} {\bibfnamefont {F.}~\bibnamefont {Liu}}, \bibinfo {author} {\bibfnamefont {H.}~\bibnamefont {Zhou}}, \bibinfo {author} {\bibfnamefont {S.-T.}\ \bibnamefont {Wang}}, \bibinfo {author} {\bibfnamefont {A.}~\bibnamefont {Sornborger}},\ and\ \bibinfo {author} {\bibfnamefont {M.}~\bibnamefont {Kornjača}},\ }\bibfield  {title} {\bibinfo {title} {Transversal architecture for megaquop-scale quantum simulation with neutral atoms},\ }\bibfield  {journal} {\bibinfo  {journal} {PRX Quantum}\ }\textbf {\bibinfo {volume} {7}},\ \href {https://doi.org/10.1103/j2fw-ccmy} {10.1103/j2fw-ccmy} (\bibinfo {year} {2026}{\natexlab{a}})\BibitemShut {NoStop}%
\bibitem [{\citenamefont {Sales~Rodriguez}\ \emph {et~al.}(2025)\citenamefont {Sales~Rodriguez}, \citenamefont {Robinson}, \citenamefont {Jepsen}, \citenamefont {He}, \citenamefont {Duckering}, \citenamefont {Zhao}, \citenamefont {Wu}, \citenamefont {Campo}, \citenamefont {Bagnall}, \citenamefont {Kwon}, \citenamefont {Karolyshyn}, \citenamefont {Weinberg}, \citenamefont {Cain}, \citenamefont {Evered}, \citenamefont {Geim}, \citenamefont {Kalinowski}, \citenamefont {Li}, \citenamefont {Manovitz}, \citenamefont {Amato-Grill}, \citenamefont {Basham}, \citenamefont {Bernstein}, \citenamefont {Braverman}, \citenamefont {Bylinskii}, \citenamefont {Choukri}, \citenamefont {DeAngelo}, \citenamefont {Fang}, \citenamefont {Fieweger}, \citenamefont {Frederick}, \citenamefont {Haines}, \citenamefont {Hamdan}, \citenamefont {Hammett}, \citenamefont {Hsu}, \citenamefont {Hu}, \citenamefont {Huber}, \citenamefont {Jia}, \citenamefont {Kedar}, \citenamefont {Kornjača}, \citenamefont {Liu}, \citenamefont {Long},
  \citenamefont {Lopatin}, \citenamefont {Lopes}, \citenamefont {Luo}, \citenamefont {Macrì}, \citenamefont {Marković}, \citenamefont {Martínez-Martínez}, \citenamefont {Meng}, \citenamefont {Ostermann}, \citenamefont {Ostroumov}, \citenamefont {Paquette}, \citenamefont {Qiang}, \citenamefont {Shofman}, \citenamefont {Singh}, \citenamefont {Singh}, \citenamefont {Sinha}, \citenamefont {Thoreen}, \citenamefont {Wan}, \citenamefont {Wang}, \citenamefont {Waxman-Lenz}, \citenamefont {Wong}, \citenamefont {Wurtz}, \citenamefont {Zhdanov}, \citenamefont {Zheng}, \citenamefont {Greiner}, \citenamefont {Keesling}, \citenamefont {Gemelke}, \citenamefont {Vuletić}, \citenamefont {Kitagawa}, \citenamefont {Wang}, \citenamefont {Bluvstein}, \citenamefont {Lukin}, \citenamefont {Lukin}, \citenamefont {Zhou},\ and\ \citenamefont {Cantú}}]{Sales_Rodriguez_2025}%
  \BibitemOpen
  \bibfield  {author} {\bibinfo {author} {\bibfnamefont {P.}~\bibnamefont {Sales~Rodriguez}}, \bibinfo {author} {\bibfnamefont {J.~M.}\ \bibnamefont {Robinson}}, \bibinfo {author} {\bibfnamefont {P.~N.}\ \bibnamefont {Jepsen}}, \bibinfo {author} {\bibfnamefont {Z.}~\bibnamefont {He}}, \bibinfo {author} {\bibfnamefont {C.}~\bibnamefont {Duckering}}, \bibinfo {author} {\bibfnamefont {C.}~\bibnamefont {Zhao}}, \bibinfo {author} {\bibfnamefont {K.-H.}\ \bibnamefont {Wu}}, \bibinfo {author} {\bibfnamefont {J.}~\bibnamefont {Campo}}, \bibinfo {author} {\bibfnamefont {K.}~\bibnamefont {Bagnall}}, \bibinfo {author} {\bibfnamefont {M.}~\bibnamefont {Kwon}}, \bibinfo {author} {\bibfnamefont {T.}~\bibnamefont {Karolyshyn}}, \bibinfo {author} {\bibfnamefont {P.}~\bibnamefont {Weinberg}}, \bibinfo {author} {\bibfnamefont {M.}~\bibnamefont {Cain}}, \bibinfo {author} {\bibfnamefont {S.~J.}\ \bibnamefont {Evered}}, \bibinfo {author} {\bibfnamefont {A.~A.}\ \bibnamefont {Geim}}, \bibinfo {author} {\bibfnamefont
  {M.}~\bibnamefont {Kalinowski}}, \bibinfo {author} {\bibfnamefont {S.~H.}\ \bibnamefont {Li}}, \bibinfo {author} {\bibfnamefont {T.}~\bibnamefont {Manovitz}}, \bibinfo {author} {\bibfnamefont {J.}~\bibnamefont {Amato-Grill}}, \bibinfo {author} {\bibfnamefont {J.~I.}\ \bibnamefont {Basham}}, \bibinfo {author} {\bibfnamefont {L.}~\bibnamefont {Bernstein}}, \bibinfo {author} {\bibfnamefont {B.}~\bibnamefont {Braverman}}, \bibinfo {author} {\bibfnamefont {A.}~\bibnamefont {Bylinskii}}, \bibinfo {author} {\bibfnamefont {A.}~\bibnamefont {Choukri}}, \bibinfo {author} {\bibfnamefont {R.~J.}\ \bibnamefont {DeAngelo}}, \bibinfo {author} {\bibfnamefont {F.}~\bibnamefont {Fang}}, \bibinfo {author} {\bibfnamefont {C.}~\bibnamefont {Fieweger}}, \bibinfo {author} {\bibfnamefont {P.}~\bibnamefont {Frederick}}, \bibinfo {author} {\bibfnamefont {D.}~\bibnamefont {Haines}}, \bibinfo {author} {\bibfnamefont {M.}~\bibnamefont {Hamdan}}, \bibinfo {author} {\bibfnamefont {J.}~\bibnamefont {Hammett}}, \bibinfo {author}
  {\bibfnamefont {N.}~\bibnamefont {Hsu}}, \bibinfo {author} {\bibfnamefont {M.-G.}\ \bibnamefont {Hu}}, \bibinfo {author} {\bibfnamefont {F.}~\bibnamefont {Huber}}, \bibinfo {author} {\bibfnamefont {N.}~\bibnamefont {Jia}}, \bibinfo {author} {\bibfnamefont {D.}~\bibnamefont {Kedar}}, \bibinfo {author} {\bibfnamefont {M.}~\bibnamefont {Kornjača}}, \bibinfo {author} {\bibfnamefont {F.}~\bibnamefont {Liu}}, \bibinfo {author} {\bibfnamefont {J.}~\bibnamefont {Long}}, \bibinfo {author} {\bibfnamefont {J.}~\bibnamefont {Lopatin}}, \bibinfo {author} {\bibfnamefont {P.~L.~S.}\ \bibnamefont {Lopes}}, \bibinfo {author} {\bibfnamefont {X.-Z.}\ \bibnamefont {Luo}}, \bibinfo {author} {\bibfnamefont {T.}~\bibnamefont {Macrì}}, \bibinfo {author} {\bibfnamefont {O.}~\bibnamefont {Marković}}, \bibinfo {author} {\bibfnamefont {L.~A.}\ \bibnamefont {Martínez-Martínez}}, \bibinfo {author} {\bibfnamefont {X.}~\bibnamefont {Meng}}, \bibinfo {author} {\bibfnamefont {S.}~\bibnamefont {Ostermann}}, \bibinfo {author}
  {\bibfnamefont {E.}~\bibnamefont {Ostroumov}}, \bibinfo {author} {\bibfnamefont {D.}~\bibnamefont {Paquette}}, \bibinfo {author} {\bibfnamefont {Z.}~\bibnamefont {Qiang}}, \bibinfo {author} {\bibfnamefont {V.}~\bibnamefont {Shofman}}, \bibinfo {author} {\bibfnamefont {A.}~\bibnamefont {Singh}}, \bibinfo {author} {\bibfnamefont {M.}~\bibnamefont {Singh}}, \bibinfo {author} {\bibfnamefont {N.}~\bibnamefont {Sinha}}, \bibinfo {author} {\bibfnamefont {H.}~\bibnamefont {Thoreen}}, \bibinfo {author} {\bibfnamefont {N.}~\bibnamefont {Wan}}, \bibinfo {author} {\bibfnamefont {Y.}~\bibnamefont {Wang}}, \bibinfo {author} {\bibfnamefont {D.}~\bibnamefont {Waxman-Lenz}}, \bibinfo {author} {\bibfnamefont {T.}~\bibnamefont {Wong}}, \bibinfo {author} {\bibfnamefont {J.}~\bibnamefont {Wurtz}}, \bibinfo {author} {\bibfnamefont {A.}~\bibnamefont {Zhdanov}}, \bibinfo {author} {\bibfnamefont {L.}~\bibnamefont {Zheng}}, \bibinfo {author} {\bibfnamefont {M.}~\bibnamefont {Greiner}}, \bibinfo {author} {\bibfnamefont
  {A.}~\bibnamefont {Keesling}}, \bibinfo {author} {\bibfnamefont {N.}~\bibnamefont {Gemelke}}, \bibinfo {author} {\bibfnamefont {V.}~\bibnamefont {Vuletić}}, \bibinfo {author} {\bibfnamefont {T.}~\bibnamefont {Kitagawa}}, \bibinfo {author} {\bibfnamefont {S.-T.}\ \bibnamefont {Wang}}, \bibinfo {author} {\bibfnamefont {D.}~\bibnamefont {Bluvstein}}, \bibinfo {author} {\bibfnamefont {M.~D.}\ \bibnamefont {Lukin}}, \bibinfo {author} {\bibfnamefont {A.}~\bibnamefont {Lukin}}, \bibinfo {author} {\bibfnamefont {H.}~\bibnamefont {Zhou}},\ and\ \bibinfo {author} {\bibfnamefont {S.~H.}\ \bibnamefont {Cantú}},\ }\bibfield  {title} {\bibinfo {title} {Experimental demonstration of logical magic state distillation},\ }\href {https://doi.org/10.1038/s41586-025-09367-3} {\bibfield  {journal} {\bibinfo  {journal} {Nature}\ }\textbf {\bibinfo {volume} {645}},\ \bibinfo {pages} {620} (\bibinfo {year} {2025})}\BibitemShut {NoStop}%
\bibitem [{\citenamefont {Terhal}(2015)}]{RevModPhys.87.307}%
  \BibitemOpen
  \bibfield  {author} {\bibinfo {author} {\bibfnamefont {B.~M.}\ \bibnamefont {Terhal}},\ }\bibfield  {title} {\bibinfo {title} {Quantum error correction for quantum memories},\ }\href {https://doi.org/10.1103/RevModPhys.87.307} {\bibfield  {journal} {\bibinfo  {journal} {Rev. Mod. Phys.}\ }\textbf {\bibinfo {volume} {87}},\ \bibinfo {pages} {307} (\bibinfo {year} {2015})}\BibitemShut {NoStop}%
\bibitem [{\citenamefont {Ryan-Anderson}\ \emph {et~al.}(2021)\citenamefont {Ryan-Anderson}, \citenamefont {Bohnet}, \citenamefont {Lee}, \citenamefont {Gresh}, \citenamefont {Hankin}, \citenamefont {Gaebler}, \citenamefont {Francois}, \citenamefont {Chernoguzov}, \citenamefont {Lucchetti}, \citenamefont {Brown}, \citenamefont {Gatterman}, \citenamefont {Halit}, \citenamefont {Gilmore}, \citenamefont {Gerber}, \citenamefont {Neyenhuis}, \citenamefont {Hayes},\ and\ \citenamefont {Stutz}}]{PhysRevX.11.041058}%
  \BibitemOpen
  \bibfield  {author} {\bibinfo {author} {\bibfnamefont {C.}~\bibnamefont {Ryan-Anderson}}, \bibinfo {author} {\bibfnamefont {J.~G.}\ \bibnamefont {Bohnet}}, \bibinfo {author} {\bibfnamefont {K.}~\bibnamefont {Lee}}, \bibinfo {author} {\bibfnamefont {D.}~\bibnamefont {Gresh}}, \bibinfo {author} {\bibfnamefont {A.}~\bibnamefont {Hankin}}, \bibinfo {author} {\bibfnamefont {J.~P.}\ \bibnamefont {Gaebler}}, \bibinfo {author} {\bibfnamefont {D.}~\bibnamefont {Francois}}, \bibinfo {author} {\bibfnamefont {A.}~\bibnamefont {Chernoguzov}}, \bibinfo {author} {\bibfnamefont {D.}~\bibnamefont {Lucchetti}}, \bibinfo {author} {\bibfnamefont {N.~C.}\ \bibnamefont {Brown}}, \bibinfo {author} {\bibfnamefont {T.~M.}\ \bibnamefont {Gatterman}}, \bibinfo {author} {\bibfnamefont {S.~K.}\ \bibnamefont {Halit}}, \bibinfo {author} {\bibfnamefont {K.}~\bibnamefont {Gilmore}}, \bibinfo {author} {\bibfnamefont {J.~A.}\ \bibnamefont {Gerber}}, \bibinfo {author} {\bibfnamefont {B.}~\bibnamefont {Neyenhuis}}, \bibinfo {author}
  {\bibfnamefont {D.}~\bibnamefont {Hayes}},\ and\ \bibinfo {author} {\bibfnamefont {R.~P.}\ \bibnamefont {Stutz}},\ }\bibfield  {title} {\bibinfo {title} {Realization of real-time fault-tolerant quantum error correction},\ }\href {https://doi.org/10.1103/PhysRevX.11.041058} {\bibfield  {journal} {\bibinfo  {journal} {Phys. Rev. X}\ }\textbf {\bibinfo {volume} {11}},\ \bibinfo {pages} {041058} (\bibinfo {year} {2021})}\BibitemShut {NoStop}%
\bibitem [{\citenamefont {Caune}\ \emph {et~al.}(2024)\citenamefont {Caune}, \citenamefont {Skoric}, \citenamefont {Blunt}, \citenamefont {Ruban}, \citenamefont {McDaniel}, \citenamefont {Valery}, \citenamefont {Patterson}, \citenamefont {Gramolin}, \citenamefont {Majaniemi}, \citenamefont {Barnes}, \citenamefont {Bialas}, \citenamefont {Buğdaycı}, \citenamefont {Crawford}, \citenamefont {Gehér}, \citenamefont {Krovi}, \citenamefont {Matekole}, \citenamefont {Topal}, \citenamefont {Poletto}, \citenamefont {Bryant}, \citenamefont {Snyder}, \citenamefont {Gillespie}, \citenamefont {Jones}, \citenamefont {Johar}, \citenamefont {Campbell},\ and\ \citenamefont {Hill}}]{caune2024demonstratingrealtimelowlatencyquantum}%
  \BibitemOpen
  \bibfield  {author} {\bibinfo {author} {\bibfnamefont {L.}~\bibnamefont {Caune}}, \bibinfo {author} {\bibfnamefont {L.}~\bibnamefont {Skoric}}, \bibinfo {author} {\bibfnamefont {N.~S.}\ \bibnamefont {Blunt}}, \bibinfo {author} {\bibfnamefont {A.}~\bibnamefont {Ruban}}, \bibinfo {author} {\bibfnamefont {J.}~\bibnamefont {McDaniel}}, \bibinfo {author} {\bibfnamefont {J.~A.}\ \bibnamefont {Valery}}, \bibinfo {author} {\bibfnamefont {A.~D.}\ \bibnamefont {Patterson}}, \bibinfo {author} {\bibfnamefont {A.~V.}\ \bibnamefont {Gramolin}}, \bibinfo {author} {\bibfnamefont {J.}~\bibnamefont {Majaniemi}}, \bibinfo {author} {\bibfnamefont {K.~M.}\ \bibnamefont {Barnes}}, \bibinfo {author} {\bibfnamefont {T.}~\bibnamefont {Bialas}}, \bibinfo {author} {\bibfnamefont {O.}~\bibnamefont {Buğdaycı}}, \bibinfo {author} {\bibfnamefont {O.}~\bibnamefont {Crawford}}, \bibinfo {author} {\bibfnamefont {G.~P.}\ \bibnamefont {Gehér}}, \bibinfo {author} {\bibfnamefont {H.}~\bibnamefont {Krovi}}, \bibinfo {author} {\bibfnamefont
  {E.}~\bibnamefont {Matekole}}, \bibinfo {author} {\bibfnamefont {C.}~\bibnamefont {Topal}}, \bibinfo {author} {\bibfnamefont {S.}~\bibnamefont {Poletto}}, \bibinfo {author} {\bibfnamefont {M.}~\bibnamefont {Bryant}}, \bibinfo {author} {\bibfnamefont {K.}~\bibnamefont {Snyder}}, \bibinfo {author} {\bibfnamefont {N.~I.}\ \bibnamefont {Gillespie}}, \bibinfo {author} {\bibfnamefont {G.}~\bibnamefont {Jones}}, \bibinfo {author} {\bibfnamefont {K.}~\bibnamefont {Johar}}, \bibinfo {author} {\bibfnamefont {E.~T.}\ \bibnamefont {Campbell}},\ and\ \bibinfo {author} {\bibfnamefont {A.~D.}\ \bibnamefont {Hill}},\ }\href {https://arxiv.org/abs/2410.05202} {\bibinfo {title} {Demonstrating real-time and low-latency quantum error correction with superconducting qubits}} (\bibinfo {year} {2024}),\ \Eprint {https://arxiv.org/abs/2410.05202} {arXiv:2410.05202 [quant-ph]} \BibitemShut {NoStop}%
\bibitem [{\citenamefont {Litinski}(2019{\natexlab{a}})}]{Litinski2019gameofsurfacecodes}%
  \BibitemOpen
  \bibfield  {author} {\bibinfo {author} {\bibfnamefont {D.}~\bibnamefont {Litinski}},\ }\bibfield  {title} {\bibinfo {title} {A {G}ame of {S}urface {C}odes: {L}arge-{S}cale {Q}uantum {C}omputing with {L}attice {S}urgery},\ }\href {https://doi.org/10.22331/q-2019-03-05-128} {\bibfield  {journal} {\bibinfo  {journal} {{Quantum}}\ }\textbf {\bibinfo {volume} {3}},\ \bibinfo {pages} {128} (\bibinfo {year} {2019}{\natexlab{a}})}\BibitemShut {NoStop}%
\bibitem [{\citenamefont {Awasthi}\ \emph {et~al.}(2026)\citenamefont {Awasthi}, \citenamefont {Sethi}, \citenamefont {Khan}, \citenamefont {Ravi},\ and\ \citenamefont {Baker}}]{awasthi2026pricepayoffnondeterminismfault}%
  \BibitemOpen
  \bibfield  {author} {\bibinfo {author} {\bibfnamefont {A.}~\bibnamefont {Awasthi}}, \bibinfo {author} {\bibfnamefont {S.}~\bibnamefont {Sethi}}, \bibinfo {author} {\bibfnamefont {S.}~\bibnamefont {Khan}}, \bibinfo {author} {\bibfnamefont {G.~S.}\ \bibnamefont {Ravi}},\ and\ \bibinfo {author} {\bibfnamefont {J.~M.}\ \bibnamefont {Baker}},\ }\href {https://arxiv.org/abs/2605.07983} {\bibinfo {title} {Price and payoff: Non-determinism in fault tolerant quantum computation}} (\bibinfo {year} {2026}),\ \Eprint {https://arxiv.org/abs/2605.07983} {arXiv:2605.07983 [quant-ph]} \BibitemShut {NoStop}%
\bibitem [{\citenamefont {Leblond}\ \emph {et~al.}(2024)\citenamefont {Leblond}, \citenamefont {Dean}, \citenamefont {Watkins},\ and\ \citenamefont {Bennink}}]{leblond_lattice_surgery}%
  \BibitemOpen
  \bibfield  {author} {\bibinfo {author} {\bibfnamefont {T.}~\bibnamefont {Leblond}}, \bibinfo {author} {\bibfnamefont {C.}~\bibnamefont {Dean}}, \bibinfo {author} {\bibfnamefont {G.}~\bibnamefont {Watkins}},\ and\ \bibinfo {author} {\bibfnamefont {R.}~\bibnamefont {Bennink}},\ }\bibfield  {title} {\bibinfo {title} {Realistic cost to execute practical quantum circuits using direct clifford+t lattice surgery compilation},\ }\bibfield  {journal} {\bibinfo  {journal} {ACM Transactions on Quantum Computing}\ }\textbf {\bibinfo {volume} {5}},\ \href {https://doi.org/10.1145/3689826} {10.1145/3689826} (\bibinfo {year} {2024})\BibitemShut {NoStop}%
\bibitem [{\citenamefont {Huggins}\ \emph {et~al.}(2025)\citenamefont {Huggins}, \citenamefont {Khattar}, \citenamefont {Xu}, \citenamefont {Harrigan}, \citenamefont {Kang}, \citenamefont {Low}, \citenamefont {Fowler}, \citenamefont {Rubin},\ and\ \citenamefont {Babbush}}]{flasq}%
  \BibitemOpen
  \bibfield  {author} {\bibinfo {author} {\bibfnamefont {W.~J.}\ \bibnamefont {Huggins}}, \bibinfo {author} {\bibfnamefont {T.}~\bibnamefont {Khattar}}, \bibinfo {author} {\bibfnamefont {A.}~\bibnamefont {Xu}}, \bibinfo {author} {\bibfnamefont {M.}~\bibnamefont {Harrigan}}, \bibinfo {author} {\bibfnamefont {C.}~\bibnamefont {Kang}}, \bibinfo {author} {\bibfnamefont {G.~H.}\ \bibnamefont {Low}}, \bibinfo {author} {\bibfnamefont {A.}~\bibnamefont {Fowler}}, \bibinfo {author} {\bibfnamefont {N.~C.}\ \bibnamefont {Rubin}},\ and\ \bibinfo {author} {\bibfnamefont {R.}~\bibnamefont {Babbush}},\ }\href {https://arxiv.org/abs/2511.08508} {\bibinfo {title} {The fluid allocation of surface code qubits {FLASQ} cost model for early fault-tolerant quantum algorithms}} (\bibinfo {year} {2025}),\ \Eprint {https://arxiv.org/abs/2511.08508} {arXiv:2511.08508 [quant-ph]} \BibitemShut {NoStop}%
\bibitem [{\citenamefont {Molavi}\ \emph {et~al.}(2025)\citenamefont {Molavi}, \citenamefont {Xu}, \citenamefont {Tannu},\ and\ \citenamefont {Albarghouthi}}]{dascot}%
  \BibitemOpen
  \bibfield  {author} {\bibinfo {author} {\bibfnamefont {A.}~\bibnamefont {Molavi}}, \bibinfo {author} {\bibfnamefont {A.}~\bibnamefont {Xu}}, \bibinfo {author} {\bibfnamefont {S.}~\bibnamefont {Tannu}},\ and\ \bibinfo {author} {\bibfnamefont {A.}~\bibnamefont {Albarghouthi}},\ }\bibfield  {title} {\bibinfo {title} {Dependency-aware compilation for surface code quantum architectures},\ }\bibfield  {journal} {\bibinfo  {journal} {Proc. ACM Program. Lang.}\ }\textbf {\bibinfo {volume} {9}},\ \href {https://doi.org/10.1145/3720416} {10.1145/3720416} (\bibinfo {year} {2025})\BibitemShut {NoStop}%
\bibitem [{\citenamefont {Kobori}\ \emph {et~al.}(2025)\citenamefont {Kobori}, \citenamefont {Suzuki}, \citenamefont {Ueno}, \citenamefont {Tanimoto}, \citenamefont {Todo},\ and\ \citenamefont {Tokunaga}}]{lsqca}%
  \BibitemOpen
  \bibfield  {author} {\bibinfo {author} {\bibfnamefont {T.}~\bibnamefont {Kobori}}, \bibinfo {author} {\bibfnamefont {Y.}~\bibnamefont {Suzuki}}, \bibinfo {author} {\bibfnamefont {Y.}~\bibnamefont {Ueno}}, \bibinfo {author} {\bibfnamefont {T.}~\bibnamefont {Tanimoto}}, \bibinfo {author} {\bibfnamefont {S.}~\bibnamefont {Todo}},\ and\ \bibinfo {author} {\bibfnamefont {Y.}~\bibnamefont {Tokunaga}},\ }\bibfield  {title} {\bibinfo {title} {{ LSQCA: Resource-Efficient Load/Store Architecture for Limited-Scale Fault-Tolerant Quantum Computing }},\ }in\ \href {https://doi.org/10.1109/HPCA61900.2025.00033} {\emph {\bibinfo {booktitle} {2025 IEEE International Symposium on High Performance Computer Architecture (HPCA)}}}\ (\bibinfo  {publisher} {IEEE Computer Society},\ \bibinfo {address} {Los Alamitos, CA, USA},\ \bibinfo {year} {2025})\ pp.\ \bibinfo {pages} {304--320}\BibitemShut {NoStop}%
\bibitem [{\citenamefont {Hsu}\ \emph {et~al.}(2021)\citenamefont {Hsu}, \citenamefont {Lin}, \citenamefont {Tseng},\ and\ \citenamefont {Chang}}]{qbrige}%
  \BibitemOpen
  \bibfield  {author} {\bibinfo {author} {\bibfnamefont {C.-H.}\ \bibnamefont {Hsu}}, \bibinfo {author} {\bibfnamefont {W.-H.}\ \bibnamefont {Lin}}, \bibinfo {author} {\bibfnamefont {W.-H.}\ \bibnamefont {Tseng}},\ and\ \bibinfo {author} {\bibfnamefont {Y.-W.}\ \bibnamefont {Chang}},\ }\bibfield  {title} {\bibinfo {title} {A bridge-based compression algorithm for topological quantum circuits},\ }in\ \href {https://doi.org/10.1109/DAC18074.2021.9586322} {\emph {\bibinfo {booktitle} {2021 58th ACM/IEEE Design Automation Conference (DAC)}}}\ (\bibinfo {year} {2021})\ pp.\ \bibinfo {pages} {457--462}\BibitemShut {NoStop}%
\bibitem [{\citenamefont {Beverland}\ \emph {et~al.}(2022)\citenamefont {Beverland}, \citenamefont {Murali}, \citenamefont {Troyer}, \citenamefont {Svore}, \citenamefont {Hoefler}, \citenamefont {Kliuchnikov}, \citenamefont {Low}, \citenamefont {Soeken}, \citenamefont {Sundaram},\ and\ \citenamefont {Vaschillo}}]{beverland2022assessingrequirementsscalepractical}%
  \BibitemOpen
  \bibfield  {author} {\bibinfo {author} {\bibfnamefont {M.~E.}\ \bibnamefont {Beverland}}, \bibinfo {author} {\bibfnamefont {P.}~\bibnamefont {Murali}}, \bibinfo {author} {\bibfnamefont {M.}~\bibnamefont {Troyer}}, \bibinfo {author} {\bibfnamefont {K.~M.}\ \bibnamefont {Svore}}, \bibinfo {author} {\bibfnamefont {T.}~\bibnamefont {Hoefler}}, \bibinfo {author} {\bibfnamefont {V.}~\bibnamefont {Kliuchnikov}}, \bibinfo {author} {\bibfnamefont {G.~H.}\ \bibnamefont {Low}}, \bibinfo {author} {\bibfnamefont {M.}~\bibnamefont {Soeken}}, \bibinfo {author} {\bibfnamefont {A.}~\bibnamefont {Sundaram}},\ and\ \bibinfo {author} {\bibfnamefont {A.}~\bibnamefont {Vaschillo}},\ }\href {https://arxiv.org/abs/2211.07629} {\bibinfo {title} {Assessing requirements to scale to practical quantum advantage}} (\bibinfo {year} {2022}),\ \Eprint {https://arxiv.org/abs/2211.07629} {arXiv:2211.07629 [quant-ph]} \BibitemShut {NoStop}%
\bibitem [{\citenamefont {Zhu}\ \emph {et~al.}(2026)\citenamefont {Zhu}, \citenamefont {Wu}, \citenamefont {Chen}, \citenamefont {He}, \citenamefont {Wu}, \citenamefont {Wang},\ and\ \citenamefont {Lao}}]{zhu2026o3lsoptimizinglatticesurgery}%
  \BibitemOpen
  \bibfield  {author} {\bibinfo {author} {\bibfnamefont {C.}~\bibnamefont {Zhu}}, \bibinfo {author} {\bibfnamefont {X.}~\bibnamefont {Wu}}, \bibinfo {author} {\bibfnamefont {J.}~\bibnamefont {Chen}}, \bibinfo {author} {\bibfnamefont {K.}~\bibnamefont {He}}, \bibinfo {author} {\bibfnamefont {J.}~\bibnamefont {Wu}}, \bibinfo {author} {\bibfnamefont {X.}~\bibnamefont {Wang}},\ and\ \bibinfo {author} {\bibfnamefont {L.}~\bibnamefont {Lao}},\ }\href {https://arxiv.org/abs/2604.15099} {\bibinfo {title} {O3ls: Optimizing lattice surgery via automatic layout searching and loose scheduling}} (\bibinfo {year} {2026}),\ \Eprint {https://arxiv.org/abs/2604.15099} {arXiv:2604.15099 [quant-ph]} \BibitemShut {NoStop}%
\bibitem [{\citenamefont {Hirano}\ \emph {et~al.}(2024)\citenamefont {Hirano}, \citenamefont {Suzuki},\ and\ \citenamefont {Fujii}}]{magicpool}%
  \BibitemOpen
  \bibfield  {author} {\bibinfo {author} {\bibfnamefont {Y.}~\bibnamefont {Hirano}}, \bibinfo {author} {\bibfnamefont {Y.}~\bibnamefont {Suzuki}},\ and\ \bibinfo {author} {\bibfnamefont {K.}~\bibnamefont {Fujii}},\ }\href {https://arxiv.org/abs/2407.07394} {\bibinfo {title} {{MagicPool}: Dealing with magic state distillation failures on large-scale fault-tolerant quantum computer}} (\bibinfo {year} {2024}),\ \Eprint {https://arxiv.org/abs/2407.07394} {arXiv:2407.07394 [quant-ph]} \BibitemShut {NoStop}%
\bibitem [{\citenamefont {Sethi}\ and\ \citenamefont {Baker}(2025)}]{rescq}%
  \BibitemOpen
  \bibfield  {author} {\bibinfo {author} {\bibfnamefont {S.}~\bibnamefont {Sethi}}\ and\ \bibinfo {author} {\bibfnamefont {J.~M.}\ \bibnamefont {Baker}},\ }\bibfield  {title} {\bibinfo {title} {{RESCQ}: Realtime scheduling for continuous angle quantum error correction architectures},\ }in\ \href {https://doi.org/10.1145/3676641.3716018} {\emph {\bibinfo {booktitle} {Proceedings of the 30th ACM International Conference on Architectural Support for Programming Languages and Operating Systems, Volume 2}}},\ \bibinfo {series and number} {ASPLOS '25}\ (\bibinfo  {publisher} {Association for Computing Machinery},\ \bibinfo {address} {New York, NY, USA},\ \bibinfo {year} {2025})\ p.\ \bibinfo {pages} {1028–1043}\BibitemShut {NoStop}%
\bibitem [{\citenamefont {Hofmeyr}\ \emph {et~al.}(2026)\citenamefont {Hofmeyr}, \citenamefont {Weiden}, \citenamefont {Kalloor}, \citenamefont {Kubiatowicz},\ and\ \citenamefont {Iancu}}]{puremagic}%
  \BibitemOpen
  \bibfield  {author} {\bibinfo {author} {\bibfnamefont {S.}~\bibnamefont {Hofmeyr}}, \bibinfo {author} {\bibfnamefont {M.}~\bibnamefont {Weiden}}, \bibinfo {author} {\bibfnamefont {J.}~\bibnamefont {Kalloor}}, \bibinfo {author} {\bibfnamefont {J.}~\bibnamefont {Kubiatowicz}},\ and\ \bibinfo {author} {\bibfnamefont {C.}~\bibnamefont {Iancu}},\ }\href {https://arxiv.org/abs/2512.06484} {\bibinfo {title} {{PureMagic}: A dynamic scheduler for lattice surgery}} (\bibinfo {year} {2026}),\ \Eprint {https://arxiv.org/abs/2512.06484} {arXiv:2512.06484 [quant-ph]} \BibitemShut {NoStop}%
\bibitem [{\citenamefont {Pflieger}\ \emph {et~al.}(2026)\citenamefont {Pflieger}, \citenamefont {Świerkowska}, \citenamefont {Giortamis},\ and\ \citenamefont {Bhatotia}}]{Harvest}%
  \BibitemOpen
  \bibfield  {author} {\bibinfo {author} {\bibfnamefont {J.}~\bibnamefont {Pflieger}}, \bibinfo {author} {\bibfnamefont {A.}~\bibnamefont {Świerkowska}}, \bibinfo {author} {\bibfnamefont {E.}~\bibnamefont {Giortamis}},\ and\ \bibinfo {author} {\bibfnamefont {P.}~\bibnamefont {Bhatotia}},\ }\href {https://arxiv.org/abs/2608.03315} {\bibinfo {title} {Harvest: Resource-aware quantum compilation for magic state protocols}} (\bibinfo {year} {2026}),\ \Eprint {https://arxiv.org/abs/2608.03315} {arXiv:2608.03315 [quant-ph]} \BibitemShut {NoStop}%
\bibitem [{\citenamefont {Dangwal}\ \emph {et~al.}(2025)\citenamefont {Dangwal}, \citenamefont {Vittal}, \citenamefont {Seifert}, \citenamefont {Chong},\ and\ \citenamefont {Ravi}}]{vqe_star}%
  \BibitemOpen
  \bibfield  {author} {\bibinfo {author} {\bibfnamefont {S.}~\bibnamefont {Dangwal}}, \bibinfo {author} {\bibfnamefont {S.}~\bibnamefont {Vittal}}, \bibinfo {author} {\bibfnamefont {L.~M.}\ \bibnamefont {Seifert}}, \bibinfo {author} {\bibfnamefont {F.~T.}\ \bibnamefont {Chong}},\ and\ \bibinfo {author} {\bibfnamefont {G.~S.}\ \bibnamefont {Ravi}},\ }\bibfield  {title} {\bibinfo {title} {Variational quantum algorithms in the era of early fault tolerance},\ }in\ \href {https://doi.org/10.1145/3695053.3731112} {\emph {\bibinfo {booktitle} {Proceedings of the 52nd Annual International Symposium on Computer Architecture}}},\ \bibinfo {series and number} {ISCA '25}\ (\bibinfo  {publisher} {Association for Computing Machinery},\ \bibinfo {address} {New York, NY, USA},\ \bibinfo {year} {2025})\ p.\ \bibinfo {pages} {1417}\BibitemShut {NoStop}%
\bibitem [{\citenamefont {Kurita}(2026)}]{kurita2026generalcircuitcompilationprotocol}%
  \BibitemOpen
  \bibfield  {author} {\bibinfo {author} {\bibfnamefont {T.}~\bibnamefont {Kurita}},\ }\href {https://arxiv.org/abs/2603.17428} {\bibinfo {title} {General circuit compilation protocol into partially fault-tolerant quantum computing architecture}} (\bibinfo {year} {2026}),\ \Eprint {https://arxiv.org/abs/2603.17428} {arXiv:2603.17428 [quant-ph]} \BibitemShut {NoStop}%
\bibitem [{\citenamefont {Akahoshi}\ \emph {et~al.}(2025)\citenamefont {Akahoshi}, \citenamefont {Toshio}, \citenamefont {Fujisaki}, \citenamefont {Oshima}, \citenamefont {Sato},\ and\ \citenamefont {Fujii}}]{93zr-1ykb}%
  \BibitemOpen
  \bibfield  {author} {\bibinfo {author} {\bibfnamefont {Y.}~\bibnamefont {Akahoshi}}, \bibinfo {author} {\bibfnamefont {R.}~\bibnamefont {Toshio}}, \bibinfo {author} {\bibfnamefont {J.}~\bibnamefont {Fujisaki}}, \bibinfo {author} {\bibfnamefont {H.}~\bibnamefont {Oshima}}, \bibinfo {author} {\bibfnamefont {S.}~\bibnamefont {Sato}},\ and\ \bibinfo {author} {\bibfnamefont {K.}~\bibnamefont {Fujii}},\ }\bibfield  {title} {\bibinfo {title} {Compilation of trotter-based time evolution for partially fault-tolerant quantum computing architecture},\ }\href {https://doi.org/10.1103/93zr-1ykb} {\bibfield  {journal} {\bibinfo  {journal} {PRX Quantum}\ }\textbf {\bibinfo {volume} {6}},\ \bibinfo {pages} {040319} (\bibinfo {year} {2025})}\BibitemShut {NoStop}%
\bibitem [{\citenamefont {Preskill}(2025)}]{beyond_NISQ}%
  \BibitemOpen
  \bibfield  {author} {\bibinfo {author} {\bibfnamefont {J.}~\bibnamefont {Preskill}},\ }\bibfield  {title} {\bibinfo {title} {Beyond {NISQ}: The megaquop machine},\ }\bibfield  {journal} {\bibinfo  {journal} {ACM Transactions on Quantum Computing}\ }\textbf {\bibinfo {volume} {6}},\ \href {https://doi.org/10.1145/3723153} {10.1145/3723153} (\bibinfo {year} {2025})\BibitemShut {NoStop}%
\bibitem [{\citenamefont {Preskill}(2018)}]{Preskill2018quantumcomputingin}%
  \BibitemOpen
  \bibfield  {author} {\bibinfo {author} {\bibfnamefont {J.}~\bibnamefont {Preskill}},\ }\bibfield  {title} {\bibinfo {title} {Quantum {C}omputing in the {NISQ} era and beyond},\ }\href {https://doi.org/10.22331/q-2018-08-06-79} {\bibfield  {journal} {\bibinfo  {journal} {{Quantum}}\ }\textbf {\bibinfo {volume} {2}},\ \bibinfo {pages} {79} (\bibinfo {year} {2018})}\BibitemShut {NoStop}%
\bibitem [{\citenamefont {Akahoshi}\ \emph {et~al.}(2024)\citenamefont {Akahoshi}, \citenamefont {Maruyama}, \citenamefont {Oshima}, \citenamefont {Sato},\ and\ \citenamefont {Fujii}}]{PRXQuantum.5.010337}%
  \BibitemOpen
  \bibfield  {author} {\bibinfo {author} {\bibfnamefont {Y.}~\bibnamefont {Akahoshi}}, \bibinfo {author} {\bibfnamefont {K.}~\bibnamefont {Maruyama}}, \bibinfo {author} {\bibfnamefont {H.}~\bibnamefont {Oshima}}, \bibinfo {author} {\bibfnamefont {S.}~\bibnamefont {Sato}},\ and\ \bibinfo {author} {\bibfnamefont {K.}~\bibnamefont {Fujii}},\ }\bibfield  {title} {\bibinfo {title} {Partially fault-tolerant quantum computing architecture with error-corrected clifford gates and space-time efficient analog rotations},\ }\href {https://doi.org/10.1103/PRXQuantum.5.010337} {\bibfield  {journal} {\bibinfo  {journal} {PRX Quantum}\ }\textbf {\bibinfo {volume} {5}},\ \bibinfo {pages} {010337} (\bibinfo {year} {2024})}\BibitemShut {NoStop}%
\bibitem [{\citenamefont {Toshio}\ \emph {et~al.}(2025)\citenamefont {Toshio}, \citenamefont {Akahoshi}, \citenamefont {Fujisaki}, \citenamefont {Oshima}, \citenamefont {Sato},\ and\ \citenamefont {Fujii}}]{PhysRevX.15.021057}%
  \BibitemOpen
  \bibfield  {author} {\bibinfo {author} {\bibfnamefont {R.}~\bibnamefont {Toshio}}, \bibinfo {author} {\bibfnamefont {Y.}~\bibnamefont {Akahoshi}}, \bibinfo {author} {\bibfnamefont {J.}~\bibnamefont {Fujisaki}}, \bibinfo {author} {\bibfnamefont {H.}~\bibnamefont {Oshima}}, \bibinfo {author} {\bibfnamefont {S.}~\bibnamefont {Sato}},\ and\ \bibinfo {author} {\bibfnamefont {K.}~\bibnamefont {Fujii}},\ }\bibfield  {title} {\bibinfo {title} {Practical quantum advantage on partially fault-tolerant quantum computer},\ }\href {https://doi.org/10.1103/PhysRevX.15.021057} {\bibfield  {journal} {\bibinfo  {journal} {Phys. Rev. X}\ }\textbf {\bibinfo {volume} {15}},\ \bibinfo {pages} {021057} (\bibinfo {year} {2025})}\BibitemShut {NoStop}%
\bibitem [{\citenamefont {Toshio}\ \emph {et~al.}(2026)\citenamefont {Toshio}, \citenamefont {Kanasugi}, \citenamefont {Fujisaki}, \citenamefont {Oshima}, \citenamefont {Sato},\ and\ \citenamefont {Fujii}}]{starv3}%
  \BibitemOpen
  \bibfield  {author} {\bibinfo {author} {\bibfnamefont {R.}~\bibnamefont {Toshio}}, \bibinfo {author} {\bibfnamefont {S.}~\bibnamefont {Kanasugi}}, \bibinfo {author} {\bibfnamefont {J.}~\bibnamefont {Fujisaki}}, \bibinfo {author} {\bibfnamefont {H.}~\bibnamefont {Oshima}}, \bibinfo {author} {\bibfnamefont {S.}~\bibnamefont {Sato}},\ and\ \bibinfo {author} {\bibfnamefont {K.}~\bibnamefont {Fujii}},\ }\href {https://arxiv.org/abs/2603.22891} {\bibinfo {title} {{STAR}-magic mutation: Even more efficient analog rotation gates for early fault-tolerant quantum computer}} (\bibinfo {year} {2026}),\ \Eprint {https://arxiv.org/abs/2603.22891} {arXiv:2603.22891 [quant-ph]} \BibitemShut {NoStop}%
\bibitem [{\citenamefont {Campbell}(2021)}]{Campbell_2022}%
  \BibitemOpen
  \bibfield  {author} {\bibinfo {author} {\bibfnamefont {E.~T.}\ \bibnamefont {Campbell}},\ }\bibfield  {title} {\bibinfo {title} {Early fault-tolerant simulations of the {Hubbard} model},\ }\href {https://doi.org/10.1088/2058-9565/ac3110} {\bibfield  {journal} {\bibinfo  {journal} {Quantum Science and Technology}\ }\textbf {\bibinfo {volume} {7}},\ \bibinfo {pages} {015007} (\bibinfo {year} {2021})}\BibitemShut {NoStop}%
\bibitem [{\citenamefont {Daley}\ \emph {et~al.}(2022)\citenamefont {Daley}, \citenamefont {Bloch}, \citenamefont {Kokail}, \citenamefont {Flannigan}, \citenamefont {Pearson}, \citenamefont {Troyer},\ and\ \citenamefont {Zoller}}]{daley2022practical}%
  \BibitemOpen
  \bibfield  {author} {\bibinfo {author} {\bibfnamefont {A.~J.}\ \bibnamefont {Daley}}, \bibinfo {author} {\bibfnamefont {I.}~\bibnamefont {Bloch}}, \bibinfo {author} {\bibfnamefont {C.}~\bibnamefont {Kokail}}, \bibinfo {author} {\bibfnamefont {S.}~\bibnamefont {Flannigan}}, \bibinfo {author} {\bibfnamefont {N.}~\bibnamefont {Pearson}}, \bibinfo {author} {\bibfnamefont {M.}~\bibnamefont {Troyer}},\ and\ \bibinfo {author} {\bibfnamefont {P.}~\bibnamefont {Zoller}},\ }\bibfield  {title} {\bibinfo {title} {Practical quantum advantage in quantum simulation},\ }\href {https://doi.org/10.1038/s41586-022-04940-6} {\bibfield  {journal} {\bibinfo  {journal} {Nature}\ }\textbf {\bibinfo {volume} {607}},\ \bibinfo {pages} {667} (\bibinfo {year} {2022})}\BibitemShut {NoStop}%
\bibitem [{\citenamefont {Lin}\ and\ \citenamefont {Tong}(2022)}]{lin2022heisenbergeftqc}%
  \BibitemOpen
  \bibfield  {author} {\bibinfo {author} {\bibfnamefont {L.}~\bibnamefont {Lin}}\ and\ \bibinfo {author} {\bibfnamefont {Y.}~\bibnamefont {Tong}},\ }\bibfield  {title} {\bibinfo {title} {Heisenberg-limited ground-state energy estimation for early fault-tolerant quantum computers},\ }\href {https://doi.org/10.1103/PRXQuantum.3.010318} {\bibfield  {journal} {\bibinfo  {journal} {PRX Quantum}\ }\textbf {\bibinfo {volume} {3}},\ \bibinfo {pages} {010318} (\bibinfo {year} {2022})}\BibitemShut {NoStop}%
\bibitem [{\citenamefont {Bravyi}\ and\ \citenamefont {Kitaev}(2005)}]{PhysRevA.71.022316}%
  \BibitemOpen
  \bibfield  {author} {\bibinfo {author} {\bibfnamefont {S.}~\bibnamefont {Bravyi}}\ and\ \bibinfo {author} {\bibfnamefont {A.}~\bibnamefont {Kitaev}},\ }\bibfield  {title} {\bibinfo {title} {Universal quantum computation with ideal clifford gates and noisy ancillas},\ }\href {https://doi.org/10.1103/PhysRevA.71.022316} {\bibfield  {journal} {\bibinfo  {journal} {Phys. Rev. A}\ }\textbf {\bibinfo {volume} {71}},\ \bibinfo {pages} {022316} (\bibinfo {year} {2005})}\BibitemShut {NoStop}%
\bibitem [{\citenamefont {Fowler}\ \emph {et~al.}(2012)\citenamefont {Fowler}, \citenamefont {Mariantoni}, \citenamefont {Martinis},\ and\ \citenamefont {Cleland}}]{PhysRevA.86.032324}%
  \BibitemOpen
  \bibfield  {author} {\bibinfo {author} {\bibfnamefont {A.~G.}\ \bibnamefont {Fowler}}, \bibinfo {author} {\bibfnamefont {M.}~\bibnamefont {Mariantoni}}, \bibinfo {author} {\bibfnamefont {J.~M.}\ \bibnamefont {Martinis}},\ and\ \bibinfo {author} {\bibfnamefont {A.~N.}\ \bibnamefont {Cleland}},\ }\bibfield  {title} {\bibinfo {title} {Surface codes: Towards practical large-scale quantum computation},\ }\href {https://doi.org/10.1103/PhysRevA.86.032324} {\bibfield  {journal} {\bibinfo  {journal} {Phys. Rev. A}\ }\textbf {\bibinfo {volume} {86}},\ \bibinfo {pages} {032324} (\bibinfo {year} {2012})}\BibitemShut {NoStop}%
\bibitem [{\citenamefont {Zhou}\ \emph {et~al.}(2025{\natexlab{a}})\citenamefont {Zhou}, \citenamefont {Duckering}, \citenamefont {Zhao}, \citenamefont {Bluvstein}, \citenamefont {Cain}, \citenamefont {Kubica}, \citenamefont {Wang},\ and\ \citenamefont {Lukin}}]{harry2025resource_estimation}%
  \BibitemOpen
  \bibfield  {author} {\bibinfo {author} {\bibfnamefont {H.}~\bibnamefont {Zhou}}, \bibinfo {author} {\bibfnamefont {C.}~\bibnamefont {Duckering}}, \bibinfo {author} {\bibfnamefont {C.}~\bibnamefont {Zhao}}, \bibinfo {author} {\bibfnamefont {D.}~\bibnamefont {Bluvstein}}, \bibinfo {author} {\bibfnamefont {M.}~\bibnamefont {Cain}}, \bibinfo {author} {\bibfnamefont {A.}~\bibnamefont {Kubica}}, \bibinfo {author} {\bibfnamefont {S.-T.}\ \bibnamefont {Wang}},\ and\ \bibinfo {author} {\bibfnamefont {M.~D.}\ \bibnamefont {Lukin}},\ }\bibfield  {title} {\bibinfo {title} {Resource analysis of low-overhead transversal architectures for reconfigurable atom arrays},\ }in\ \href {https://doi.org/10.1145/3695053.3731039} {\emph {\bibinfo {booktitle} {Proceedings of the 52nd Annual International Symposium on Computer Architecture}}},\ \bibinfo {series and number} {ISCA '25}\ (\bibinfo  {publisher} {Association for Computing Machinery},\ \bibinfo {address} {New York, NY, USA},\ \bibinfo {year} {2025})\ p.\ \bibinfo {pages}
  {1432–1448}\BibitemShut {NoStop}%
\bibitem [{\citenamefont {Zhou}\ \emph {et~al.}(2025{\natexlab{b}})\citenamefont {Zhou}, \citenamefont {Zhao}, \citenamefont {Cain}, \citenamefont {Bluvstein}, \citenamefont {Maskara}, \citenamefont {Duckering}, \citenamefont {Hu}, \citenamefont {Wang}, \citenamefont {Kubica},\ and\ \citenamefont {Lukin}}]{Zhou_2025}%
  \BibitemOpen
  \bibfield  {author} {\bibinfo {author} {\bibfnamefont {H.}~\bibnamefont {Zhou}}, \bibinfo {author} {\bibfnamefont {C.}~\bibnamefont {Zhao}}, \bibinfo {author} {\bibfnamefont {M.}~\bibnamefont {Cain}}, \bibinfo {author} {\bibfnamefont {D.}~\bibnamefont {Bluvstein}}, \bibinfo {author} {\bibfnamefont {N.}~\bibnamefont {Maskara}}, \bibinfo {author} {\bibfnamefont {C.}~\bibnamefont {Duckering}}, \bibinfo {author} {\bibfnamefont {H.-Y.}\ \bibnamefont {Hu}}, \bibinfo {author} {\bibfnamefont {S.-T.}\ \bibnamefont {Wang}}, \bibinfo {author} {\bibfnamefont {A.}~\bibnamefont {Kubica}},\ and\ \bibinfo {author} {\bibfnamefont {M.~D.}\ \bibnamefont {Lukin}},\ }\bibfield  {title} {\bibinfo {title} {Low-overhead transversal fault tolerance for universal quantum computation},\ }\href {https://doi.org/10.1038/s41586-025-09543-5} {\bibfield  {journal} {\bibinfo  {journal} {Nature}\ }\textbf {\bibinfo {volume} {646}},\ \bibinfo {pages} {303–308} (\bibinfo {year} {2025}{\natexlab{b}})}\BibitemShut {NoStop}%
\bibitem [{\citenamefont {Litinski}(2019{\natexlab{b}})}]{Litinski2019magicstate}%
  \BibitemOpen
  \bibfield  {author} {\bibinfo {author} {\bibfnamefont {D.}~\bibnamefont {Litinski}},\ }\bibfield  {title} {\bibinfo {title} {Magic {S}tate {D}istillation: {N}ot as {C}ostly as {Y}ou {T}hink},\ }\href {https://doi.org/10.22331/q-2019-12-02-205} {\bibfield  {journal} {\bibinfo  {journal} {{Quantum}}\ }\textbf {\bibinfo {volume} {3}},\ \bibinfo {pages} {205} (\bibinfo {year} {2019}{\natexlab{b}})}\BibitemShut {NoStop}%
\bibitem [{\citenamefont {Bravyi}\ and\ \citenamefont {Haah}(2012)}]{PhysRevA.86.052329}%
  \BibitemOpen
  \bibfield  {author} {\bibinfo {author} {\bibfnamefont {S.}~\bibnamefont {Bravyi}}\ and\ \bibinfo {author} {\bibfnamefont {J.}~\bibnamefont {Haah}},\ }\bibfield  {title} {\bibinfo {title} {Magic-state distillation with low overhead},\ }\href {https://doi.org/10.1103/PhysRevA.86.052329} {\bibfield  {journal} {\bibinfo  {journal} {Phys. Rev. A}\ }\textbf {\bibinfo {volume} {86}},\ \bibinfo {pages} {052329} (\bibinfo {year} {2012})}\BibitemShut {NoStop}%
\bibitem [{\citenamefont {Gidney}\ and\ \citenamefont {Fowler}(2019)}]{Gidney2019efficientmagicstate}%
  \BibitemOpen
  \bibfield  {author} {\bibinfo {author} {\bibfnamefont {C.}~\bibnamefont {Gidney}}\ and\ \bibinfo {author} {\bibfnamefont {A.~G.}\ \bibnamefont {Fowler}},\ }\bibfield  {title} {\bibinfo {title} {Efficient magic state factories with a catalyzed {$|CCZ\rangle$} to {$2|T\rangle$} transformation},\ }\href {https://doi.org/10.22331/q-2019-04-30-135} {\bibfield  {journal} {\bibinfo  {journal} {{Quantum}}\ }\textbf {\bibinfo {volume} {3}},\ \bibinfo {pages} {135} (\bibinfo {year} {2019})}\BibitemShut {NoStop}%
\bibitem [{\citenamefont {O'Gorman}\ and\ \citenamefont {Campbell}(2017)}]{PhysRevA.95.032338}%
  \BibitemOpen
  \bibfield  {author} {\bibinfo {author} {\bibfnamefont {J.}~\bibnamefont {O'Gorman}}\ and\ \bibinfo {author} {\bibfnamefont {E.~T.}\ \bibnamefont {Campbell}},\ }\bibfield  {title} {\bibinfo {title} {Quantum computation with realistic magic-state factories},\ }\href {https://doi.org/10.1103/PhysRevA.95.032338} {\bibfield  {journal} {\bibinfo  {journal} {Phys. Rev. A}\ }\textbf {\bibinfo {volume} {95}},\ \bibinfo {pages} {032338} (\bibinfo {year} {2017})}\BibitemShut {NoStop}%
\bibitem [{\citenamefont {Ross}\ and\ \citenamefont {Selinger}(2016)}]{rzsynthesis}%
  \BibitemOpen
  \bibfield  {author} {\bibinfo {author} {\bibfnamefont {N.~J.}\ \bibnamefont {Ross}}\ and\ \bibinfo {author} {\bibfnamefont {P.}~\bibnamefont {Selinger}},\ }\bibfield  {title} {\bibinfo {title} {Optimal ancilla-free clifford+t approximation of z-rotations},\ }\href@noop {} {\bibfield  {journal} {\bibinfo  {journal} {Quantum Info. Comput.}\ }\textbf {\bibinfo {volume} {16}},\ \bibinfo {pages} {901} (\bibinfo {year} {2016})}\BibitemShut {NoStop}%
\bibitem [{\citenamefont {Tan}\ \emph {et~al.}(2025)\citenamefont {Tan}, \citenamefont {Lin},\ and\ \citenamefont {Cong}}]{tan2024enola}%
  \BibitemOpen
  \bibfield  {author} {\bibinfo {author} {\bibfnamefont {D.~B.}\ \bibnamefont {Tan}}, \bibinfo {author} {\bibfnamefont {W.-H.}\ \bibnamefont {Lin}},\ and\ \bibinfo {author} {\bibfnamefont {J.}~\bibnamefont {Cong}},\ }\bibfield  {title} {\bibinfo {title} {Compilation for dynamically field-programmable qubit arrays with efficient and provably near-optimal scheduling},\ }in\ \href@noop {} {\emph {\bibinfo {booktitle} {Proceedings of the 30th Asia and South Pacific Design Automation Conference}}}\ (\bibinfo {year} {2025})\ pp.\ \bibinfo {pages} {921--929}\BibitemShut {NoStop}%
\bibitem [{\citenamefont {Lin}\ \emph {et~al.}(2025)\citenamefont {Lin}, \citenamefont {Tan},\ and\ \citenamefont {Cong}}]{zac}%
  \BibitemOpen
  \bibfield  {author} {\bibinfo {author} {\bibfnamefont {W.-H.}\ \bibnamefont {Lin}}, \bibinfo {author} {\bibfnamefont {D.~B.}\ \bibnamefont {Tan}},\ and\ \bibinfo {author} {\bibfnamefont {J.}~\bibnamefont {Cong}},\ }\bibfield  {title} {\bibinfo {title} {Reuse-aware compilation for zoned quantum architectures based on neutral atoms},\ }in\ \href {https://doi.org/10.1109/HPCA61900.2025.00021} {\emph {\bibinfo {booktitle} {2025 IEEE International Symposium on High Performance Computer Architecture (HPCA)}}}\ (\bibinfo {year} {2025})\ pp.\ \bibinfo {pages} {127--142}\BibitemShut {NoStop}%
\bibitem [{\citenamefont {Jonker}\ and\ \citenamefont {Volgenant}(1987)}]{jonker1988shortest}%
  \BibitemOpen
  \bibfield  {author} {\bibinfo {author} {\bibfnamefont {R.}~\bibnamefont {Jonker}}\ and\ \bibinfo {author} {\bibfnamefont {A.}~\bibnamefont {Volgenant}},\ }\bibfield  {title} {\bibinfo {title} {A shortest augmenting path algorithm for dense and sparse linear assignment problems},\ }\href {https://doi.org/10.1007/BF02278710} {\bibfield  {journal} {\bibinfo  {journal} {Computing}\ }\textbf {\bibinfo {volume} {38}},\ \bibinfo {pages} {325–340} (\bibinfo {year} {1987})}\BibitemShut {NoStop}%
\bibitem [{\citenamefont {Crouse}(2016)}]{2dassignment}%
  \BibitemOpen
  \bibfield  {author} {\bibinfo {author} {\bibfnamefont {D.~F.}\ \bibnamefont {Crouse}},\ }\bibfield  {title} {\bibinfo {title} {On implementing {2D} rectangular assignment algorithms},\ }\href {https://doi.org/10.1109/TAES.2016.140952} {\bibfield  {journal} {\bibinfo  {journal} {IEEE Transactions on Aerospace and Electronic Systems}\ }\textbf {\bibinfo {volume} {52}},\ \bibinfo {pages} {1679} (\bibinfo {year} {2016})}\BibitemShut {NoStop}%
\bibitem [{\citenamefont {Jellum}\ \emph {et~al.}(2022)\citenamefont {Jellum}, \citenamefont {Orlandi{\'c}}, \citenamefont {Brekke}, \citenamefont {Johansen},\ and\ \citenamefont {Bryne}}]{10.1145/3546072}%
  \BibitemOpen
  \bibfield  {author} {\bibinfo {author} {\bibfnamefont {E.}~\bibnamefont {Jellum}}, \bibinfo {author} {\bibfnamefont {M.}~\bibnamefont {Orlandi{\'c}}}, \bibinfo {author} {\bibfnamefont {E.}~\bibnamefont {Brekke}}, \bibinfo {author} {\bibfnamefont {T.}~\bibnamefont {Johansen}},\ and\ \bibinfo {author} {\bibfnamefont {T.}~\bibnamefont {Bryne}},\ }\bibfield  {title} {\bibinfo {title} {Solving sparse assignment problems on fpgas},\ }\bibfield  {journal} {\bibinfo  {journal} {ACM Trans. Archit. Code Optim.}\ }\textbf {\bibinfo {volume} {19}},\ \href {https://doi.org/10.1145/3546072} {10.1145/3546072} (\bibinfo {year} {2022})\BibitemShut {NoStop}%
\bibitem [{\citenamefont {Ismail}\ \emph {et~al.}(2026{\natexlab{b}})\citenamefont {Ismail}, \citenamefont {Kornjača}, \citenamefont {Hu}, \citenamefont {Maskara}, \citenamefont {Wang}, \citenamefont {Zhou},\ and\ \citenamefont {Zhao}}]{ismail2026fastparallelhighratestar}%
  \BibitemOpen
  \bibfield  {author} {\bibinfo {author} {\bibfnamefont {R.}~\bibnamefont {Ismail}}, \bibinfo {author} {\bibfnamefont {M.}~\bibnamefont {Kornjača}}, \bibinfo {author} {\bibfnamefont {H.-Y.}\ \bibnamefont {Hu}}, \bibinfo {author} {\bibfnamefont {N.}~\bibnamefont {Maskara}}, \bibinfo {author} {\bibfnamefont {S.-T.}\ \bibnamefont {Wang}}, \bibinfo {author} {\bibfnamefont {H.}~\bibnamefont {Zhou}},\ and\ \bibinfo {author} {\bibfnamefont {C.}~\bibnamefont {Zhao}},\ }\href {https://arxiv.org/abs/2606.25011} {\bibinfo {title} {Fast and parallel high-rate star architecture for megaquop quantum simulation}} (\bibinfo {year} {2026}{\natexlab{b}}),\ \Eprint {https://arxiv.org/abs/2606.25011} {arXiv:2606.25011 [quant-ph]} \BibitemShut {NoStop}%
\bibitem [{\citenamefont {Yang}\ \emph {et~al.}(2026)\citenamefont {Yang}, \citenamefont {Chadwick}, \citenamefont {Teo}, \citenamefont {Viszlai},\ and\ \citenamefont {Chong}}]{yang2026spacetimeefficienthardwarecompatiblecomplexquantum}%
  \BibitemOpen
  \bibfield  {author} {\bibinfo {author} {\bibfnamefont {W.}~\bibnamefont {Yang}}, \bibinfo {author} {\bibfnamefont {J.}~\bibnamefont {Chadwick}}, \bibinfo {author} {\bibfnamefont {M.~H.}\ \bibnamefont {Teo}}, \bibinfo {author} {\bibfnamefont {J.}~\bibnamefont {Viszlai}},\ and\ \bibinfo {author} {\bibfnamefont {F.}~\bibnamefont {Chong}},\ }\href {https://arxiv.org/abs/2602.14273} {\bibinfo {title} {Spacetime-efficient and hardware-compatible complex quantum logic units in {qLDPC} codes}} (\bibinfo {year} {2026}),\ \Eprint {https://arxiv.org/abs/2602.14273} {arXiv:2602.14273 [quant-ph]} \BibitemShut {NoStop}%
\bibitem [{\citenamefont {Bhardwaj}\ \emph {et~al.}(2026)\citenamefont {Bhardwaj}, \citenamefont {Ma}, \citenamefont {Meister}, \citenamefont {King}, \citenamefont {Bluvstein}, \citenamefont {Preskill}, \citenamefont {Cain}, \citenamefont {Xu},\ and\ \citenamefont {Huang}}]{bhardwaj2026highrateqldpcprocessors}%
  \BibitemOpen
  \bibfield  {author} {\bibinfo {author} {\bibfnamefont {A.}~\bibnamefont {Bhardwaj}}, \bibinfo {author} {\bibfnamefont {M.}~\bibnamefont {Ma}}, \bibinfo {author} {\bibfnamefont {N.}~\bibnamefont {Meister}}, \bibinfo {author} {\bibfnamefont {R.}~\bibnamefont {King}}, \bibinfo {author} {\bibfnamefont {D.}~\bibnamefont {Bluvstein}}, \bibinfo {author} {\bibfnamefont {J.}~\bibnamefont {Preskill}}, \bibinfo {author} {\bibfnamefont {M.}~\bibnamefont {Cain}}, \bibinfo {author} {\bibfnamefont {Q.}~\bibnamefont {Xu}},\ and\ \bibinfo {author} {\bibfnamefont {H.-Y.}\ \bibnamefont {Huang}},\ }\href {https://arxiv.org/abs/2607.28795} {\bibinfo {title} {High-rate {qLDPC} processors}} (\bibinfo {year} {2026}),\ \Eprint {https://arxiv.org/abs/2607.28795} {arXiv:2607.28795 [quant-ph]} \BibitemShut {NoStop}%
\bibitem [{\citenamefont {Xu}\ \emph {et~al.}(2024)\citenamefont {Xu}, \citenamefont {Bonilla~Ataides}, \citenamefont {Pattison}, \citenamefont {Raveendran}, \citenamefont {Bluvstein}, \citenamefont {Wurtz}, \citenamefont {Vasi{\'c}}, \citenamefont {Lukin}, \citenamefont {Jiang},\ and\ \citenamefont {Zhou}}]{Xu2024ConstantOverhead}%
  \BibitemOpen
  \bibfield  {author} {\bibinfo {author} {\bibfnamefont {Q.}~\bibnamefont {Xu}}, \bibinfo {author} {\bibfnamefont {J.~P.}\ \bibnamefont {Bonilla~Ataides}}, \bibinfo {author} {\bibfnamefont {C.~A.}\ \bibnamefont {Pattison}}, \bibinfo {author} {\bibfnamefont {N.}~\bibnamefont {Raveendran}}, \bibinfo {author} {\bibfnamefont {D.}~\bibnamefont {Bluvstein}}, \bibinfo {author} {\bibfnamefont {J.}~\bibnamefont {Wurtz}}, \bibinfo {author} {\bibfnamefont {B.}~\bibnamefont {Vasi{\'c}}}, \bibinfo {author} {\bibfnamefont {M.~D.}\ \bibnamefont {Lukin}}, \bibinfo {author} {\bibfnamefont {L.}~\bibnamefont {Jiang}},\ and\ \bibinfo {author} {\bibfnamefont {H.}~\bibnamefont {Zhou}},\ }\bibfield  {title} {\bibinfo {title} {Constant-overhead fault-tolerant quantum computation with reconfigurable atom arrays},\ }\href {https://doi.org/10.1038/s41567-024-02479-z} {\bibfield  {journal} {\bibinfo  {journal} {Nature Physics}\ }\textbf {\bibinfo {volume} {20}},\ \bibinfo {pages} {1084} (\bibinfo {year} {2024})}\BibitemShut {NoStop}%
\bibitem [{\citenamefont {Gu}\ \emph {et~al.}(2026)\citenamefont {Gu}, \citenamefont {Liu}, \citenamefont {Quintavalle}, \citenamefont {Xu}, \citenamefont {Eisert},\ and\ \citenamefont {Roffe}}]{gu2026qgpuparallellogicquantum}%
  \BibitemOpen
  \bibfield  {author} {\bibinfo {author} {\bibfnamefont {B.}~\bibnamefont {Gu}}, \bibinfo {author} {\bibfnamefont {A.~Z.}\ \bibnamefont {Liu}}, \bibinfo {author} {\bibfnamefont {A.~O.}\ \bibnamefont {Quintavalle}}, \bibinfo {author} {\bibfnamefont {Q.}~\bibnamefont {Xu}}, \bibinfo {author} {\bibfnamefont {J.}~\bibnamefont {Eisert}},\ and\ \bibinfo {author} {\bibfnamefont {J.}~\bibnamefont {Roffe}},\ }\href {https://arxiv.org/abs/2603.05398} {\bibinfo {title} {{QGPU}: Parallel logic in quantum {LDPC} codes}} (\bibinfo {year} {2026}),\ \Eprint {https://arxiv.org/abs/2603.05398} {arXiv:2603.05398 [quant-ph]} \BibitemShut {NoStop}%
\bibitem [{\citenamefont {Yoder}\ \emph {et~al.}(2025)\citenamefont {Yoder}, \citenamefont {Schoute}, \citenamefont {Rall}, \citenamefont {Pritchett}, \citenamefont {Gambetta}, \citenamefont {Cross}, \citenamefont {Carroll},\ and\ \citenamefont {Beverland}}]{yoder2025tourgrossmodularquantum}%
  \BibitemOpen
  \bibfield  {author} {\bibinfo {author} {\bibfnamefont {T.~J.}\ \bibnamefont {Yoder}}, \bibinfo {author} {\bibfnamefont {E.}~\bibnamefont {Schoute}}, \bibinfo {author} {\bibfnamefont {P.}~\bibnamefont {Rall}}, \bibinfo {author} {\bibfnamefont {E.}~\bibnamefont {Pritchett}}, \bibinfo {author} {\bibfnamefont {J.~M.}\ \bibnamefont {Gambetta}}, \bibinfo {author} {\bibfnamefont {A.~W.}\ \bibnamefont {Cross}}, \bibinfo {author} {\bibfnamefont {M.}~\bibnamefont {Carroll}},\ and\ \bibinfo {author} {\bibfnamefont {M.~E.}\ \bibnamefont {Beverland}},\ }\href {https://arxiv.org/abs/2506.03094} {\bibinfo {title} {Tour de gross: A modular quantum computer based on bivariate bicycle codes}} (\bibinfo {year} {2025}),\ \Eprint {https://arxiv.org/abs/2506.03094} {arXiv:2506.03094 [quant-ph]} \BibitemShut {NoStop}%
\bibitem [{\citenamefont {Liu}\ \emph {et~al.}(2026)\citenamefont {Liu}, \citenamefont {Foxman}, \citenamefont {Anselmetti},\ and\ \citenamefont {Ding}}]{liu2026assessingcapabilitiesbottlenecksearly}%
  \BibitemOpen
  \bibfield  {author} {\bibinfo {author} {\bibfnamefont {K.}~\bibnamefont {Liu}}, \bibinfo {author} {\bibfnamefont {B.}~\bibnamefont {Foxman}}, \bibinfo {author} {\bibfnamefont {G.-L.~R.}\ \bibnamefont {Anselmetti}},\ and\ \bibinfo {author} {\bibfnamefont {Y.}~\bibnamefont {Ding}},\ }\href {https://arxiv.org/abs/2604.20013} {\bibinfo {title} {Assessing system capabilities and bottlenecks of an early fault-tolerant bicycle architecture}} (\bibinfo {year} {2026}),\ \Eprint {https://arxiv.org/abs/2604.20013} {arXiv:2604.20013 [quant-ph]} \BibitemShut {NoStop}%
\bibitem [{\citenamefont {Webster}\ \emph {et~al.}(2026)\citenamefont {Webster}, \citenamefont {Berent}, \citenamefont {Chandra}, \citenamefont {Hockings}, \citenamefont {Baspin}, \citenamefont {Thomsen}, \citenamefont {Smith},\ and\ \citenamefont {Cohen}}]{webster2026pinnaclearchitecturereducingcost}%
  \BibitemOpen
  \bibfield  {author} {\bibinfo {author} {\bibfnamefont {P.}~\bibnamefont {Webster}}, \bibinfo {author} {\bibfnamefont {L.}~\bibnamefont {Berent}}, \bibinfo {author} {\bibfnamefont {O.}~\bibnamefont {Chandra}}, \bibinfo {author} {\bibfnamefont {E.~T.}\ \bibnamefont {Hockings}}, \bibinfo {author} {\bibfnamefont {N.}~\bibnamefont {Baspin}}, \bibinfo {author} {\bibfnamefont {F.}~\bibnamefont {Thomsen}}, \bibinfo {author} {\bibfnamefont {S.~C.}\ \bibnamefont {Smith}},\ and\ \bibinfo {author} {\bibfnamefont {L.~Z.}\ \bibnamefont {Cohen}},\ }\href {https://arxiv.org/abs/2602.11457} {\bibinfo {title} {The {Pinnacle} architecture: Reducing the cost of breaking {RSA}-2048 to 100000 physical qubits using quantum {LDPC} codes}} (\bibinfo {year} {2026}),\ \Eprint {https://arxiv.org/abs/2602.11457} {arXiv:2602.11457 [quant-ph]} \BibitemShut {NoStop}%
\bibitem [{\citenamefont {Ekert}\ and\ \citenamefont {Jozsa}(1996)}]{RevModPhys.68.733}%
  \BibitemOpen
  \bibfield  {author} {\bibinfo {author} {\bibfnamefont {A.}~\bibnamefont {Ekert}}\ and\ \bibinfo {author} {\bibfnamefont {R.}~\bibnamefont {Jozsa}},\ }\bibfield  {title} {\bibinfo {title} {Quantum computation and shor's factoring algorithm},\ }\href {https://doi.org/10.1103/RevModPhys.68.733} {\bibfield  {journal} {\bibinfo  {journal} {Rev. Mod. Phys.}\ }\textbf {\bibinfo {volume} {68}},\ \bibinfo {pages} {733} (\bibinfo {year} {1996})}\BibitemShut {NoStop}%
\bibitem [{\citenamefont {Proos}\ and\ \citenamefont {Zalka}(2004)}]{proos2004shorsdiscretelogarithmquantum}%
  \BibitemOpen
  \bibfield  {author} {\bibinfo {author} {\bibfnamefont {J.}~\bibnamefont {Proos}}\ and\ \bibinfo {author} {\bibfnamefont {C.}~\bibnamefont {Zalka}},\ }\href {https://arxiv.org/abs/quant-ph/0301141} {\bibinfo {title} {Shor's discrete logarithm quantum algorithm for elliptic curves}} (\bibinfo {year} {2004}),\ \Eprint {https://arxiv.org/abs/quant-ph/0301141} {arXiv:quant-ph/0301141 [quant-ph]} \BibitemShut {NoStop}%
\bibitem [{\citenamefont {Roetteler}\ \emph {et~al.}(2017)\citenamefont {Roetteler}, \citenamefont {Naehrig}, \citenamefont {Svore},\ and\ \citenamefont {Lauter}}]{10.1007/978-3-319-70697-9_9}%
  \BibitemOpen
  \bibfield  {author} {\bibinfo {author} {\bibfnamefont {M.}~\bibnamefont {Roetteler}}, \bibinfo {author} {\bibfnamefont {M.}~\bibnamefont {Naehrig}}, \bibinfo {author} {\bibfnamefont {K.~M.}\ \bibnamefont {Svore}},\ and\ \bibinfo {author} {\bibfnamefont {K.}~\bibnamefont {Lauter}},\ }\bibfield  {title} {\bibinfo {title} {Quantum resource estimates for computing elliptic curve discrete logarithms},\ }in\ \href@noop {} {\emph {\bibinfo {booktitle} {Advances in Cryptology -- ASIACRYPT 2017}}},\ \bibinfo {editor} {edited by\ \bibinfo {editor} {\bibfnamefont {T.}~\bibnamefont {Takagi}}\ and\ \bibinfo {editor} {\bibfnamefont {T.}~\bibnamefont {Peyrin}}}\ (\bibinfo  {publisher} {Springer International Publishing},\ \bibinfo {address} {Cham},\ \bibinfo {year} {2017})\ pp.\ \bibinfo {pages} {241--270}\BibitemShut {NoStop}%
\bibitem [{\citenamefont {Low}\ and\ \citenamefont {Chuang}(2019)}]{low2019hamiltonian}%
  \BibitemOpen
  \bibfield  {author} {\bibinfo {author} {\bibfnamefont {G.~H.}\ \bibnamefont {Low}}\ and\ \bibinfo {author} {\bibfnamefont {I.~L.}\ \bibnamefont {Chuang}},\ }\bibfield  {title} {\bibinfo {title} {Hamiltonian simulation by qubitization},\ }\href@noop {} {\bibfield  {journal} {\bibinfo  {journal} {Quantum}\ }\textbf {\bibinfo {volume} {3}},\ \bibinfo {pages} {163} (\bibinfo {year} {2019})}\BibitemShut {NoStop}%
\bibitem [{\citenamefont {Dilworth}(1990)}]{Dilworth1990}%
  \BibitemOpen
  \bibfield  {author} {\bibinfo {author} {\bibfnamefont {R.~P.}\ \bibnamefont {Dilworth}},\ }\bibinfo {title} {A decomposition theorem for partially ordered sets},\ in\ \href {https://doi.org/10.1007/978-1-4899-3558-8_1} {\emph {\bibinfo {booktitle} {The Dilworth Theorems: Selected Papers of Robert P. Dilworth}}}\ (\bibinfo  {publisher} {Birkh{\"a}user Boston},\ \bibinfo {address} {Boston, MA},\ \bibinfo {year} {1990})\ pp.\ \bibinfo {pages} {7--12}\BibitemShut {NoStop}%
\bibitem [{\citenamefont {Edmonds}\ and\ \citenamefont {Karp}(1972)}]{edmonds1972}%
  \BibitemOpen
  \bibfield  {author} {\bibinfo {author} {\bibfnamefont {J.}~\bibnamefont {Edmonds}}\ and\ \bibinfo {author} {\bibfnamefont {R.~M.}\ \bibnamefont {Karp}},\ }\bibfield  {title} {\bibinfo {title} {Theoretical improvements in algorithmic efficiency for network flow problems},\ }\href {https://doi.org/10.1145/321694.321699} {\bibfield  {journal} {\bibinfo  {journal} {J. ACM}\ }\textbf {\bibinfo {volume} {19}},\ \bibinfo {pages} {248–264} (\bibinfo {year} {1972})}\BibitemShut {NoStop}%
\end{thebibliography}%

\end{document}